\documentclass[11pt]{article}%
\usepackage{amsmath}
\usepackage{amsfonts}
\usepackage{amssymb}
\usepackage{graphics}
\usepackage{graphicx}
\usepackage{float}
\usepackage{fullpage}%
\providecommand{\U}[1]{\protect\rule{.1in}{.1in}}
\begin{document}

\title{The Ghost in the Quantum Turing Machine}
\author{Scott Aaronson\thanks{MIT. \ Email: aaronson@csail.mit.edu. \ \ This material
is based upon work supported by the National Science Foundation under Grant
No.\ 0844626, as well as by an NSF STC grant, a TIBCO
Chair, a Sloan Fellowship, and an Alan T.\ Waterman Award.}}
\date{}
\maketitle

\begin{abstract}
In honor of Alan Turing's hundredth birthday, I unwisely set out some thoughts
about one of Turing's obsessions throughout his life, the question of physics
and free will. \ I focus relatively narrowly on a notion that I call
\textquotedblleft Knightian freedom\textquotedblright: a certain kind of
in-principle physical unpredictability that goes beyond probabilistic
unpredictability. Other, more metaphysical aspects of free will\ I regard as
possibly outside the scope of science.

I examine a viewpoint, suggested independently by Carl Hoefer, Cristi Stoica,
and even Turing himself, that tries to find scope for \textquotedblleft
freedom\textquotedblright\ in the universe's boundary conditions rather than
in the dynamical laws. \ Taking this viewpoint seriously leads to many
interesting conceptual problems. \ I investigate how far one can go toward
solving those problems, and along the way, encounter (among other things) the
No-Cloning Theorem, the measurement problem, decoherence, chaos, the arrow of
time, the holographic principle, Newcomb's paradox, Boltzmann brains,
algorithmic information theory, and the Common Prior Assumption.\ \ I also
compare the viewpoint explored here to the more radical\ speculations of Roger Penrose.

The result of all this is an unusual perspective on time, quantum mechanics,
and causation, of which I myself remain skeptical, but which has several
appealing features. \ Among other things, it suggests interesting empirical
questions in neuroscience, physics, and cosmology; and takes a millennia-old
philosophical\ debate into some underexplored territory.

\end{abstract}
\tableofcontents

\pagebreak

\vspace*{2in}

%TCIMACRO{\FRAME{ftbpF}{3.1531in}{2.5071in}{0pt}{}{\Qlb{unseen}}{unseen.jpg}%
%width 3.1531in;  height 2.5071in;  depth 0pt;  original-width 6.25in;
%original-height 4.958in;  cropleft "0";  croptop "1";  cropright "1";
%cropbottom "0";  file_alias "1";  file-properties "XNPEU"; }} }%
%BeginExpansion
\begin{figure}[H]%
\centering
\includegraphics[
natheight=4.958000in,
natwidth=6.250000in,
height=2.5071in,
width=3.1531in
]%
{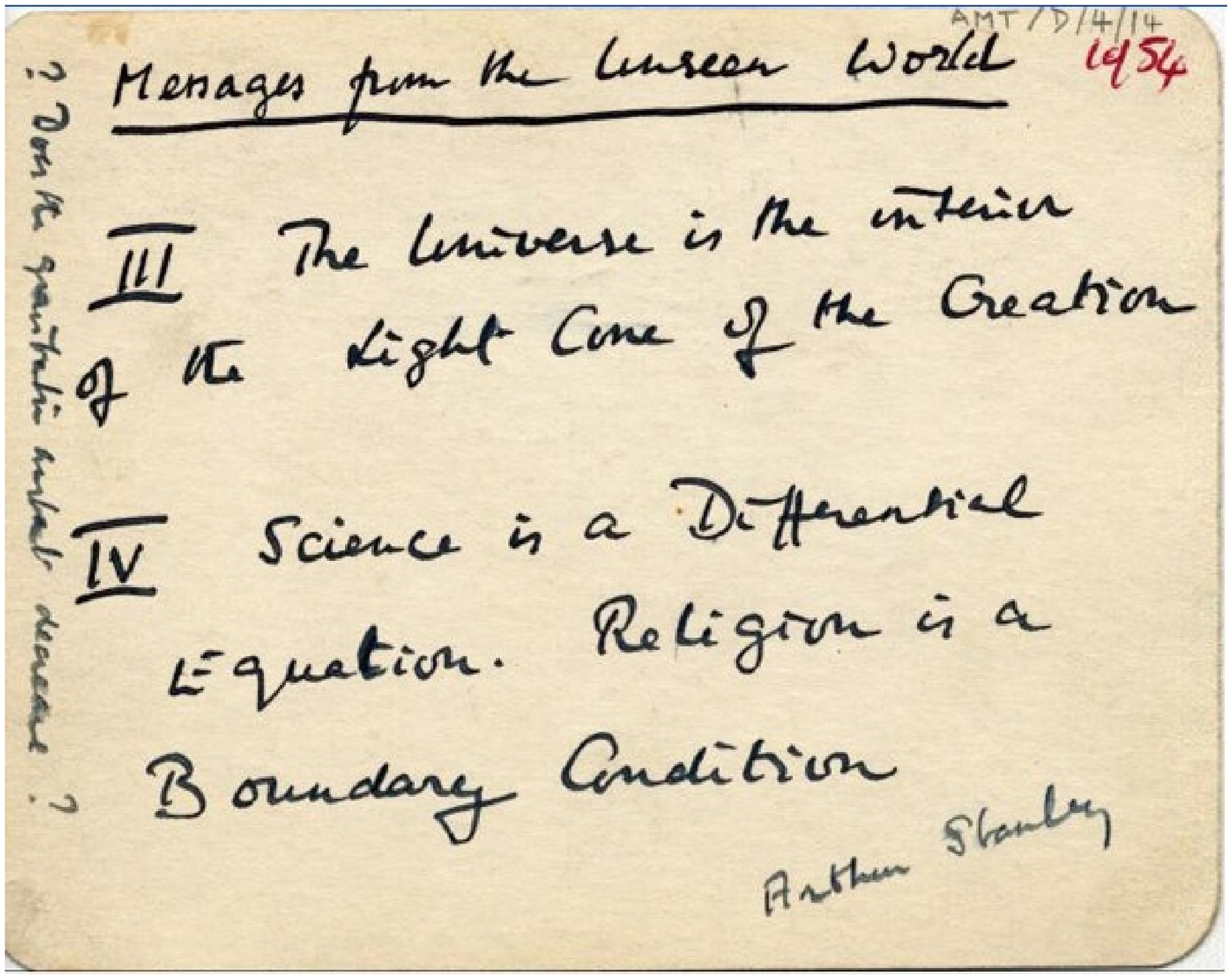}%
\label{unseen}%
\end{figure}
%EndExpansion

Postcard from Alan M.\ Turing to Robin Gandy, March 1954 (reprinted in Hodges
\cite{hodges})

It reads, in part:

\begin{quotation}
\underline{Messages from the Unseen World}

The Universe is the interior of the Light Cone of the Creation

Science is a Differential Equation. \ Religion is a Boundary Condition
\end{quotation}

``Arthur Stanley'' refers to Arthur Stanley Eddington, whose books were a major early influence on Turing.

\pagebreak

\section{Introduction\label{INTRO}}

When I was a teenager, Alan Turing was at the top of my pantheon of scientific
heroes, above even Darwin, Ramanujan, Einstein, and Feynman. \ Some of the
reasons were obvious: the founding of computer science, the proof of the
unsolvability of the Entscheidungsproblem, the breaking of the Nazi Enigma
code, the unapologetic nerdiness and the near-martyrdom for human rights.
\ But beyond the facts of his biography, I idolized Turing as an
\textquotedblleft\"{u}ber-reductionist\textquotedblright: the scientist who
had gone further than anyone before him to reveal the mechanistic nature of
reality. \ Through his discovery of computational universality, as well as the
Turing Test criterion for intelligence, Turing finally unmasked the
pretensions of anyone who claimed there was anything more to mind, brain, or
the physical world than the unfolding of an immense computation. \ After
Turing, it seemed to me, one could assert with confidence that all our hopes,
fears, sensations, and choices were just evanescent patterns in some sort of
\textit{cellular automaton}: that is, a huge array of bits, different in
detail but not in essence from Conway's famous Game of Life,\footnote{Invented
by the mathematician John Conway in 1970, the Game of Life involves a large
two-dimensional array of pixels, with each pixel either \textquotedblleft
live\textquotedblright\ or \textquotedblleft dead.\textquotedblright\ \ At
each (discrete) time step, the pixels get updated via a deterministic rule:
each live pixel \textquotedblleft dies\textquotedblright\ if less than $2$\ or
more than $3$ of its $8$ neighbors were alive, and each dead pixel
\textquotedblleft comes alive\textquotedblright\ if exactly $3$ of its $8$
neighbors were alive. \ `Life' is famous for the complicated, unpredictable
patterns that typically arise from a simple starting configuration and
repeated application of the rules. \ Conway (see \cite{levy}) has expressed
certainty that, on a large enough Life board, living beings would arise, who
would then start squabbling over territory and writing learned PhD theses!
\ Note that, with an \textit{exponentially}-large Life board (and, say, a
uniformly-random initial configuration), Conway's claim is vacuously true, in
the sense that one could find essentially any regularity one wanted just by
chance. \ But one assumes that Conway meant something stronger.} getting
updated in time by simple, local, mechanistic rules.

So it's striking that Turing's own views about these issues, as revealed in
his lectures as well as private correspondence, were much more complicated
than my early caricature. \ As a teenager, Turing devoured the popular books
of Sir Arthur Eddington, who was one of the first (though not, of course, the
last!) to speculate about the implications of the then-ongoing quantum
revolution in physics for ancient questions about mind and free will. \ Later,
as a prize from his high school in 1932, Turing selected John von Neumann's
just-published \textit{Mathematische Grundlagen der Quantenmechanik}
\cite{vonneumann:qm}: a treatise on quantum mechanics famous for its
mathematical rigor, but also for its perspective that the collapse of the
wavefunction ultimately involves the experimenter's mental state. \ As
detailed by Turing biographer Andrew Hodges \cite{hodges}, these early
readings had a major impact on Turing's intellectual preoccupations throughout
his life, and probably even influenced his 1936 work on the theory of computing.

Turing also had a more personal reason for worrying about these
\textquotedblleft deep\textquotedblright\ questions. \ In 1930, Christopher
Morcom---Turing's teenage best friend, scientific peer, and (probably)
unrequited love---died from tuberculosis, sending a grief-stricken Turing into
long ruminations about the nature of personal identity and consciousness.
\ Let me quote from a remarkable disquisition, entitled \textquotedblleft
Nature of Spirit,\textquotedblright\ that the 19-year-old Turing sent in 1932
to Christopher Morcom's mother.

\begin{quotation}
\noindent It used to be supposed in Science that if everything was known about
the Universe at any particular moment then we can predict what it will be
through all the future. \ This idea was really due to the great success of
astronomical prediction. \ More modern science however has come to the
conclusion that when we are dealing with atoms and electrons we are quite
unable to know the exact state of them; our instruments being made of atoms
and electrons themselves. \ The conception then of being able to know the
exact state of the universe then really must break down on the small scale.
\ This means then that the theory which held that as eclipses etc.\ are
predestined so were all our actions breaks down too. \ We have a will which is
able to determine the action of the atoms probably in a small portion of the
brain, or possibly all over it. \ The rest of the body acts so as to amplify
this. \ (Quoted in Hodges \cite{hodges})
\end{quotation}

The rest of Turing's letter discusses the prospects for the survival of the
\textquotedblleft spirit\textquotedblright\ after death, a topic with obvious
relevance to Turing at that time. \ In later years, Turing would eschew that
sort of mysticism. \ Yet even in a 1951 radio address \textit{defending} the
possibility of human-level artificial intelligence, Turing still brought up
Eddington, and the possible limits on prediction of human brains imposed by
the uncertainty principle:

\begin{quotation}
\noindent If it is accepted that real brains, as found in animals, and in
particular in men, are a sort of machine it will follow that our digital
computer suitably programmed, will behave like a brain. \ [But the argument
for this conclusion] involves several assumptions which can quite reasonably
be challenged. \ [It is] necessary that this machine should be of the sort
whose behaviour is in principle predictable by calculation. \ We certainly do
not know how any such calculation should be done, and it was even argued by
Sir Arthur Eddington that on account of the indeterminacy principle in quantum
mechanics no such prediction is even theoretically possible.\footnote{As
Hodges (personal communication) points out, it's interesting to contrast these
remarks with a view Turing had expressed just a year earlier, in
\textquotedblleft Computing Machinery and Intelligence\textquotedblright%
\ \cite{turing:ai}: \textquotedblleft It is true that a discrete-state machine
must be different from a continuous machine. \ But if we adhere to the
conditions of the imitation game, the interrogator will not be able to take
any advantage of this difference.\textquotedblright\ \ Note that there's no
actual contradiction between this statement and the one about the uncertainty
principle, especially if we distinguish (as I will) between simulating a
\textit{particular} brain and simulating \textit{some} brain-like entity able
to pass the Turing test. \ However, I'm not aware of any place where Turing
explicitly makes that distinction.} (Reprinted in Shieber \cite{shieber:book})
\end{quotation}

Finally, two years after his sentencing for \textquotedblleft homosexual
indecency,\textquotedblright\ and a few months before his tragic death by
self-poisoning, Turing wrote the striking aphorisms that I\ quoted earlier:
\textquotedblleft The universe is the interior of the light-cone of the
Creation. \ Science is a differential equation. \ Religion is a boundary
condition.\textquotedblright

The reason I'm writing this essay is that I \textit{think} I now understand
what Turing could have meant by these remarks. \ Building on ideas of Hoefer
\cite{hoefer}, Stoica \cite{stoica}, and others, I'll examine a
perspective---which I call the \textquotedblleft
freebit\ perspective,\textquotedblright\ for reasons to be explained
later---that locates a nontrivial sort of freedom\ in the universe's boundary
conditions, even while embracing the mechanical nature of the time-evolution
laws. \ We'll find that a central question, for this perspective, is how well
complicated biological systems like human brains can \textit{actually} be
predicted: not by hypothetical Laplace demons, but by prediction devices
compatible with the laws of physics. \ It's in the discussion of this
predictability question (and \textit{only} there) that quantum mechanics
enters the story.

Of course, the idea that quantum mechanics might have \textit{something} to do
with free will is not new; neither are the problems with that idea or the
controversy surrounding it. \ While I chose Turing's postcard for the opening
text of this essay, I also could have chosen a striking claim by Niels Bohr,
from a 1932 lecture about the implications of Heisenberg's uncertainty principle:

\begin{quotation}
\noindent[W]e should doubtless kill an animal if we tried to carry the
investigation of its organs so far that we could tell the part played by the
single atoms in vital functions. \ In every experiment on living organisms
there must remain some uncertainty as regards the physical conditions to which
they are subjected, and the idea suggests itself that the minimal freedom we
must allow the organism will be just large enough to permit it, so to say, to
hide its ultimate secrets from us. \ (Reprinted in \cite{bohr})
\end{quotation}

Or this, from the physicist Arthur Compton:

\begin{quotation}
\noindent A set of known physical conditions is not adequate to specify
precisely what a forthcoming event will be. \ These conditions, insofar as
they can be known, define instead a range of possible events from among which
some particular event will occur. \ When one exercises freedom, by his act of
choice he is himself adding a factor not supplied by the physical conditions
and is thus himself determining what will occur. \ That he does so is known
only to the person himself. \ From the outside one can see in his act only the
working of physical law. \cite{compton}
\end{quotation}

I want to know:

\begin{quotation}
\noindent\textit{Were Bohr and Compton right or weren't they? \ Does quantum
mechanics (specifically, say, the No-Cloning Theorem or the uncertainty
principle) put interesting limits on an external agent's ability to scan,
copy, and predict human brains and other complicated biological systems, or
doesn't it?}
\end{quotation}

Of course, one needs to spell out carefully what one means by
\textquotedblleft interesting limits,\textquotedblright\ an \textquotedblleft
external agent,\textquotedblright\ the \textquotedblleft ability to scan,
copy, and predict,\textquotedblright\ and so forth.\footnote{My own attempt to
do so is in Appendix \ref{MEAN}.} \ But once that's done, I regard the above
as an unsolved scientific question, and a big one. \ Many people seem to think
the answer is obvious (though they disagree on what it is!), or else they
reject the question as meaningless, unanswerable, or irrelevant. In this essay
I'll argue strongly for a different perspective:\ that we can easily imagine
worlds consistent with quantum mechanics (and all other known physics and
biology) where the answer to the question is yes, and other such worlds where
the answer is no. \ And we don't yet know which kind we live in. \ The most we
can say is that, like $\mathsf{P}$ versus $\mathsf{NP}$ or nature of quantum
gravity, the question is well beyond our \textit{current} ability to answer.

Furthermore, the two kinds of world lead, not merely to different
philosophical stances, but to different visions of the remote future. \ Will
our descendants all choose to upload themselves into a digital hive-mind,
after a \textquotedblleft technological singularity\textquotedblright\ that
makes such things possible? \ Will they then multiply themselves into
trillions of perfect computer-simulated replicas, living in various simulated
worlds of their own invention, inside of which there might be further
simulated worlds with still more replicated minds? \ What will it be
\textit{like} to exist in so many manifestations: will each copy have its own
awareness, or will they comprise a single awareness that experiences trillions
of times more than we do? \ Supposing all this to be possible, is there any
reason why our descendants might want to hold back on it?

Now, if it turned out that Bohr and Compton were wrong---that human brains
were as probabilistically predictable by external agents as ordinary digital
computers equipped with random-number generators---then the freebit picture
that I explore in this essay would be falsified, to whatever extent it says
anything interesting. \ It should go without saying that I see the freebit
picture's vulnerability to future empirical findings as a feature rather than
a bug.

In summary, I'll make no claim to show here that the freebit picture\ is
\textit{true}. \ I'll confine myself to two weaker claims:

\begin{enumerate}
\item[(1)] That the picture is \textit{sensible} (or rather, not obviously
much crazier than the alternatives): many considerations one might think would
immediately make a hash of this picture, fail to do so for interesting reasons.

\item[(2)] That the picture is \textit{falsifiable}: there are open empirical
questions that need to turn out one way rather than another, for this picture
to stand even a \textit{chance} of working.
\end{enumerate}

I ask others to take this essay as I do: as an exercise in what physicists
call model-building. \ I want to see \textit{how far I can get} in thinking
about time, causation, predictability, and quantum mechanics in a certain
unusual but apparently-consistent way. \ \textquotedblleft
Resolving\textquotedblright\ the millennia-old free will debate isn't even on
the table! \ The most I can hope for, if I'm lucky, is to construct a model
whose strengths and weaknesses help to move the debate slightly forward.

\subsection{\textquotedblleft Free Will\textquotedblright\ Versus
\textquotedblleft Freedom\textquotedblright\label{FWFREEDOM}}

There's one terminological issue that experience has shown I need to dispense
with before anything else. \ In this essay, I'll sharply distinguish between
\textquotedblleft free will\textquotedblright\ and another concept that I'll
call \textquotedblleft freedom,\textquotedblright\ and will mostly concentrate
on the latter.

By \textquotedblleft free will,\textquotedblright\ I'll mean a metaphysical
attribute that I hold to be largely outside the scope of science---and which I
can't even \textit{define} clearly, except to say that, if there's an
otherwise-undefinable thing that people have tried to get at for centuries
with the phrase \textquotedblleft free will,\textquotedblright\ then free will
is that thing! \ More seriously, as many philosophers have pointed out,
\textquotedblleft free will\textquotedblright\ seems to combine two distinct
ideas: first, that your choices are \textquotedblleft free\textquotedblright%
\ from any kind of external constraint; and second, that your choices are not
arbitrary or capricious, but are \textquotedblleft willed by
you.\textquotedblright\ \ The second idea---that of being \textquotedblleft
willed by you\textquotedblright---is the one I consider outside the scope of
science, for the simple reason that no matter what the empirical facts were, a
skeptic could always deny that a given decision was \textquotedblleft
really\textquotedblright\ yours, and hold the true decider to have been God,
the universe, an impersonating demon, etc. \ I see no way to formulate, in
terms of observable concepts, what it would even mean for such a skeptic to be
right or wrong.

But crucially, the situation seems different if we set aside the
\textquotedblleft will\textquotedblright\ part of free will,\ and consider
only the \textquotedblleft free\textquotedblright\ part. \ Throughout, I'll
use the term \textit{freedom},\ or \textit{Knightian freedom},\ to mean a
certain strong kind of physical unpredictability: a lack of determination,
even probabilistic determination, by knowable external\ factors. \ That is, a
physical system will be \textquotedblleft free\textquotedblright\ if and only
if it's unpredictable in a sufficiently strong sense, and \textquotedblleft
freedom\textquotedblright\ will simply be that property possessed by free
systems. \ A system that's not \textquotedblleft free\textquotedblright\ will
be called \textquotedblleft mechanistic.\textquotedblright

Many issues arise when we try to make the above notions more precise. \ For
one thing, we need a definition of unpredictability that does \textit{not}
encompass the \textquotedblleft merely probabilistic\textquotedblright%
\ unpredictability of (say) a photon or a radioactive atom---since, as I'll
discuss in Section \ref{KNIGHTIANPHYS}, I accept the often-made point that
\textit{that} kind of unpredictability has nothing to do with what most people
would call \textquotedblleft freedom,\textquotedblright\ and is fully
compatible with a system's being \textquotedblleft
mechanistic.\textquotedblright\ \ Instead, we'll want what economists call
\textquotedblleft Knightian\textquotedblright\ unpredictability, meaning
unpredictability that we lack a reliable way even to quantify using
probability distributions. \ Ideally, our criteria for Knightian
unpredictability will be so stringent that they won't encompass systems like
the Earth's weather---for which, despite the presence of chaos, we arguably
\textit{can} give very well-calibrated probabilistic forecasts.

A second issue is that unpredictability seems observer-relative: a system
that's unpredictable to one observer might be perfectly predictable to
another. \ This is the reason why, throughout this essay, I'll be interested
less in particular methods of prediction than in the \textit{best} predictions
that could ever be made, consistent both with the laws of physics and with the
need not to destroy the system being studied.

This brings us to a third issue: it's not obvious what should count as
\textquotedblleft destroying\textquotedblright\ the system, or which
interventions a would-be predictor should or shouldn't be allowed to perform.
\ For example, in order to ease the prediction of a human brain, should a
predictor first\ be allowed to replace each neuron by a \textquotedblleft
functionally-equivalent\textquotedblright\ microchip? \ How would we decide
whether the microchip \textit{was} functionally equivalent to the neuron?

I'll offer detailed thoughts about these issues in Appendix \ref{MEAN}. \ For
now, though, the point I want to make is that, once we \textit{do} address
these issues, it seems to me that \textquotedblleft freedom\textquotedblright%
---in the sense of Knightian unpredictability by any external physical
observer---is perfectly within the scope of science. \ We're no longer talking
about ethics, metaphysics, or the use of language: only about whether
such-and-such a system is or isn't physically predictable in the relevant way!
\ A similar point was recently made forcefully by the philosopher Mark
Balaguer \cite{balaguer}, in his interesting book \textit{Free Will as an Open
Scientific Problem}. \ (However, while I strongly agree with Balaguer's basic
thesis, as discussed above I reject any connection between freedom and
\textquotedblleft merely probabilistic\textquotedblright\ unpredictability,
whereas Balaguer seems to accept such a connection.)

Surprisingly, my experience has been that many scientifically-minded people
will happily accept that humans plausibly \textit{are} physically
unpredictable in the relevant sense. \ Or at least, they'll accept my own
position, that whether humans \textit{are or aren't} so predictable is an
empirical question whose answer is neither obvious nor known.
\ Again and again, I've found, people will concede that chaos, the No-Cloning
Theorem, or some other phenomenon might make human brains physically
unpredictable---indeed, they'll seem oddly indifferent to the question of
whether they do or don't! \ But they'll never fail to add: \textquotedblleft
even if so, who cares? \ we're just talking about unpredictability! \ that
obviously has nothing to do with \textit{free will}!\textquotedblright

For my part, I grant that free will can't be \textit{identified} with
unpredictability, without doing violence to the usual meanings of those
concepts. \ Indeed, it's precisely because I grant this that I write,
throughout the essay, about \textquotedblleft freedom\textquotedblright\ (or
\textquotedblleft Knightian freedom\textquotedblright)\ rather than
\textquotedblleft free will.\textquotedblright\ \ I insist, however, that
unpredictability has \textit{something} to do with free will---in roughly the
same sense that verbal intelligence has \textit{something} to do with
consciousness, or optical physics has \textit{something} to do with
subjectively-perceived colors. \ That is, some people might see
unpredictability as a pale empirical shadow\ of the \textquotedblleft
true\textquotedblright\ metaphysical quality, free will, that we really want
to understand. \ But the great lesson of the scientific revolution, going back
to Galileo, is that understanding the \textquotedblleft empirical
shadow\textquotedblright\ of something is \textit{vastly} better than not
understanding the thing at all! \ Furthermore, the former might already be an
immense undertaking, as understanding human intelligence and the physical
universe turned out to be (even setting aside the \textquotedblleft
mysteries\textquotedblright\ of consciousness and metaphysics). \ Indeed I
submit that, for the past four centuries, \textquotedblleft start with the
shadow\textquotedblright\ has been a spectacularly fruitful approach to
unravelling the mysteries of the universe: one that's succeeded where greedy
attempts to go behind the shadow have failed. \ If one likes, the goal of this
essay is to explore what happens when one applies a \textquotedblleft start
with the shadow\textquotedblright\ approach to the free-will debate.

Personally, I'd go even further than claiming a vague connection between
unpredictability and free will. \ Just as displaying intelligent behavior (by
passing the Turing Test or some other means) might be thought a
\textit{necessary condition} for consciousness if not a sufficient one, so I
tend to see Knightian unpredictability as a necessary condition for free will.
\ In other words, if a system were completely predictable (even
probabilistically) by an outside entity---not merely in principle\ but in
practice---then I find it hard to understand why we'd still want to ascribe
\textquotedblleft free will\textquotedblright\ to the system. \ Why not admit
that we now fully understand what makes this system tick?

However, I'm aware that many people sharply reject the idea that
unpredictability is a necessary condition for free will. \ Even if a computer
in another room perfectly predicted all of their actions, days in advance,
these people would still call their actions \textquotedblleft
free,\textquotedblright\ so long as \textquotedblleft they themselves
chose\textquotedblright\ the actions that the computer also predicted for
them. \ In Section \ref{UPLOADING}, I'll explore some of the difficulties that
this position leads to when carried to science-fiction conclusions. \ For now,
though, it's not important to dispute the point. \ I'll happily settle for the
weaker claim that unpredictability has \textit{something} to do with free
will, just as intelligence has something to do with consciousness. \ More
precisely: in both cases, even when people \textit{think} they're asking
purely philosophical questions about the latter concept, much of what they
want to know often turns out to hinge on \textquotedblleft grubby empirical
questions\textquotedblright\ about the former concept!\footnote{A perfect
example of this phenomenon is provided by the countless people who claim that
even if a computer program passed the Turing Test, it still wouldn't be
conscious---and then, without batting an eye, defend that claim using
arguments that presuppose that the program \textit{couldn't} pass the Turing
Test after all! \ (\textquotedblleft Sure, the program might solve math
problems, but it could never write love poetry,\textquotedblright\ etc.\ etc.)
\ The temptation to hitch metaphysical claims to empirical ones, without even
realizing the chasm one is crossing, seems incredibly strong.} \ So if the
philosophical questions seem too ethereally inaccessible, then we might as
well focus for a while on the scientific ones.

\subsection{Note on the Title\label{TITLE}}

The term \textquotedblleft ghost in the machine\textquotedblright\ was
introduced in 1949 by Gilbert Ryle \cite{ryle}. \ His purpose was to ridicule
the notion of a \textquotedblleft mind-substance\textquotedblright: a
mysterious entity that exists outside of space and ordinary physical
causation; has no size, weight, or other material properties; is knowable by
its possessor with absolute certainty (while the minds of others are
\textit{not} so knowable); and somehow receives signals from the brain and
influences the brain's activity, even though it's nowhere to be found
\textit{in} the brain. \ Meanwhile, a \textit{quantum Turing machine}, defined
by Deutsch \cite{deutsch:qc} (see also Bernstein and Vazirani \cite{bv}), is a
Turing machine able to exploit the principle of quantum superposition. \ As
far as anyone knows today \cite{aar:np}, our universe seems to be efficiently
simulable by---or even \textquotedblleft isomorphic to\textquotedblright---a
quantum Turing machine, which would take as input the universe's quantum
initial state (say, at the Big Bang), then run the evolution equations forward.

\subsection{Level\label{LEVEL}}

Most of this essay should be accessible to any educated reader. \ In a few
sections, though, I assume familiarity with basic concepts from quantum
mechanics, or (less often) relativity, thermodynamics, Bayesian probability,
or theoretical computer science. \ When I \textit{do} review concepts from
those fields, I usually focus only on the ones most relevant to whatever point
I'm making. \ To do otherwise would make the essay even more absurdly long
than it already is! \ Readers seeking an accessible introduction to some of
the established theories invoked in this essay might enjoy my recent book
\textit{Quantum Computing Since Democritus} \cite{aar:qcsd}.\footnote{For the
general reader, other good background reading for this essay might include
\textit{From Eternity to Here} by Sean Carroll \cite{carroll:eternity},
\textit{The Beginning of Infinity} by David Deutsch \cite{deutsch:infinity},
\textit{The Emperor's New Mind} by Roger Penrose \cite{penrose}, or
\textit{Free Will as an Open Scientific Problem} by Mark Balaguer
\cite{balaguer}. \ Obviously, none of these authors necessarily endorse
everything I say (or vice versa)! \ What the books have in common is simply
that they explain one or more concepts invoked in this essay in much more
detail than I do.}

In the main text, I've tried to keep the discussion extremely informal. \ I've
found that, with a contentious subject like free will, mathematical rigor (or
the pretense of it) can easily obfuscate more than it clarifies. \ However,
for interested readers, I did put some more technical material into
appendices: a suggested formalization of \textquotedblleft Knightian
freedom\textquotedblright\ in Appendix \ref{MEAN}; some observations about
prediction, Kolmogorov complexity, and the universal prior in Appendix
\ref{KOLMOG}; and a suggested formalization of the notion of \textquotedblleft
freebits\textquotedblright\ in Appendix \ref{KNIGHTIAN}.

\section{FAQ\label{FAQ}}

In discussing a millennia-old conundrum like free will, a central difficulty
is that \textit{almost everyone already knows what he or she thinks}---even if
the certainties that one person brings to the discussion are completely at
odds with someone else's. \ One practical consequence is that, no matter how I
organize this essay, I'm bound to make a large fraction of readers impatient;
some will accuse me of dodging the real issues by dwelling on irrelevancies.
\ So without further ado, I'll now offer a \textquotedblleft Frequently Asked
Questions\textquotedblright\ list. \ In the thirteen questions below, I'll
engage determinists, compatibilists, and others who might have strong
\textit{a priori} reasons to be leery of my whole project. \ I'll try to
clarify my areas of agreement and disagreement, and hopefully convince the
skeptics to read further. \ Then, after developing my own ideas in Sections
\ref{KNIGHTIANPHYS}\ and \ref{FIO}, I'll come back and address still further
objections in Section \ref{OBJECTIONS}.

\subsection{Narrow Scientism\label{SCIENTISM}}

\textbf{For thousands of years, the free-will debate has encompassed moral,
legal, phenomenological, and even theological questions. \ You seem to want to
sweep all of that away, and focus exclusively on what would some would
consider a narrow scientific issue having to do with physical predictability.
\ Isn't that presumptuous?}

On the contrary, it seems presumptuous \textit{not} to limit my scope! \ Since
it's far beyond my aims and abilities to address all aspects of the free-will
debate,\ as discussed in Section \ref{FWFREEDOM} I decided to focus on one
issue: the physical and technological questions surrounding how well human and
animal brains can ever be predicted, in principle, by external entities that
also want to keep those brains alive. \ I focus on this for several reasons:
because it seems underexplored; because I might have something to say about
it; and because even if what I say is wrong, the predictability issue has the
appealing property that \textit{progress} on it seems possible. \ Indeed, even
if one granted---which I don't---that the predictability issue had nothing to
do with the \textquotedblleft true\textquotedblright\ mystery of free will,
I'd still care about the former at least as much as I cared about the latter!

However, in the interest of laying all my cards on the table, let me offer
some brief remarks on the moral, legal, phenomenological, and theological
aspects of free will.

On the moral and legal aspects, my own view is summed up beautifully by the
Ambrose Bierce poem:

\begin{quotation}
There's no free will, says the philosopher

To hang is most unjust.

There's no free will, assent the officers

We hang because we must. \cite{bierce}
\end{quotation}

For the foreseeable future, I can't see that the legal or
practical\ implications of the free-will debate are nearly as great as many
commentators have made them out to be, for the simple reason that (as Bierce
points out) any implications would apply \textquotedblleft
symmetrically,\textquotedblright\ to accused and accuser alike.

But I would go further: I've found many discussions about free will and legal
responsibility to be downright \textit{patronizing}. \ The subtext of such
discussions usually seems to be:

\begin{quotation}
\noindent We, the educated, upper-class people having this conversation,
\textit{should} accept that the entire concept of \textquotedblleft
should\textquotedblright\ is quaint and misguided, when it comes to the
uneducated, lower-class sorts of people who commit crimes. \ Those poor dears'
upbringing, brain chemistry, and so forth absolve them of any real
responsibility for their crimes: the notion that they had the
\textquotedblleft free will\textquotedblright\ to choose otherwise is just
na\"{\i}ve. \ My friends and I are \textit{right} because we accept that
enlightened stance, while other educated people are \textit{wrong} because
they fail to accept it. \ For us educated people, of course, the relevance of
the categories \textquotedblleft right\textquotedblright\ and
\textquotedblleft wrong\textquotedblright\ requires no justification or argument.
\end{quotation}

Or conversely:

\begin{quotation}
\noindent Whatever the truth, we educated people \textit{should} maintain that
all people are responsible for their choices---since otherwise, we'd have no
basis to punish the criminals and degenerates in our midst, and civilization
would collapse. \ For \textit{us}, of course, the meaningfulness of the word
\textquotedblleft should\textquotedblright\ in the previous sentence is not
merely a useful fiction, but is clear as day.
\end{quotation}

On the phenomenological aspects of free will: if someone claimed to know, from
introspection, either that free will exists or that it doesn't exist, then of
course I could never refute that person to his or her satisfaction. \ But
precisely because one can't decide between conflicting introspective reports,
in this essay I'll be exclusively interested in what can be learned from
scientific observation and argument. \ Appeals to inner experience---including
my own and the reader's---will be out of bounds. \ Likewise, while it might be
impossible to avoid grazing the \textquotedblleft mystery of
consciousness\textquotedblright\ in a discussion of human predictability, I'll
do my best to avoid addressing that mystery head-on.

On the theological aspects of free will: probably the most relevant thing to
say is that, even if there existed an omniscient God who knew all of our
future choices, that fact wouldn't concern us in this essay, \textit{unless}
God's knowledge could somehow be made manifest in the physical world, and used
to \textit{predict} our choices. \ In that case, however, we'd no longer be
talking about \textquotedblleft theological\textquotedblright\ aspects of free
will, but simply again about scientific aspects.

\subsection{Bait-and-Switch\label{BAITSWITCH}}

\textbf{Despite everything you said in Section \ref{FWFREEDOM}, I'm still not
convinced that we can learn anything about free will from an analysis of
unpredictability. \ Isn't that a shameless \textquotedblleft
bait-and-switch\textquotedblright?}

Yes, but it's a shameless bait-and-switch with a distinguished history! \ I
claim that, \textit{whenever} it's been possible to make definite progress on
ancient philosophical problems, such progress has almost always involved a
similar \textquotedblleft bait-and-switch.\textquotedblright\ \ In other
words: one replaces an unanswerable philosophical riddle $Q$ by a
\textquotedblleft merely\textquotedblright\ scientific or mathematical
question $Q^{\prime}$, which captures \textit{part} of what people have wanted
to know when they've asked $Q$. \ Then, with luck, one solves $Q^{\prime}$.

Of course, even if $Q^{\prime}$ is solved, centuries later philosophers might
still be debating the exact relation between $Q$ and $Q^{\prime}$! \ And
further exploration might lead to \textit{other} scientific or mathematical
questions---$Q^{\prime\prime}$, $Q^{\prime\prime\prime}$, and so on---which
capture aspects of $Q$ that $Q^{\prime}$\ left untouched. \ But from my
perspective, this process of \textquotedblleft breaking off\textquotedblright%
\ answerable parts of unanswerable riddles, then trying to answer those parts,
is the closest thing to philosophical progress that there is.

Successful examples of this breaking-off\ process fill intellectual history.
\ The use of calculus to treat infinite series, the link between mental
activity and nerve impulses, natural selection, set theory and first-order
logic, special relativity, G\"{o}del's theorem, game theory, information
theory, computability and complexity theory, the Bell inequality, the theory
of common knowledge, Bayesian causal networks---each of these advances
addressed questions that could rightly have been called \textquotedblleft
philosophical\textquotedblright\ before the advance was made. \ And after each
advance, there was \textit{still} plenty for philosophers to debate about
truth and provability and infinity, space and time and causality, probability
and information and life and mind. \ But crucially, it seems to me that the
technical advances transformed the philosophical discussion as philosophical
discussion \textit{itself} rarely transforms it! \ And therefore, if such
advances don't count as \textquotedblleft philosophical
progress,\textquotedblright\ then it's not clear that anything should.

Appropriately for this essay, perhaps the \textit{best} precedent for my
bait-and-switch is the Turing Test. \ Turing began his famous 1950 paper
\textquotedblleft Computing Machinery and Intelligence\textquotedblright%
\ \cite{turing:ai} with the words:

\begin{quotation}
\noindent I propose to consider the question, \textquotedblleft Can machines
think?\textquotedblright
\end{quotation}

But after a few pages of ground-clearing, he wrote:

\begin{quotation}
\noindent The original question, \textquotedblleft Can machines
think?\textquotedblright\ I believe to be too meaningless to deserve discussion.
\end{quotation}

So with legendary abruptness, Turing simply \textit{replaced} the original
question by a different one: \textquotedblleft Are there imaginable digital
computers which would do well in the imitation game\textquotedblright---i.e.,
which would successfully fool human interrogators in a teletype conversation
into \textit{thinking} they were human? \ Though some writers would later
accuse Turing of conflating intelligence with the \textquotedblleft mere
simulation\textquotedblright\ of it, Turing was perfectly clear about what he
was doing:

\begin{quotation}
\noindent I shall replace the question by another, which is closely related to
it and is expressed in relatively unambiguous words ... We cannot altogether
abandon the original form of the problem, for opinions will differ as to the
appropriateness of the substitution and we must at least listen to what has to
be said in this connexion [sic].
\end{quotation}

The claim is not that the new question, about the imitation game, is
\textit{identical} to the original question about machine intelligence.\ \ The
claim, rather, is that the new question is a worthy candidate for what we
\textit{should} have asked or \textit{meant} to have asked, if our goal was to
learn something new rather than endlessly debating definitions. \ In math and
science, the process of revising one's original question is often the core of
a research project, with the actual answering of the revised question being
the relatively easy part!

A good replacement question $Q^{\prime}$ should satisfy two properties:

\begin{itemize}
\item[(a)] $Q^{\prime}$ should capture some \textit{aspect} of the original
question $Q$---so that an answer to $Q^{\prime}$ would be \textit{hard to
ignore} in any subsequent discussion of $Q$.

\item[(b)] $Q^{\prime}$ should be precise enough that one can see what it
would mean to make \textit{progress} on $Q^{\prime}$: what experiments one
would need to do, what theorems one would need to prove, etc.
\end{itemize}

The Turing Test, I think, captured people's imaginations precisely because it
succeeded so well at (a) and (b). \ Let me put it this way: if a digital
computer were built that aced the imitation game, then \textit{it's hard to
see what more science could possibly say} in support of machine intelligence
being possible. \ Conversely, if digital computers were proved unable to win
the imitation game, then it's hard to see what more science could say in
support of machine intelligence \textit{not} being possible. \ Either way,
though, we're no longer \textquotedblleft slashing air,\textquotedblright%
\ trying to pin down the true meanings of words like \textquotedblleft
machine\textquotedblright\ and \textquotedblleft think\textquotedblright%
:\ we've hit the relatively-solid ground of a science and engineering problem.
\ Now if we want to go further we need to \textit{dig}\ (that is, do research
in cognitive science, machine learning, etc). \ This digging might take
centuries of backbreaking work; we have no idea if we'll ever reach the
bottom. \ But at least it's something humans know how to do and have done
before. \ Just as important, diggers (unlike air-slashers) tend to uncover
countless treasures besides the ones they were looking for.

By analogy, in this essay I advocate replacing the question of whether humans
have free will, by the question of how accurately their choices can be
predicted, in principle, by external agents compatible with the laws of
physics. \ And while I don't pretend that the \textquotedblleft
replacement\textquotedblright\ question is identical to the original, I do
claim the following: if humans turned out to be arbitrarily predictable in the
relevant sense, then \textit{it's hard to see what more science could possibly
say in support of \textquotedblleft free will being a
chimera.\textquotedblright}\ \ Conversely, if a fundamental reason were
discovered why the appropriate \textquotedblleft prediction
game\textquotedblright\ \textit{couldn't} be won, then it's hard to see what
more science could say in support of \textquotedblleft free will being
real.\textquotedblright

Either way, I'll try to sketch the research program that confronts us if we
take the question seriously: a program that spans neuroscience, chemistry,
physics, and even cosmology. \ Not surprisingly, much of this program consists
of problems that scientists in the relevant fields are already working on, or
longstanding puzzles of which they're well aware. \ But there are also
questions---for example, about the \textquotedblleft past macroscopic
determinants\textquotedblright\ of the quantum states occurring in
nature---which as far as I know \textit{haven't} been asked in the form they
take here.

\subsection{Compatibilism\label{COMPAT}}

\textbf{Like many scientifically-minded people, I'm a \textit{compatibilist}:
someone who believes free will can exist even in a mechanistic universe. \ For
me, \textquotedblleft free will is as real as baseball,\textquotedblright\ as
the physicist Sean Carroll memorably put it.\footnote{See
\ blogs.discovermagazine.com/cosmicvariance/2011/07/13/free-will-is-as-real-as-baseball/}
\ That is, the human capacity to weigh options and make a decision
\textquotedblleft exists\textquotedblright\ in the same sense as Sweden,
caramel corn, anger, or other complicated notions that might interest us, but
that no one expects to play a role in the fundamental laws of the
universe.\ \ As for the fundamental laws, I believe them to be completely
mechanistic and impersonal: as far as they know or care, a human brain is just
one more evanescent pattern of computation, along with sunspots and
hurricanes. \ Do you dispute any of that? \ What, if anything, can a
compatibilist take from your essay?}

I have a lot of sympathy for compatibilism---certainly more than for an
incurious mysticism that doesn't even try to reconcile itself with a
scientific worldview. \ So I hope compatibilists will find much of what I have
say \textquotedblleft compatible\textquotedblright\ with their own views!

Let me first clear up a terminological confusion. \ Compatibilism is often
defined as the belief that free will is compatible with \textit{determinism}.
\ But as far as I can see, the question of determinism versus indeterminism
has almost nothing to do with what compatibilists actually believe. \ After
all, most compatibilists happily accept quantum mechanics, with its strong
indeterminist implications (see Question \ref{QMHV}), but regard it as having
almost no bearing on their position. \ No doubt some compatibilists find it
important to stress that \textit{even if classical physics had been right},
there still would have been no difficulty for free will. \ But it seems to me
that one can be a \textquotedblleft compatibilist\textquotedblright\ even
while denying that point, or remaining agnostic about it. \ In this essay,
I'll simply define \textquotedblleft compatibilism\textquotedblright\ to be
the belief that free will is compatible with a \textit{broadly mechanistic
worldview}---that is, with a universe governed by impersonal mathematical laws
of \textit{some} kind. \ Whether it's important that those laws be
probabilistic (or chaotic, or computationally universal, or whatever else),
I'll regard as internal disputes within compatibilism.

I can now come to the question: is my perspective compatible with
compatibilism? \ Alas, at the risk of sounding lawyerly, I can't answer
without a further distinction! \ Let's define \textit{strong compatibilism} to
mean the belief that the statement \textquotedblleft Alice has free
will\textquotedblright\ is compatible with the actual, physical existence of a
machine that predicts all of Alice's future choices---a machine whose
predictions Alice herself can read and verify after the fact. \ (Where by
\textquotedblleft predict,\textquotedblright\ we mean \textquotedblleft in
roughly the same sense that quantum mechanics predicts the behavior of a
radioactive atom\textquotedblright: that is, by\ giving arbitrarily-accurate
probabilities, in cases where deterministic prediction is physically
impossible.) \ By contrast, let's define \textit{weak compatibilism} to mean
the belief that \textquotedblleft Alice has free will\textquotedblright\ is
compatible with Alice living in a mechanistic, law-governed universe---but
\textit{not} necessarily with her living in a universe where the prediction
machine can be built.

Then my perspective is compatible with weak compatibilism, but incompatible
with strong compatibilism. \ My perspective \textit{embraces} the mechanical
nature of the universe's time-evolution laws, and in that sense is proudly
\textquotedblleft compatibilist.\textquotedblright\ \ On the other hand, I
care whether our choices can \textit{actually} be mechanically predicted---not
by hypothetical Laplace demons but by physical machines. \ I'm troubled if
they are, and I take seriously the possibility that they aren't (e.g., because
of chaotic amplification of unknowable details of the initial conditions).

\subsection{Quantum Flapdoodle\label{FLAPDOODLE}}

\textbf{The usual motivation for mentioning quantum mechanics and mind in the
same breath has been satirized as \textquotedblleft quantum mechanics is
mysterious, the mind is also mysterious, ergo they must be related
somehow\textquotedblright!\ \ Aren't you worried that, merely by writing an
essay that \textit{seems} to take such a connection seriously, you'll fan the
flames of pseudoscience? \ That any subtleties and caveats in your position
will quickly get lost?}

Yes! \ Even though I can only take responsibility for what I write, not for
what various Internet commentators, etc. might mistakenly \textit{think} I
wrote, it would be distressing to see this essay twisted to support credulous
doctrines that I abhor. \ So for the record, let me state the following:

\begin{itemize}
\item[(a)] I don't think quantum mechanics, or anything else, lets us
\textquotedblleft bend the universe to our will,\textquotedblright\ except
through interacting with our external environments in the ordinary causal
ways. \ Nor do I think that quantum mechanics says \textquotedblleft
everything is holistically connected to everything else\textquotedblright%
\ (whatever that means). \ Proponents of these ideas usually invoke the
phenomenon of \textit{quantum entanglement} between particles, which can
persist no matter how far apart the particles are. \ But contrary to a
widespread misunderstanding encouraged by generations of \textquotedblleft
quantum mystics,\textquotedblright\ it's an elementary fact that entanglement
does \textit{not} allow instantaneous communication.\ \ More precisely,
quantum mechanics is \textquotedblleft local\textquotedblright\ in the
following sense: if Alice and Bob share a pair of entangled particles, that
nothing that Alice does to her particle only can affect the probability of any
outcome of any measurement that Bob performs on his particle
only.\footnote{Assuming, of course, that we don't \textit{condition} on
Alice's knowledge---something that could change Bob's probabilities even in
the case of mere classical correlation between the particles.} \ Because of
the famous Bell inequality, it's crucial that we don't interpret the concept
of \textquotedblleft locality\textquotedblright\ to mean \textit{more} than
that! \ But quantum mechanics' revision to our concept of locality is so
subtle that neither scientists, mystics, nor anyone else anticipated it beforehand.

\item[(b)] I don't think quantum mechanics has vindicated Eastern religious
teachings, any more than (say) Big Bang cosmology has vindicated the Genesis
account of creation. \ In both cases, while there are interesting parallels, I
find it dishonest to seek out only the points of convergence while ignoring
the many inconvenient parts that don't fit! \ Personally, I'd say that the
quantum picture of the world---as a complex unit vector $\left\vert
\psi\right\rangle $\ evolving linearly in a Hilbert space---is not a close
match to \textit{any} pre-$20^{th}$-century conception of reality.

\item[(c)] I don't think quantum mechanics has overthrown Enlightenment ideals
of science and rationality.\ \ Quantum mechanics does overthrow the
\textquotedblleft na\"{\i}ve realist\textquotedblright\ vision of particles
with unambiguous trajectories through space, and it does raise profound
conceptual problems that will concern us later on. \ On the other hand, the
point is still to describe a physical world external to our minds by positing
a \textquotedblleft state\textquotedblright\ for that world, giving precise
mathematical rules for the evolution of the state, and testing the results
against observation. \ Compared to classical physics, the reliance on
mathematics has only increased; while the Enlightenment ideal of describing
Nature as we find it to be, rather than as intuition says it \textquotedblleft
must\textquotedblright\ be, is emphatically upheld.

\item[(d)] I don't think the human brain is a quantum computer in any
interesting sense. \ As I explain in \cite{aar:qcsd}, at least three
considerations lead me to this opinion. \ First, it would be nearly miraculous
if complicated entangled states---which today, can generally survive for at
most a few seconds in near-absolute-zero laboratory conditions---could last
for any appreciable time\ in the hot, wet environment of the brain. \ (Many
researchers have made some version of that argument, but see Tegmark
\cite{tegmark:qmbrain} for perhaps the most detailed version.) \ Second, the
sorts of tasks quantum computers are known to be good at (for example,
factoring large integers and simulating quantum systems) seem like a terrible
fit to the sorts of tasks that \textit{humans} seem be good at, or that could
have plausibly had survival value on the African savannah! \ Third, and most
importantly, I don't see anything that the brain being a quantum computer
would plausibly help to \textit{explain}. \ For example, why would a conscious
quantum computer be any less mysterious than a conscious \textit{classical}
computer? \ My conclusion is that, \textit{if} quantum effects play any role
in the brain, then such effects are almost certainly short-lived and
microscopic.\footnote{This is not to say, of course, that the brain's activity
might not \textit{amplify} such effects to the macroscopic, classical
scale---a possibility that will certainly concern us later on.} \ At the
\textquotedblleft macro\textquotedblright\ level of most interest to
neuroscience, the evidence is overwhelming that the brain's computation and
information storage are classical. \ (See Section \ref{PENROSE} for further
discussion of these issues in the context of Roger Penrose's views.)

\item[(e)] I don't think consciousness is in any sense necessary to bring
about the \textquotedblleft reduction of the wavefunction\textquotedblright%
\ in quantum measurement. \ And I say that, despite freely confessing to
unease with all existing accounts of quantum measurement! \ My position is
that, to whatever extent the reduction of the wavefunction is a real process
at all (as opposed to an artifact of observers' limited perspectives, as in
the Many-Worlds Interpretation), it must be a process that can occur even in
interstellar space, with no conscious observers anywhere around. \ For
otherwise, we're forced to the absurd conclusion that the universe's quantum
state evolved linearly via the Schr\"{o}dinger equation for billions of years,
\textit{until} the first observers arose (who: humans? monkeys? aliens?) and
looked around them---at which instant the state suddenly and violently collapsed!
\end{itemize}

If one likes, whatever I \textit{do} say about quantum mechanics and mind in
this essay will be said in the teeth of the above points. \ In other words,
I'll regard points (a)-(e) as sufficiently well-established to serve as useful
\textit{constraints}, which a new proposal ought to satisfy as a prerequisite
to being taken seriously.

\subsection{Brain-Uploading: Who Cares?\label{UPLOADING}}

\textbf{Suppose it were possible to \textquotedblleft upload\textquotedblright%
\ a human brain to a computer, and thereafter predict the brain unlimited
accuracy. \ Who cares? \ Why should anyone even worry that that would create a
problem for free will or personal identity?}

For me, the problem comes from the observation that it seems impossible to
give any operational difference\ between a perfect predictor of your actions,
and a second copy or instantiation of yourself. \ If there are two entities,
both of which respond to every situation exactly as \textquotedblleft
you\textquotedblright\ would, then by what right can we declare that only one
such entity is the \textquotedblleft real\textquotedblright\ you, and the
other is just a predictor, simulation, or model? \ But having multiple copies
of you in the same universe\ seems to open a Pandora's box of science-fiction
paradoxes. \ Furthermore, these paradoxes aren't \textquotedblleft merely
metaphysical\textquotedblright: they concern \textit{how you should do
science} knowing there might be clones of yourself, and which predictions and
decisions you should make.

Since this point is important, let me give some examples. \ Planning a
dangerous mountain-climbing trip? \ Before you go, make a backup of
yourself---or two or three---so that if tragedy should strike, you can restore
from backup and then continue life as if you'd never left. \ Want to visit
Mars? \ Don't bother with the perilous months-long journey through space; just
use a brain-scanning machine to \textquotedblleft fax
yourself\textquotedblright\ there as pure information, whereupon another
machine on Mars will construct a new body for you, functionally identical to
the original.

Admittedly, some awkward questions arise. \ For example, after you've been
faxed to Mars, what should be done with the \textquotedblleft
original\textquotedblright\ copy of you left on Earth? \ Should it be
destroyed with a quick, painless gunshot to the head? \ Would \textit{you}
agree to be \textquotedblleft faxed\textquotedblright\ to Mars, knowing that
that's what would be done to the original? \ Alternatively, if the original
were left alive, then what makes you sure you would \textquotedblleft wake
up\textquotedblright\ as the copy on Mars? \ At best, wouldn't you have 50/50
odds of still finding yourself on Earth? \ Could that problem be
\textquotedblleft solved\textquotedblright\ by putting a thousand copies of
you on Mars, while leaving only one copy on Earth? \ Likewise, suppose you
return unharmed from your mountain-climbing trip, and decide that the backup
copies you made before you left are now an expensive nuisance. \ If you
destroy them, are you guilty of murder? \ Or is it more like suicide? \ Or neither?

Here's a \textquotedblleft purer\textquotedblright\ example of such a puzzle,
which I've adapted from the philosopher Nick Bostrom \cite{bostrom}. \ Suppose
an experimenter flips a fair coin while you lie anesthetized in a white,
windowless hospital room. \ If the coin lands heads, then she'll create a
thousand copies of you, place them in a thousand identical rooms, and wake
each one up. \ If the coin lands tails, then she'll wake you up without
creating any copies. \ You wake up in a white, windowless room just like the
one you remember. \ Knowing the setup of the experiment, at what odds should
you be willing to bet that the coin landed heads? \ Should your odds just be
50/50, since the coin was fair? \ Or should they be biased 1000:1 in favor of
the coin having landed heads---since if it \textit{did} land heads, then there
are a thousand of you confronting the same situation, compared to only one if
the coin landed tails?

Many people immediately respond\ that the odds should be 50/50: they consider
it a metaphysical absurdity to adjust the odds based on the number of copies
of yourself in existence. \ (Are we to imagine a \textquotedblleft warehouse
full of souls,\textquotedblright\ with the odds of any particular soul being
taken out of the warehouse proportional to the number of suitable bodies for
it?) \ However, those who consider 50/50 the obvious answer should consider a
slight variant of the puzzle. \ Suppose that, if the coin lands tails, then as
before the experimenter leaves a single copy of you in a white room. \ If the
coin lands heads, then the experimenter creates a thousand copies of you and
places them in a thousand windowless rooms. \ Now, though, 999 of the rooms
are painted blue; only one of the rooms is white like you remember.

You wake up from the anesthesia and find yourself in a white room.
\ \textit{Now} what posterior probability should you assign to the coin having
landed heads? \ \textit{If} you answered 50/50 to the first puzzle, then a
simple application of Bayes' rule implies that, in the \textit{second} puzzle,
you should consider it overwhelmingly likely that the coin landed tails. \ For
if the coin landed heads, then presumably you had a 99.9\% probability of
being one of the 999 copies who woke up in a blue room. \ So the fact that you
woke up in a white room furnishes powerful evidence about the coin. \ Not
surprisingly, many people find \textit{this} result just as metaphysically
unacceptable as the 1000:1 answer to the first puzzle! \ Yet as Bostrom points
out, it seems mathematically inconsistent to insist on 50/50 as the answer to
\textit{both} puzzles.

Probably the most famous \textquotedblleft paradox of
brain-copying\textquotedblright\ was invented by Simon Newcomb, then
popularized by Robert Nozick \cite{nozick}\ and Martin Gardner
\cite{gardner:newcomb}. \ In \textit{Newcomb's paradox}, a superintelligent
\textquotedblleft Predictor\textquotedblright\ presents you with two closed
boxes, and offers you a choice between opening the first box only or opening
both boxes. \ Either way, you get to keep whatever you find in the box or
boxes that you open. \ The contents of the first box can vary---sometimes it
contains \$1,000,000, sometimes nothing---but the second box always contains \$1,000.

Just from what was already said, it seems that it must be preferable to open
both boxes. \ For whatever you would get by opening the first box only, you
can get \$1,000 more by opening the second box as well. \ But here's the
catch: using a detailed brain model, the Predictor has already foreseen your
choice. \ \textit{If} it predicted that you would open both boxes, then the
Predictor left the first box empty; while if it predicted that you would open
the first box only, then the Predictor put \$1,000,000 in the first box.
\ Furthermore, the Predictor has played this game hundreds of times before,
both with you and with other people, and its predictions have been right every
time. \ Everyone who opened the first box ended up with \$1,000,000, while
everyone who opened both boxes ended up with only \$1,000. \ Knowing all of
this, what do you do?

Some people dismiss the problem as contradictory---arguing that, if the
assumed Predictor exists, then you have no free will, so there's no use
fretting over how many boxes to open since your choice is already
predetermined anyway. \ Among those willing to play along, opinion has been
split for decades between \textquotedblleft one-boxers\textquotedblright\ and
\textquotedblleft two-boxers.\textquotedblright\ \ Lately, though, the
one-boxers seem to have been gaining the upper hand---and reasonably so in my
opinion, since by the assumptions of the thought experiment, the one-boxers
\textit{do} always walk away richer!

As I see it, the real problem is to \textit{explain} how one-boxing could
possibly be rational, given that, at the time you're contemplating your
decision, the million dollars are either in the first box or not. \ Can a
last-minute decision to open both boxes somehow \textquotedblleft reach
backwards in time,\textquotedblright\ causing the million dollars that
\textquotedblleft would have been\textquotedblright\ in the first box to
disappear? \ Do we need to distinguish between your \textquotedblleft
actual\textquotedblright\ choices and your \textquotedblleft
dispositions,\textquotedblright\ and say that, while one-boxing is admittedly
irrational, \textit{making yourself into the sort of person }who one-boxes is rational?

While I consider myself a one-boxer, the only justification for one-boxing
that makes sense to me goes as follows.\footnote{I came up with this
justification around 2002, and set it out in a blog post in 2006: see
www.scottaaronson.com/blog/?p=30. \ Later, I learned that Radford Neal
\cite{neal}\ had independently proposed similar ideas.} \ In principle, you
could base your decision of whether to one-box or two-box on anything you
like: for example, on whether the name of some obscure childhood friend had an
even or odd number of letters. \ However, this suggests that the problem of
predicting whether you will one-box or two-box is \textquotedblleft
you-complete.\textquotedblright\footnote{In theoretical computer science, a
problem belonging to a class $C$ is called $C$-\textit{complete}, if solving
the problem would suffice to solve any other problem in $C$.} \ In other
words,\ if the Predictor can solve this problem reliably, then it seems to me
that it must possess a simulation of you so detailed as to constitute another
\textit{copy} of you (as discussed previously).

But in that case, to whatever extent we want to think about Newcomb's paradox
in terms of a freely-willed decision at all, we need to imagine \textit{two}
entities separated in space and time---the \textquotedblleft flesh-and-blood
you,\textquotedblright\ and the simulated version being run by the
Predictor---that are nevertheless \textquotedblleft tethered
together\textquotedblright\ and share common interests. \ \textit{If} we think
this way, then we can easily explain why one-boxing can be rational, even
without backwards-in-time causation. \ Namely, as you contemplate whether to
open one box or two, who's to say that you're not \textquotedblleft
actually\textquotedblright\ the simulation? \ If you are, then of course your
decision can affect what the Predictor does in an ordinary, causal way.

For me, the takeaway is this. \ \textit{If} any of these
technologies---brain-uploading, teleportation, the Newcomb predictor,
etc.---were actually realized, then all sorts of \textquotedblleft woolly
metaphysical questions\textquotedblright\ about personal identity and free
will would start to have \textit{practical consequences}. \ Should you fax
yourself to Mars or not? \ Sitting in the hospital room, should you bet that
the coin landed heads or tails? \ Should you expect to \textquotedblleft wake
up\textquotedblright\ as one of your backup copies, or as a simulation being
run by the Newcomb Predictor? \ These questions all seem \textquotedblleft
empirical,\textquotedblright\ yet one can't answer them without taking an
implicit stance on questions that many people would prefer to regard as
outside the scope of science.

Thus, the idea that we can \textquotedblleft escape all that philosophical
crazy-talk\textquotedblright\ by declaring that the human mind is a computer
program running on the hardware of the brain, and that's all there is to it,
strikes me as ironically backwards. \ Yes, we can say that, and we might even
be right. \ But far from bypassing all philosophical perplexities, such a move
lands in a \textit{swamp} of them! \ For now we need to give some account of
how a rational agent ought to make decisions and scientific predictions, in
situations where it knows it's only one of several exact copies of itself
inhabiting the same universe.

Many will try to escape the problem, by saying that such an agent, being (by
assumption) \textquotedblleft just a computer program,\textquotedblright%
\ simply does whatever its code determines it does given the relevant initial
conditions. \ For example, if a piece of code says to bet heads in a certain
game, then all agents running that code will bet heads; if the code says to
bet tails, then the agents will bet tails. \ Either way, an \textit{outside}
observer who knew the code could easily calculate the probability that the
agents will win or lose their bet. \ So what's the philosophical problem?

For me, the problem with this response is simply that it gives up on science
as \textit{something agents can use to predict their future experiences.}
\ The agents wanted science to tell them, \textquotedblleft given
such-and-such physical conditions, here's what you \textit{should} expect to
see, and why.\textquotedblright\ \ Instead they're getting the worthless
tautology, \textquotedblleft if your internal code causes you to expect to see
$X$, then you expect to see $X$, while if your internal code causes you to
expect to see $Y$, then you expect to see $Y$.\textquotedblright\ \ But the
same could be said about \textit{anything}, with no scientific understanding
needed! \ To paraphrase Democritus,\footnote{In Democritus's famous dialogue
between the intellect and the senses, the intellect declares:
\textquotedblleft By convention there is sweetness, by convention bitterness,
by convention color, in reality only atoms and the void.\textquotedblright%
\ \ To which the senses reply: \textquotedblleft Foolish intellect! Do you
seek to overthrow us, while it is from us that you take your evidence? \ Your
victory is your defeat.\textquotedblright} it seems like the ultimate victory
of the mechanistic worldview is also its defeat.

As far as I can see, the only hope for avoiding these difficulties is
if---because of chaos, the limits of quantum measurement, or whatever other
obstruction---minds \textit{can't} be copied perfectly from one physical
substrate to another, as can programs on standard digital computers. \ So
that's a possibility that this essay explores at some length. \ To clarify, we
can't use any philosophical difficulties that would arise if minds were
copyable, as evidence for the empirical claim that they're \textit{not}
copyable. \ The universe has never shown any particular tendency to cater to
human philosophical prejudices! \ But I'd say the difficulties provide more
than enough reason to \textit{care} about the copyability question.

\subsection{Determinism versus Predictability\label{DETERPREDICT}}

\textbf{I'm a determinist:\ I believe,\ not only that humans lack free will,
but that everything that happens is completely determined by prior causes.
\ So why should an analysis of \textquotedblleft mere
unpredictability\textquotedblright\ change my thinking at all? \ After all, I
readily admit that, despite being metaphysically determined, many future
events are unpredictable in practice. \ But for me, the fact that we can't
predict something is \textit{our} problem, not Nature's!}

There's an observation that doesn't get made often enough in free-will
debates, but that seems extremely pertinent here. \ Namely: \textit{if you
stretch the notion of \textquotedblleft determination\textquotedblright\ far
enough, then events become \textquotedblleft determined\textquotedblright\ so
trivially that the notion itself becomes vacuous.}

For example, a religious person might maintain that all events are
predetermined by God's Encyclopedia, which of course only God can read.
\ Another, secular person might maintain that \textit{by definition},
\textquotedblleft the present state of the universe\textquotedblright%
\ contains all the data needed to determine future events, even if those
future events (such as quantum measurement outcomes) aren't \textit{actually}
predictable via present-day measurements. \ In other words: if, in a given
conception of physics, the present state does \textit{not} fully determine all
the future states, then such a person will simply add \textquotedblleft hidden
variables\textquotedblright\ to the present state until it does so.

Now, if our hidden-variable theorist isn't careful, and piles on additional
requirements like spatial locality, then she'll quickly find herself
contradicting one or more of quantum mechanics' no-hidden-variable theorems
(such as the Bell \cite{bell}, Kochen-Specker \cite{ks}, or PBR \cite{pbr}%
\ theorems). \ But the bare assertion that \textquotedblleft everything is
determined by the current state\textquotedblright\ is no more disprovable than
the belief in God's Encyclopedia.

To me, this immunity from any possible empirical discovery\ shows just how
\textit{feeble} a concept \textquotedblleft determinism\textquotedblright%
\ really is, unless it's supplemented by further concepts like locality,
simplicity, or (best of all) actual predictability. \ A form of
\textquotedblleft determinism\textquotedblright\ that applies not merely to
our universe, but to any\textit{ logically possible }universe, is not a
determinism that has \textquotedblleft fangs,\textquotedblright\ or that could
credibly threaten any notion of free will worth talking about.

\subsection{Predictability in the Eye of the Beholder\label{BEHOLDER}}

\textbf{A system that's predictable to one observer might be unpredictable to
another. \ Given that predictability is such a relative notion, how could it
possibly play the fundamental role you need it to?}

This question was already briefly addressed in Section \ref{FWFREEDOM}, but
since it arises so frequently, it might be worth answering again. \ In this
essay, I call a physical system $S$ \textquotedblleft
predictable\textquotedblright\ if (roughly speaking) there's \textit{any
possible technology},\textit{ consistent with the laws of physics}, that would
allow an external observer to gain enough information about $S$, without
destroying $S$, to calculate well-calibrated probabilities (to any desired
accuracy) for the outcomes of all possible future measurements on $S$ within
some allowed set. \ Of course, this definition introduces many concepts that
require further clarification: for example, what do we mean by
\textquotedblleft destroying\textquotedblright\ $S$? \ What does it mean for
probabilities to be \textquotedblleft well-calibrated\textquotedblright?
\ Which measurements on $S$ are \textquotedblleft allowed\textquotedblright%
?\ \ What can the external observer be assumed to know about $S$ before
encountering it? \ For that matter, what exactly counts as an
\textquotedblleft external observer,\textquotedblright\ or a \textquotedblleft
physical system\textquotedblright? \ I set out my thoughts about these
questions, and even suggest a tentative formal definition of \textquotedblleft
Knightian freedom\textquotedblright\ in terms of other concepts, in Appendix
\ref{MEAN}.

For now, though, the main point is that, whenever I talk about whether a
system \textquotedblleft can\textquotedblright\ be predicted, the word
\textquotedblleft can\textquotedblright\ has basically the same meaning as
when physicists talk about whether information \textquotedblleft
can\textquotedblright\ get from point $A$ to point $B$. \ Just like in the
latter case, we don't care whether the two points are \textit{literally}
connected by a phone line, so too in the former case, we don't care whether
the requisite prediction machine has actually been built, or could plausibly
be built in the next millennium. \ Instead, we're allowed to imagine
\textit{arbitrarily-advanced technologies}, so long as our imaginations are
constrained by the laws of physics.\footnote{By \textquotedblleft the laws of
physics,\textquotedblright\ I mean the currently-accepted
laws---\textit{unless} we're explicitly tinkering with those laws to see what
happens. \ Of course, whenever known physics imposes an inconvenient limit
(for example, no faster-than-light communication), the convention in most
science-fiction writing is simply to stipulate that physicists of the future
will discover some way around the limit (such as wormholes, tachyons,
\textquotedblleft hyperdrive,\textquotedblright\  etc). \ In this essay I take
a different approach, trying to be as conservative as I can about fundamental
physics.}

(Observe that, were our imaginations not constrained \textit{even} by physics,
we'd have to say that \textit{anything whatsoever} \textquotedblleft
can\textquotedblright\ happen, except outright logical contradictions. \ So in
particular, we'd reach the uninteresting conclusion that \textit{any} system
can be perfectly predicted---by God, for example, or by magical demons. \ For
more see Question \ref{DETERPREDICT}.)

\subsection{Quantum Mechanics and Hidden Variables\label{QMHV}}

\textbf{Forget about free will or Knightian uncertainty: I deny even that
\textit{probability} plays any fundamental role in physics. \ For me, like for
Einstein, the much-touted \textquotedblleft randomness\textquotedblright\ of
quantum mechanics merely shows that we humans haven't yet discovered the
underlying deterministic rules. \ Can you prove that I'm wrong?}

With minimal use of Occam's Razor, yes, I can! \ In 1926, when Einstein wrote
his famous aphorism about God and dice, the question of whether quantum events
were \textquotedblleft truly\textquotedblright\ random or merely pseudorandom
could still be considered metaphysical. \ After all, common sense suggests we
can never say with confidence that \textit{anything} is random: the most we
can ever say is that \textit{we} failed to find a pattern in it.

But common sense is flawed here. \ A large body of work, starting with that of
Bell in the 1960s \cite{bell}, has furnished evidence that quantum measurement
outcomes \textit{can't} be governed by any hidden pattern, but must be random
in just the way quantum mechanics says they are. \ Crucially, this evidence
doesn't circularly assume that quantum mechanics is the final theory of
nature. \ Instead, it assumes just a few general principles (such as spatial
locality and \textquotedblleft no cosmic conspiracies\textquotedblright),
together with the results of specific experiments that have already been done.
\ Since these points are often misunderstood, it might be worthwhile to spell
them out in more detail.

Consider the \textit{Bell inequality}, whose violation by entangled particles
(in accord with quantum mechanics) has been experimentally demonstrated more
and more firmly since the 1980s \cite{aspect}. \ From a modern perspective,
Bell simply showed that certain games, played by two cooperating but
non-communicating players Alice and Bob, can be won with greater probability
if Alice and Bob share entangled particles than if they merely share
correlated \textit{classical} information.\footnote{The standard example is
the \textit{CHSH game} \cite{chsh}. \ Here Alice and Bob are given bits $x$
and $y$ respectively, which are independent and uniformly random. \ Their goal
is for Alice to output a bit $a$, and Bob a bit $b$, such that $a+b\left(
\operatorname{mod}2\right)  =xy$. \ Alice and Bob can agree on a strategy in
advance, but can't communicate after receiving $x$ and $y$. \ Classically,
it's easy to see that the best they can do is always to output $a=b=0$, in
which case they win the game with probability $3/4$. \ By contrast, if Alice
and Bob own one qubit each of the entangled state $\frac{1}{\sqrt{2}}\left(
\left\vert 00\right\rangle +\left\vert 11\right\rangle \right)  $, then there
exists a strategy by which they can win with probability $\cos^{2}\left(
\pi/8\right)  \approx0.85$. \ That strategy has the following form: Alice
measures her qubit in a way that depends on $x$ and outputs the result as $a$,
while Bob measures his qubit in a way that depends on $y$ and outputs the
result as $b$.} \ Bell's theorem is usually presented as ruling out a class of
theories called \textit{local hidden-variable theories}.\ \ Those theories
sought to explain Alice and Bob's measurement results in terms of ordinary
statistical correlations between two random variables $X$ and $Y$, which are
somehow associated with Alice's and Bob's particles respectively, and which
have the properties that nothing Alice does can affect $Y$ and that nothing
Bob does can affect $X$. \ (One can imagine the particles flipping a coin at
the moment of their creation, whereupon one of them declares,
\textquotedblleft OK, if anyone asks, I'll be spinning up and you'll be
spinning down!\textquotedblright)

In popular treatments, Bell's theorem is usually presented as demonstrating
the reality of what Einstein called \textquotedblleft spooky action at a
distance.\textquotedblright\footnote{The reason why many people (including me)
cringe at that sort of talk is the \textit{no-communication theorem}, which
explains why, despite Bell's theorem, entanglement \textit{can't} be used to
send actual messages faster than light. \ (Indeed, if it could, then quantum
mechanics would flat-out contradict special relativity.)
\par
The situation is this: \textit{if} one wanted to violate the Bell inequality
using classical physics, \textit{then} one would need faster-than-light
communication. \ But that doesn't imply that quantum mechanics' violation of
the same inequality should \textit{also} be understood in terms of
faster-than-light communication! \ We're really dealing with an intermediate
case here---\textquotedblleft more than classical locality, but less than
classical nonlocality\textquotedblright---which I don't think anyone even
recognized as a logical possibility until quantum mechanics forced it on
them.} \ However, as many people have pointed out over the years---see, for
example, my\ 2002 critique \cite{aar:rev} of Stephen Wolfram's \textit{A New
Kind of Science} \cite{wolfram}---one can also see Bell's theorem in a
different way: as using the \textit{assumption} of no instantaneous
communication to address the even more basic issue of \textit{determinism}.
\ From this perspective, Bell's theorem says the following:

\begin{quotation}
\noindent\textit{Unless} Alice and Bob's particles communicate faster than
light, the results of all possible measurements that Alice and Bob could make
on those particles \textit{cannot have been determined prior to measurement}%
---not even by some bizarre, as-yet-undiscovered uncomputable law---assuming
the statistics of all the possible measurements agree with the quantum
predictions. \ Instead, the results \textit{must} be \textquotedblleft
generated randomly on-the-fly\textquotedblright\ in response to whichever
measurement is made, just as quantum mechanics says they are.
\end{quotation}

The above observation was popularized in 2006 by John Conway and Simon Kochen,
who called it the \textquotedblleft Free Will Theorem\textquotedblright%
\ \cite{conwaykochen}. \ Conway and Kochen put the point as follows: if
there's no faster-than-light communication, \textit{and} Alice and Bob have
the \textquotedblleft free will\textquotedblright\ to choose how to measure
their respective particles, then the particles must have their own
\textquotedblleft free will\textquotedblright\ to choose how to respond to the measurements.

Alas, Conway and Kochen's use of the term \textquotedblleft free
will\textquotedblright\ has generated confusion. \ For the
record,\ \textit{what Conway and Kochen mean by \textquotedblleft free
will\textquotedblright\ has only the most tenuous connection to what most
people (including me, in this essay) mean by it!} \ Their result might more
accurately be called the \textquotedblleft freshly-generated randomness
theorem.\textquotedblright\footnote{Or the \textquotedblleft Free Whim
Theorem,\textquotedblright\ as Conway likes to suggest when people point out
the irrelevance of human free will to the theorem.} \ For the indeterminism
that's relevant here is \textquotedblleft only\textquotedblright%
\ probabilistic: indeed, Alice and Bob could be replaced by simple
dice-throwing or quantum-state-measuring automata without affecting the
theorem at all.\footnote{This point was recently brought out by Fritz, in his
paper \textquotedblleft Bell's theorem without free will\textquotedblright%
\ \cite{fritz}. \ Fritz replaces the so-called \textquotedblleft free will
assumption\textquotedblright\ of the Bell inequality---that is, the assumption
that Alice and Bob get to choose which measurements to perform---by an
assumption about the \textit{independence} of separated physical devices.}

Another recent development has made the conflict between quantum mechanics and
determinism particularly vivid. \ It's now known how to exploit Bell's theorem
to generate so-called \textquotedblleft Einstein-certified random
bits\textquotedblright\ for use in cryptographic
applications\ \cite{pironio,vaziranividick}, starting from a much smaller
number of \textquotedblleft seed\textquotedblright\ bits that are known to be
random.\footnote{The protocol of Vazirani and Vidick \cite{vaziranividick}%
\ needs only $O\left(  \log n\right)  $ seed bits to generate $n$%
\ Einstein-certified output bits.} \ Here \textquotedblleft
Einstein-certified\textquotedblright\ means that, \textit{if} the bits pass
certain statistical tests, then they \textit{must} be close to
uniformly-random, unless nature resorted to \textquotedblleft cosmic
conspiracies\textquotedblright\ between separated physical devices to bias the bits.

Thus, if one wants to restore determinism while preserving the empirical
success of quantum mechanics, then one has to posit a conspiracy\ in which
every elementary particle, measuring device, and human brain potentially
colludes. \ Furthermore, this conspiracy needs to be so diabolical as to leave
essentially no trace of its existence! \ For example, in order to explain why
we can't exploit the conspiracy to send faster-than-light signals, one has to
imagine that the conspiracy prevents our own brains (or the quantum-mechanical
random number generators in our computers, etc.) from making the choices that
would cause those signals to be sent. \ To my mind, this is no better than the
creationists' God, who planted fossils in the ground to confound the paleontologists.

I should say that at least one prominent physicist, Gerard 't Hooft, actually
advocates such a cosmic conspiracy \cite{thooft:freewill} (under the name
\textquotedblleft superdeterminism\textquotedblright); he speculates that a
yet-undiscovered replacement for quantum mechanics will reveal its
workings.\footnote{Some people might argue that Bohmian mechanics \cite{bohm},
the interpretation of quantum mechanics that originally inspired Bell's
theorem, is also \textquotedblleft superdeterministic.\textquotedblright%
\ \ But Bohmian mechanics is empirically equivalent to standard quantum
mechanics---from which fact it follows immediately that the \textquotedblleft
determinism\textquotedblright\ of Bohm's theory is a formal construct that,
whatever else one thinks about it, has no actual consequences for prediction.
\ To put it differently: at least in its standard version, Bohmian mechanics
buys its \textquotedblleft determinism\textquotedblright\ via the mathematical
device of pushing all the randomness back to the beginning of time. \ It then
accepts the nonlocality that such a tactic inevitably entails because of
Bell's theorem.} \ For me, though, the crux is that\ once we start positing
conspiracies between distant regions of spacetime, or between the particles we
measure and our own instruments or brains, determinism becomes consistent with
\textit{any} possible scientific discovery, and therefore retreats into
vacuousness. \ As the extreme case, as pointed out in Question
\ref{DETERPREDICT}, someone could always declare that everything that happens
was \textquotedblleft determined\textquotedblright\ by God's unknowable book
listing everything that will ever happen! \ That sort of determinism can never
be falsified, but has zero predictive or explanatory power.

In summary, I think it's fair to say that \textit{physical indeterminism is
now a settled fact to roughly the same extent as evolution, heliocentrism, or
any other discovery in science}. \ So \textit{if} that fact is considered
relevant to the free-will debate, then all sides might as well just accept it
and move on! \ (Of course, we haven't yet touched the question of whether
physical indeterminism \textit{is} relevant to the free-will debate.)

\subsection{The Consequence Argument\label{CONSEQUENCE}}

\textbf{How does your perspective respond to Peter van Inwagen's Consequence
Argument?}

Some background for non-philosophers:\ the Consequence Argument \cite{inwagen}
is an attempt to formalize most people's intuition for why free will is
incompatible with determinism. \ The argument consists of the following steps:

\begin{enumerate}
\item[(i)] If determinism is true, then our choices today are determined by
whatever the state of the universe was (say) $100$ million years ago, when
dinosaurs roamed the earth.

\item[(ii)] The state of the universe $100$ million years ago is
clearly\ outside our ability to alter.

\item[(iii)] Therefore, if determinism is true, then our choices today are
outside our ability to alter.

\item[(iv)] Therefore, if determinism is true, then we don't have free will.
\end{enumerate}

(A side note: as discussed in Question \ref{COMPAT}, the traditional obsession
with \textquotedblleft determinism\textquotedblright\ here seems unfortunate
to me. \ What people really mean to ask, I think, is whether free will is
compatible with \textit{any} mechanistic account of the universe, regardless
of whether the account happens to be deterministic or probabilistic. \ On the
other hand, one could easily rephrase the Consequence Argument to allow for
this, with the state of the universe $100$ million years ago now fully
determining the \textit{probabilities} of our choices today, if not the
choices themselves. \ And I don't think that substitution would make any
essential difference to what follows.)

One can classify beliefs about free will according to how they respond to the
Consequence Argument. \ If you accept the argument as well as its starting
premise of determinism (or mechanism), and hence also the conclusion of no
free will, then you're a \textit{hard determinist (or mechanist)}. \ If you
accept the argument, but reject the conclusion by denying the starting premise
of determinism or mechanism, then you're a \textit{metaphysical libertarian}.
\ If you reject the argument by denying that steps (iii) or (iv) follow from
the previous steps, then you're a \textit{compatibilist}.

What about the perspective I explore here? \ It denies step (ii)---or rather,
it denies the usual notion of \textquotedblleft the state of the universe
$100$ million years ago,\textquotedblright\ insisting on a distinction between
\textquotedblleft macrofacts\textquotedblright\ and \textquotedblleft
microfacts\textquotedblright\ about that state. \ It agrees that the past
\textit{macro}facts---such as whether a dinosaur kicked a particular
stone---have an objective reality that is outside our ability to alter. \ But
it denies that we can always speak in straightforward causal terms about the
past \textit{micro}facts, such as the quantum state $\left\vert \psi
\right\rangle $\ of some particular photon impinging on the dinosaur's tail.

As such, my perspective can be seen as an example of a little-known view that
Fischer \cite{fischer}\ calls \textit{multiple-pasts compatibilism}. \ As I'd
put it, multiple-pasts compatibilism agrees that the past microfacts about the
world determine its future, and it also agrees that the past \textit{macro}%
facts are outside our ability to alter.\ \ However, it maintains that there
might be many possible settings of the past microfacts---the polarizations of
individual photons, etc.---that all coexist in the same \textquotedblleft
past-ensemble.\textquotedblright\ \ By definition, such microfacts can't
possibly have made it into the history books, and a multiple-pasts
compatibilist would deny that they're necessarily \textquotedblleft outside
our ability to alter.\textquotedblright\ \ Instead, our choices today might
play a role in selecting \textit{one} past from a giant ensemble of
macroscopically-identical but microscopically-different pasts.

While I take the simple idea above as a starting point, there are two main
ways in which I try to go further than it. \ First, I insist that whether we
\textit{can} make the needed distinction between \textquotedblleft
microfacts\textquotedblright\ and\ \textquotedblleft
macrofacts\textquotedblright\ is a \textit{scientific} question---one that can
only be addressed through detailed investigation of the quantum decoherence
process and other aspects of physics. \ Second, I change the focus from
unanswerable metaphysical conundrums about what \textquotedblleft
determines\textquotedblright\ what, toward empirical questions about the
actual \textit{predictability} of our choices (or at least the probabilities
of those choices) given the past macrofacts. \ I argue that, by making
progress on the predictability question, we can learn something about whether
multiple-pasts compatibilism is a \textit{viable} response to the Consequence
Argument, even if we can never know for certain whether it's the
\textit{right} response.

\subsection{Paradoxes of Prediction\label{PARADOXES}}

\textbf{You say you're worried about the consequences for rational
decision-making, Bayesian inference, and so forth if our choices were all
mechanically predictable. \ Why isn't it reassurance enough that, logically,
predictions of an agent's behavior can never be known ahead of time to the
agent itself? \ For if they \textit{were} known ahead of time, then in real
life---as opposed to Greek tragedies, or stories like Philip K.\ Dick's
\textit{Minority Report} \cite{dick}---the agent could simply defy the
prediction by doing something else! \ Likewise, why isn't enough that, because
of the Time Hierarchy Theorem from computational complexity, predicting an
agent's choices might require as much computational power as the agent itself
expends in making those choices?}

The obvious problem with such computational or self-referential arguments is
that they can't possibly prevent one agent, say Alice, from predicting the
behavior of a \textit{different} agent, say Bob. \ And in order to do that,
Alice doesn't need unlimited computational power: \textit{she only needs a bit
more computational power than Bob has.}\footnote{This is analogous to how, in
computational complexity theory, there exists a program that that uses (say)
$n^{2.0001}$\ time steps, and that simulates any $n^{2}$-step program provided
to it as input. \ The time hierarchy theorem, which is close to tight, only
rules out the simulation of $n^{2}$-step programs using significantly
\textit{less} than $n^{2}$\ time steps.}\ \ Furthermore, Bob's free will
actually seems \textit{more} threatened by Alice predicting his actions than
by Bob predicting his own actions, supposing the latter were possible! \ This
explains why I won't be concerned in this essay with computational obstacles
to prediction, but only with obstacles that arise from Alice's physical
inability to gather the requisite information about Bob.

Admittedly, as MacKay \cite{mackay}, Lloyd \cite{lloyd:freewill},\ and others
have stressed, if Alice wants to predict Bob's choices, then she needs to be
careful not to \textit{tell} Bob her predictions before they come true!\ \ And
that does indeed make predictions of Bob's actions an unusual sort of
\textquotedblleft knowledge\textquotedblright: knowledge that can be falsified
by the very fact of Bob's learning it.

But unlike some authors, I don't make much of these observations. \ For even
if Alice can't tell Bob what he's going to do, it's easy enough for her to
demonstrate to him afterwards that \textit{she} knew. \ For example, Alice
could put her prediction into a sealed envelope, and let Bob open the envelope
only after the prediction came true.\ \ Or she could send Bob a
\textit{cryptographic commitment} to her prediction, withholding the
decryption key until afterward. \ If Alice could do these things reliably,
then it seems likely that Bob's self-conception would change just as radically
as if he knew the predictions in advance.

\subsection{Singulatarianism\label{SINGULAR}}

\textbf{How could it possibly make a difference to anyone's life whether his
or her neural computations were buffeted around by microscopic events subject
to Knightian uncertainty? \ Suppose that only \textquotedblleft
ordinary\textquotedblright\ quantum randomness and classical chaos turned out
to be involved: how on earth could that matter, outside the narrow confines of
free-will debates? \ Is the variety of free will that apparently interests
you---one based on the physical unpredictability of our choices---really a
variety \textquotedblleft worth wanting,\textquotedblright\ in Daniel
Dennett's famous phrase \cite{dennett:elbow}?}

As a first remark,\ if there's \textit{anything} in this debate that all sides
can agree on, hopefully they can agree that the truth (whatever it is) doesn't
care what we want, consider \textquotedblleft worth wanting,\textquotedblright%
\ or think is necessary or sufficient to make our lives meaningful!

But to lay my cards on the table, my interest in the issues discussed in this
essay was sparked, in part, by considering arguments of the so-called
Singulatarians. \ These are people who look at current technological
trends---including advances in neuroscience, nanotechnology, and AI, as well
as the dramatic increases in computing power---and foresee a \textquotedblleft
technological singularity\textquotedblright\ perhaps $50$-$200$ years in our
future (not surprisingly, the projected timescales vary). \ By this, they mean
not a mathematical singularity, but a rapid \textquotedblleft phase
transition,\textquotedblright\ perhaps analogous to the appearance of the
first life on earth, the appearance of humanity, or the invention of
agriculture or of writing. \ In Singulatarians' view, the next such change
will happen around the time when humans manage to build artificial
intelligences that are smarter than the smartest humans. \ It stands to
reason, the Singulatarians say, that such AIs will realize they can best
further their goals (whatever those might be) by building AIs even smarter
than themselves---and then the super-AIs will build yet smarter AIs, and so
on, presumably until the fundamental physical limits on intelligence (whatever
they are) are reached.

Following the lead of science-fiction movies, of course one might wonder about
the role of humans in the resulting world. \ Will the AIs wipe us out,
treating us as callously as humans have treated each other and most animal
species? \ Will the AIs keep us around as pets, or as revered (if rather
dimwitted) creators and ancestors? \ Will humans be invited to upload their
brains to the post-singularity computer cloud---with each human, perhaps,
living for billions of years in his or her own simulated paradise? \ Or will
the humans simply merge their consciousnesses into the AI hive-mind, losing
their individual identities but becoming part of something unimaginably greater?

Hoping to find out, many Singulatarians have signed up to have their brains
cryogenically frozen upon their (biological) deaths, so that some future
intelligence (before or after the singularity) might be able to use the
information therein to revive them.\footnote{The largest cryonics organization
is Alcor, www.alcor.org.} \ One leading Singulatarian, Eliezer Yudkowsky, has
written at length about the irrationality of people who \textit{don't} sign up
for cryonics: how they value social conformity and \textquotedblleft not being
perceived as weird\textquotedblright\ over a non-negligible probability of
living for billions of years.\footnote{See, for example,
lesswrong.com/lw/mb/lonely\_dissent/}

With some notable exceptions, academics in neuroscience and other relevant
fields have tended to dismiss Singulatarians as nerds with hyperactive
imaginations: people who have no idea how great the difficulties are in
modeling the human brain or building a general-purpose AI. \ Certainly, one
could argue that the Singulatarians' \textit{timescales} might be wildly off.
\ And even if one accepts their timescales, one could argue that (almost by
definition) the unknowns in their scenario are so great as to negate any
\textit{practical} consequences for the humans alive now. \ For example,
suppose we conclude---as many Singulatarians have---that the greatest problem
facing humanity today is how to ensure that, when superhuman AIs are finally
built, those AIs will be \textquotedblleft friendly\textquotedblright\ to
human concerns. \ The difficulty is: \textit{given our current ignorance about
AI, how on earth should we act on that conclusion?} \ Indeed, how could we
have any confidence that whatever steps we \textit{did} take wouldn't
backfire, and increase the probability of an \textit{un}friendly AI?

Yet on questions of principle---that is, of what the laws of physics could
ultimately allow---I think the uncomfortable truth is that it's the
Singulatarians who are the scientific conservatives, while those who reject
their vision as fantasy are scientific radicals. \ For at some level, all the
Singulatarians are doing is taking conventional thinking about physics and the
brain to its logical conclusion. \ If the brain is a \textquotedblleft meat
computer,\textquotedblright\ then given the right technology, why
\textit{shouldn't} we be able to copy its program from one physical substrate
to another? \ And why couldn't we then run multiple copies of the program in
parallel, leading to all the philosophical perplexities discussed in Section
\ref{UPLOADING}?

Maybe the conclusion is that we should all become Singulatarians! \ But given
the stakes, it seems worth exploring the possibility that there are scientific
reasons why human minds \textit{can't} be casually treated as copyable
computer programs: not just practical difficulties, or the sorts of
question-begging appeals to human specialness that are child's-play for
Singulatarians to demolish. \ If one likes, the origin of this essay was my
own refusal to accept the lazy cop-out position, which answers the question of
whether the Singulatarians' ideas are \textit{true} by repeating that their
ideas are crazy and weird. \ If uploading our minds to digital computers is
indeed a fantasy, then I demand to know what it is about the physical universe
that \textit{makes} it a fantasy.

\subsection{The Libet Experiments\label{LIBET}}

\textbf{Haven't neuroscience experiments already proved that our choices
aren't nearly as unpredictable as we imagine? \ Seconds before a subject is
conscious of making a decision, EEG recordings can already detect the neural
buildup to that decision. \ Given that empirical fact, isn't any attempt to
ground freedom in unpredictability doomed from the start?}

It's important to understand what experiments have and haven't shown, since
the details tend to get lost in popularization. \ The celebrated experiments
by Libet (see \cite{libet}) from the 1970s used EEGs to detect a
\textquotedblleft readiness potential\textquotedblright\ building up in a
subject's brain up to a second and a half before the subject made the
\textquotedblleft freely-willed decision\textquotedblright\ to flick her
finger. \ The implications of this finding for free will were avidly
discussed---especially since\ the subject might not have been
\textit{conscious} of any intention to flick her finger until (say) half a
second before the act. \ So, did the prior appearance of the readiness
potential prove that what we perceive as \textquotedblleft conscious
intentions\textquotedblright\ are just window-dressing, which our brains add
after the fact?

However, as Libet acknowledged, an important gap in the experiment was that it
had inadequate \textquotedblleft control.\textquotedblright\ \ That is, how
often did the readiness potential form, \textit{without} the subject flicking
her finger (which might indicate a decision that was \textquotedblleft vetoed
at the last instant\textquotedblright)? \ Because of this gap, it was unclear
to what extent the signal Libet found could actually be used for prediction.

More recent experiments---see especially Soon et al.\ \cite{soonetal}---have
tried to close this gap, by using fMRI scans to predict \textit{which} of two
buttons a person would press. \ Soon et al.\ \cite{soonetal} report that they
were able to do so four or more seconds in advance, with success probability
significantly better than chance (around $60\%$). \ The question is, how much
should we read into these findings?

My own view is that the quantitative aspects are crucial when discussing these
experiments. \ For compare a (hypothetical) ability to predict human decisions
a full minute in advance, to an ability to predict the same decisions $0.1$
seconds in advance, in terms of the intuitive \textquotedblleft
threat\textquotedblright\ to free will. \ The two cases seem utterly
different! \ A minute seems like clearly enough time for a deliberate choice,
while $0.1$ seconds seems like clearly \textit{not} enough time; on the latter
scale, we are only talking about physiological reflexes. \ (For intermediate
scales like $1$\ second, intuition---or at least my intuition---is more conflicted.)

Similarly, compare a hypothetical ability to predict human decisions with 99\%
accuracy, against an ability to predict them with 51\% accuracy. \ I expect
that only the former, and not the latter, would strike anyone as threatening
or even uncanny. \ For it's obvious that human decisions are \textit{somewhat}
predictable:\ if they weren't, there would be nothing for advertisers,
salespeople, seducers, or demagogues to exploit! \ Indeed, with zero
predictability, we couldn't even talk about \textit{personality} or
\textit{character} as having any stability over time. \ So
\textit{better-than-chance} predictability is just too low a bar for clearing
it to have any relevance to the free-will debate. \ One wants to know:
\textit{how much} better than chance? \ Is the accuracy better than what my
grandmother, or an experienced cold-reader, could achieve?

Even within the limited domain of button-pressing, years ago I wrote a program
that invited the user to press the `G' or `H' keys in any
sequence---`GGGHGHHHHGHG'---and that tried to predict which key the user would
press next. \ The program used only the crudest
pattern-matching---\textquotedblleft in the past, was the subsequence GGGH
more likely to be followed by G or H?\textquotedblright\ \ Yet humans are so
poor at generating \textquotedblleft random\textquotedblright\ digits that the
program regularly achieved prediction accuracies of around $70\%$---no fMRI
scans needed!\footnote{Soon et al.\ \cite{soonetal} (see the Supplementary
Material, p.\ 14-15) argue that, by introducing a long delay between trials
and other means, they were able to rule out the possibility that their
prediction accuracy was due purely to \textquotedblleft
carryover\textquotedblright\ between successive trials. \ They also found that
the probability of a run of $N$\ successive presses of the same button
decreased exponentially with $N$, as would be expected if the choices were
independently random. \ However, it would be interesting for future research
to compare fMRI-based prediction \textquotedblleft
head-to-head\textquotedblright\ against prediction using carefully designed
machine learning algorithms that see only the sequence of previous button
presses.}

In summary, I believe neuroscience might \textit{someday} advance to the point
where it completely rewrites the terms of the free-will debate, by showing
that the human brain is \textquotedblleft physically predictable by outside
observers\textquotedblright\ in the same sense as a digital computer. \ But it
seems nowhere close to that point today. \ Brain-imaging experiments have
succeeded in demonstrating predictability with better-than-chance accuracy, in
limited contexts and over short durations. \ Such experiments are deeply
valuable and worth trying to push as far as possible. \ On the other hand, the
mere \textit{fact} of limited predictability is something that humans knew
millennia before brain-imaging technologies became available.

\subsection{Mind and Morality\label{MIND}}

\textbf{Notwithstanding your protests that you won't address the
\textquotedblleft mystery of consciousness,\textquotedblright\ your entire
project seems geared toward the old, disreputable quest to find some sort of
dividing line between \textquotedblleft real, conscious,
biological\textquotedblright\ humans and digital computer programs, even
supposing that the latter could perfectly emulate the former. \ Many thinkers
have sought such a line before, but most scientifically-minded people regard
the results as dubious. \ For Roger Penrose, the dividing line involves neural
structures called microtubules harnessing exotic quantum-gravitational effects
(see Section \ref{PENROSE}). \ For the philosopher John Searle
\cite{searle:redis}, the line involves the brain's unique \textquotedblleft
biological causal powers\textquotedblright:\ powers whose existence Searle
loudly asserts but never actually explains. \ For you, the line seems to
involve hypothesized limits on predictability imposed by the No-Cloning
Theorem. \ But regardless of where the line gets drawn, let's discuss where
the rubber meets the road. \ Suppose that in the far future, there are
trillions of emulated brains (or \textquotedblleft ems\textquotedblright)
running on digital computers. \ In such a world, would you consider it
acceptable to \textquotedblleft murder\textquotedblright\ an em (say, by
deleting its file), simply because it lacked\ the \textquotedblleft Knightian
unpredictability\textquotedblright\ that biological brains might or might not
derive from amplified quantum fluctuations? \ If so, then isn't that a cruel,
arbitrary, \textquotedblleft meatist\textquotedblright\ double standard---one
that violates the most basic principles of your supposed hero, Alan Turing?}

For me, this \textit{moral} objection to my project is possibly the most
pressing objection of all. \ Will \textit{I} be the one to shoot a humanoid
robot pleading for its life, simply because the robot lacks the
supposedly-crucial \textquotedblleft freebits,\textquotedblright\ or
\textquotedblleft Knightian unpredictability,\textquotedblright\ or whatever
else the magical stuff is supposed to be that separates humans from machines?

Thus, perhaps my most important reason to take the freebit picture seriously
is that it \textit{does} suggest a reply to the objection: one that strikes me
as both intellectually consistent and moral. \ I simply need to adopt the
following ethical stance: \textit{I'm against any irreversible destruction of
knowledge, thoughts, perspectives, adaptations, or ideas, except possibly by
their owner.} \ Such destruction is worse the more valuable the thing
destroyed, the longer it took to create, and the harder it is to replace.
\ From this basic revulsion to \textit{irreplaceable loss}, hatred of murder,
genocide, the hunting of endangered species to extinction, and even (say) the
burning of the Library of Alexandria can all be derived as consequences.

Now, what about the case of \textquotedblleft deleting\textquotedblright\ an
emulated human brain from a computer memory? \ The same revulsion applies in
full force---\textit{if the copy deleted is the last copy in existence}. \ If,
however, there are other extant copies, then the deleted copy can always be
\textquotedblleft restored from backup,\textquotedblright\ so deleting it
seems at worst like property damage. \ For biological brains, by contrast,
whether such backup copies \textit{can} be physically created is of course
exactly what's at issue, and the freebit picture conjectures a negative answer.

These considerations suggest that the moral status of ems really
\textit{could} be different than that of organic humans, but for
straightforward practical reasons that have nothing to do with
\textquotedblleft meat chauvinism,\textquotedblright\ or with question-begging
philosophical appeals to human specialness.\ \ The crucial point is that even
a program that passed the Turing Test would revert to looking
\textquotedblleft crude and automaton-like,\textquotedblright\ \textit{if you
could read, trace, and copy its code}. \ And whether the code \textit{could}
be read and copied might depend strongly on the machine's physical substrate.
\ Destroying something that's both as complex as a human being \textit{and}
one-of-a-kind could be regarded as an especially heinous crime.

I see it as the great advantage of this reply that it makes no direct
reference to the first-person experience of \textit{anyone}, neither
biological humans nor ems. \ \ On this account, we don't \textit{need} to
answer probably-unanswerable questions about whether or not ems would be
conscious, in order to constructed a workable moral code that applies to them.
\ Deleting the last copy of an em in existence should be prosecuted as murder,
\textit{not} because doing so snuffs out some inner light of consciousness
(who is anyone else to know?), but rather because it deprives the rest of
society of a unique, irreplaceable store of knowledge and experiences,
precisely as murdering a human would.

\section{Knightian Uncertainty and Physics\label{KNIGHTIANPHYS}}

Having spent almost half the essay answering \textit{a priori} objections to
the investigation I want to undertake, I'm finally ready to start the
investigation itself! \ In this section, I'll set out and defend two
propositions, both of which are central to what follows.

The first proposition is that \textit{probabilistic uncertainty (like that of
quantum-mechanical measurement outcomes) can't possibly, by itself, provide
the sort of \textquotedblleft indeterminism\textquotedblright\ that could be
relevant to the free-will debate.} \ In other words, \textit{if} we see a
conflict between free will and the deterministic predictability of human
choices, then we should see the same conflict between free will and
\textit{probabilistic} predictability, assuming the probabilistic predictions
are as accurate as those of quantum mechanics. \ Conversely, if we hold free
will to be compatible with \textquotedblleft quantum-like
predictability,\textquotedblright\ then we might as well hold free will to be
compatible with \textit{deterministic} predictability also. \ In my
perspective, for a form of uncertainty to be relevant to free will, a
necessary condition is that it be what the economists call \textit{Knightian
uncertainty}.\ \ Knightian uncertainty simply refers to uncertainty that we
lack a clear, agreed-upon way to quantify---like our uncertainty about
existence of extraterrestrial life, as opposed to our uncertainty about the
outcome of a coin toss.

The second proposition is that, in current physics, there appears to be only
one source of Knightian uncertainty that could possibly be both fundamental
and relevant to human choices. \ That source is \textit{uncertainty about the
microscopic, quantum-mechanical details of the universe's initial conditions
(or the initial conditions of our local region of the universe)}. \ In
classical physics, there's no known fundamental principle that prevents a
predictor from learning the relevant initial conditions to whatever precision
it likes, without disturbing the system to be predicted. \ But in quantum
mechanics there \textit{is} such a principle, namely the uncertainty
principle\ (or from a more \textquotedblleft modern\textquotedblright%
\ standpoint, the No-Cloning Theorem). \ It's crucial to understand that this
source of uncertainty is separate from the randomness of quantum measurement
\textit{outcomes}: the latter is much more often invoked in free-will
speculations, but in my opinion it shouldn't be. \ If we know a system's
quantum state $\rho$, then quantum mechanics lets us calculate the probability
of any outcome of any measurement that might later be made on the system.
\ But if we \textit{don't} know the state $\rho$, then $\rho$\ itself can be
thought of as subject to Knightian uncertainty.

In the next two subsections, I'll expand on and justify the above claims.

\subsection{Knightian Uncertainty\label{KNIGHTIAN}}

A well-known argument maintains that the very concept of free will is\textit{
}logically incoherent. \ The argument goes like this:

\begin{quotation}
\noindent Any event is either determined by earlier events\ (like the return
of Halley's comet),\ or else not determined by earlier events\ (like the decay
of a radioactive atom).\ \ If the event is determined, then clearly it isn't
\textquotedblleft free.\textquotedblright\ \ But if the event is
\textit{un}determined, it isn't \textquotedblleft free\textquotedblright%
\ either: it's merely arbitrary, capricious, and random. \ Therefore no event
can be \textquotedblleft free.\textquotedblright
\end{quotation}

As I'm far from the first to point out, the above argument has a gap,
contained in the vague phrase \textquotedblleft arbitrary, capricious, and
random.\textquotedblright\ \ An event can be \textquotedblleft
arbitrary,\textquotedblright\ in the sense of being undetermined by previous
events, without being random\ in the narrower technical sense of being
generated by some known or knowable probabilistic process. \ The distinction
between arbitrary\ and random\ is not just word-splitting: it plays a huge
practical role in computer science, economics, and other fields. \ To
illustrate, consider the following two events:

\begin{quotation}
\noindent$E_{1}=$ Three weeks from today, the least significant digit of the
Dow Jones average will be even

\noindent$E_{2}=$ Humans will make contact with an extraterrestrial
civilization within the next $500$\ years
\end{quotation}

For both events, we are ignorant about whether they will occur, but we are
ignorant in completely different ways. \ For\ $E_{1}$, we can quantify our
ignorance by a probability distribution, in such a way that almost any
reasonable person would agree with our quantification. \ For $E_{2}$, we can't.

For another example, consider a computer program, which has a bug that only
appears when a call to the random number generator\ returns the result $3456$.
\ That's not necessarily a big deal---since with high probability, the program
would need to be run thousands of times before the bug reared its head.
\ Indeed, today many problems are solved using \textit{randomized algorithms}
(such as Monte-Carlo simulation), which \textit{do} have a small but nonzero
probability of failure.\footnote{The probability of failure can be made
arbitrarily small by the simple expedient of running the algorithm over and
over and taking a majority vote.} \ On the other hand, if the program has a
bug that only occurs when the user \textit{inputs} $3456$, that's a much more
serious problem. \ For how can the programmer know, in advance, whether
$3456$\ is an input (maybe even the \textit{only} input) that the user cares
about? \ So a programmer \textit{must} treat the two types of uncertainty
differently: she can't just toss them both into a bin labeled
\textquotedblleft arbitrary, capricious, and random.\textquotedblright\ \ And
indeed, the difference between the two types of uncertainty shows up
constantly in theoretical computer science and information
theory.\footnote{For example, it shows up in the clear distinctions made
between random and adversarial noise, between probabilistic and
nondeterministic Turing machines, and between average-case and worst-case
analysis of algorithms.}

In\ economics, the \textquotedblleft second\ type\textquotedblright\ of
uncertainty---the type that can't be objectively quantified using
probabilities---is called \textit{Knightian uncertainty}, after Frank Knight,
who wrote about it extensively in the 1920s \cite{knight}. \ Knightian
uncertainty has been invoked to explain phenomena from risk-aversion in
behavioral economics to the 2008 financial crisis (and was popularized by
Taleb \cite{taleb} under the name \textquotedblleft black
swans\textquotedblright). \ An agent in a state of Knightian uncertainty might
describe its beliefs using a convex set of probability distributions, rather
than a single distribution.\footnote{In Appendix \ref{FREESTATES}, I briefly
explain why I think convex sets provide the \textquotedblleft
right\textquotedblright\ representation of Knightian uncertainty, though this
point doesn't much matter for the rest of the essay.} \ For example, it might
say that a homeowner will default on a mortgage with some probability between
$0.1$\ and $0.3$, but within that interval, be unwilling to quantify its
uncertainty further. \ The notion of probability intervals leads naturally to
various generalizations of probability theory, of which the best-known is the
\textit{Dempster-Shafer theory of belief} \cite{shafer}.

What does any of this have to do with the free-will debate? \ As I said in
Section \ref{FWFREEDOM}, from my personal perspective Knightian uncertainty
seems like a \textit{precondition} for free will as I understand the latter.
\ In other words, I \textit{agree} with the free-will-is-incoherent camp when
it says that a random event can't be considered \textquotedblleft
free\textquotedblright\ in any meaningful sense. \ Several writers, such as
Kane \cite{fourviews}, Balaguer \cite{balaguer}, Satinover \cite{satinover},
and Koch \cite{koch}, have speculated that the randomness inherent in
quantum-mechanical wavefunction collapse, were it relevant to brain function,
could provide all the scope for free will that's needed. \ But I think those
writers are mistaken on this point.

For me, the bottom line is simply that it seems like a sorry and pathetic
\textquotedblleft free will\textquotedblright\ that's ruled by ironclad,
externally-knowable statistical laws, and that retains only a
\textquotedblleft ceremonial\textquotedblright\ role, analogous to spinning a
roulette wheel or shuffling a deck of cards. \ Should we say that a
radioactive nucleus has \textquotedblleft free will,\textquotedblright\ just
because (according to quantum mechanics) we can't predict exactly when it will
decay, but can only calculate a precise probability distribution over the
decay times? \ That seems perverse---especially since given \textit{many}
nuclei, we can predict almost perfectly what \textit{fraction} will have
decayed by such-and-such a time. \ Or imagine an artificially-intelligent
robot that used nuclear decays as a source of random numbers. \ Does anyone
seriously maintain that, if we swapped out the actual decaying nuclei for a
\textquotedblleft mere\textquotedblright\ pseudorandom computer simulation of
such nuclei\ (leaving all other components unchanged), the robot would
suddenly be robbed of its free will? \ While I'm leery of \textquotedblleft
arguments from obviousness\textquotedblright\ in this subject, it really
\textit{does} seem to me that if we say the robot has free will\ in the first
case then we should also say so in the second.

And thus, I think that the free-will-is-incoherent camp would be right,
\textit{if} all uncertainty were probabilistic. \ But I consider it far from
obvious that all uncertainty \textit{is} (usefully regarded as) probabilistic.
\ Some uncertainty strikes me as Knightian, in the sense that rational people
might never even reach agreement about how to assign the probabilities. \ And
while Knightian uncertainty might or might not be relevant to predicting human
choices, I definitely (for reasons I'll discuss later) don't think that
current knowledge of physics or neuroscience lets us exclude the possibility.

At this point, though, we'd better hear from those who reject the entire
concept of Knightian uncertainty. \ Some thinkers---I'll call them
\textit{Bayesian fundamentalists}---hold that Bayesian probability theory
provides the only sensible way to represent uncertainty. \ On that view,
\textquotedblleft Knightian uncertainty\textquotedblright\ is just a fancy
name for someone's failure to carry a probability analysis far enough.

In support of their position, Bayesian fundamentalists often invoke the
so-called \textit{Dutch book arguments} (see for example Savage \cite{savage}%
), which say that any rational agent satisfying a few axioms must behave
\textit{as if} its beliefs were organized using probability theory.
\ Intuitively, even if you claim not to have any opinion whatsoever about
(say) the probability of life being discovered on Mars, I can still
\textquotedblleft elicit\textquotedblright\ a probabilistic prediction from
you, by observing which bets about the question you will or won't accept.

However, a central assumption on which the Dutch book arguments
rely---basically, that a rational agent shouldn't mind taking at least one
side of any bet---has struck many commentators as dubious. \ And if we drop
that assumption, then the path is open to Knightian uncertainty (involving,
for example, convex \textit{sets} of probability distributions).

Even if we accept the standard derivations of probability theory, the bigger
problem is that Bayesian agents can have different \textquotedblleft
priors.\textquotedblright\ \ If one strips away the philosophy, Bayes' rule is
just an elementary mathematical fact about how one should update one's prior
beliefs in light of new evidence. \ So one can't use Bayesianism to justify a
belief in the existence of \textit{objective} probabilities underlying all
events, unless one is also prepared to defend the existence of an
\textquotedblleft objective prior.\textquotedblright\ \ In economics, the idea
that all rational agents can be assumed to start with the same prior is called
the \textit{Common Prior Assumption}, or CPA. \ Assuming the CPA leads to some
wildly counterintuitive consequences, most famously \textit{Aumann's agreement
theorem} \cite{aumann}. \ That theorem says that two rational agents with
common priors (but differing information) can never \textquotedblleft agree to
disagree\textquotedblright: as soon as their opinions on any subject become
common knowledge, their opinions must be equal.

The CPA has long been controversial; see Morris \cite{morris} for a summary of
arguments for and against. \ To my mind, though, the real question is:
\textit{what could possibly have led anyone to take the CPA seriously in the
first place?}

Setting aside methodological arguments in economics\footnote{E.g., that even
if the CPA is wrong, we should assume it because economic theorizing would be
too unconstrained without it, or because many interesting theorems need it as
a hypothesis.} (which don't concern us here), I'm aware of two substantive
arguments in favor of the CPA. \ The first argument\ is that, if two rational
agents (call them Alice and Bob) have different priors, then Alice will
realize that \textit{if she had been born Bob}, she would have had Bob's
prior, and Bob will realize that if he had been born Alice, he would have had
Alice's. \ But if Alice and Bob are indeed rational, then why should they
assign any weight to personal accidents of their birth, which are clearly
irrelevant to the objective state of the universe? \ (See Cowen and Hanson
\cite{cowenhanson} for a detailed discussion of this argument.)

The simplest reply is that, even if Alice and Bob accepted this reasoning,
they would \textit{still} generally end up with different priors, unless they
furthermore shared the same \textit{reference class}: that is, the set of all
agents who they imagine they \textquotedblleft could have
been\textquotedblright\ if they weren't themselves. \ For example, if Alice
includes all humans in her reference class, while Bob includes only those
humans capable of understanding Bayesian reasoning such as he and Alice are
engaging in now, then their beliefs will differ. \ But requiring agreement on
the reference class makes the argument circular---presupposing, as it does, a
\textquotedblleft God's-eye perspective\textquotedblright\ transcending the
individual agents, the very thing whose existence or relevance was in
question. \ Section \ref{INDEXFREE} will go into more detail about
\textquotedblleft indexical\textquotedblright\ puzzles (that is, puzzles
concerning the probability of your own existence, the likelihood of having
been born at one time rather than another, etc). \ But I hope this discussion
already makes clear how much debatable metaphysics lurks in the assumption
that a single Bayesian prior governs (or \textit{should} govern) every
probability judgment of every rational being.

The second argument for the CPA is more ambitious: it seeks to tell us
\textit{what} the true prior is, not merely that it exists. \ According to
this argument, any sufficiently intelligent being ought to use what's called
the \textit{universal prior} from algorithmic information theory. \ This is
basically a distribution that assigns a probability proportional to\ $2^{-n}%
$\ to every possible universe describable by an $n$-bit computer program. \ In
Appendix \ref{KOLMOG},\ I'll examine this notion further, explain why some
people \cite{hutter,schmidhuber}\ have advanced it as a candidate for the
\textquotedblleft true\textquotedblright\ prior, but also explain why, despite
its mathematical interest, I don't think it can fulfill that role. \ (Briefly,
a predictor using the universal prior can be thought of as a superintelligent
entity that figures out the right probabilities almost as fast as is
information-theoretically possible. \ But that's conceptually very different
from an entity that \textit{already knows} the probabilities.)

\subsection{Quantum Mechanics and the No-Cloning Theorem\label{NOCLONE}}

While defending the meaningfulness of Knightian uncertainty, the last section
left an obvious question unanswered: where, in a law-governed universe, could
we possibly \textit{find} Knightian uncertainty?

Granted, in almost any part of science, it's easy to find systems that are
\textquotedblleft effectively\textquotedblright\ subject to Knightian
uncertainty, in that we don't yet have models for the systems that capture all
the important components and their probabilistic interactions. \ The earth's
climate, a country's economy, a forest ecosystem, the early universe, a
high-temperature superconductor, or even a flatworm or a cell are examples.
\ Even if our probabilistic models of many of these systems are improving over
time, none of them come anywhere close to (say) the quantum-mechanical model
of the hydrogen atom: a model that answers essentially \textit{everything} one
could think to ask within its domain, modulo an unavoidable (but
precisely-quantifiable) random element.

However, in all of these cases (the earth's climate, the flatworm, etc.), the
question arises: what grounds could we ever have to think that Knightian
uncertainty was \textit{inherent} to the system, rather than an artifact of
our own ignorance? \ Of course, one could have asked the same question about
probabilistic uncertainty, before the discovery of quantum mechanics and its
no-hidden-variable theorems (see Section \ref{QMHV}). \ But the fact remains
that today, we don't have any physical theory that demands Knightian
uncertainty in anything like the way that quantum mechanics demands
probabilistic uncertainty. \ And as I said in Section \ref{KNIGHTIAN}, I
insist that the \textquotedblleft merely probabilistic\textquotedblright%
\ aspect of quantum mechanics can't do the work that many free-will advocates
have wanted it to do for a century.

On the other hand, no matter how much we've learned about the dynamics of the
physical world, there remains one enormous source of Knightian uncertainty
that's been \textquotedblleft hiding in plain sight,\textquotedblright\ and
that receives surprisingly little attention in the free-will debate. \ This is
our ignorance of the relevant \textit{initial conditions}. \ By this I mean
both the initial conditions of the entire universe or multiverse (say, at the
Big Bang), and the \textquotedblleft indexical conditions,\textquotedblright%
\ which characterize the part of the universe or multiverse in which
\textit{we} happen to reside. \ To make a prediction, of course one\ needs
initial conditions as well as dynamical laws: indeed, outside of idealized toy
problems, the initial conditions are typically the \textquotedblleft
larger\textquotedblright\ part of the input. \ Yet leaving aside recent
cosmological speculations (and \textquotedblleft genericity\textquotedblright%
\ assumptions, like those of thermodynamics), the specification of initial
conditions is normally not even considered the \textit{task} of physics. \ So,
if there are no laws that fix the initial conditions, or even a distribution
over possible initial conditions---if there aren't even especially promising
\textit{candidates} for such laws---then why isn't this just what free-will
advocates have been looking for?

It will be answered immediately\ that there's an excellent reason why not.
\ Namely, whatever else we do or don't know about the universe's initial state
(e.g., at the Big Bang), clearly nothing about it was determined by any of
\textit{our} choices! \ (This is the assumption made explicit in step (ii) of
van Inwagen's \textquotedblleft Consequence Argument\textquotedblright\ from
Section \ref{CONSEQUENCE}.)

The above answer might strike the reader as conclusive. \ Yet \textit{if} our
interest is in actual, physical predictability---rather than in the
metaphysical concept of \textquotedblleft determination\textquotedblright%
---then notice that it's no longer conclusive at all. \ For we still need to
ask: how much can we \textit{learn} about the initial state by making
measurements? \ This, of course, is where quantum mechanics might become relevant.

It's actually easiest for our purposes to forget the famous
\textit{uncertainty principle}, and talk instead about the \textit{No-Cloning
Theorem}. \ The latter is simply the statement that there's no physical
procedure, consistent with quantum mechanics, that takes as input a system in
an arbitrary quantum state $\left\vert \psi\right\rangle $,\footnote{Here we
assume for simplicity that we're talking about pure states, not mixed states.}
and outputs two systems \textit{both} in the state $\left\vert \psi
\right\rangle $.\footnote{The word \textquotedblleft
arbitrary\textquotedblright\ is needed because, if we knew how $\left\vert
\psi\right\rangle $\ was prepared, then of course we could simply run the
preparation procedure a second time. \ The No-Cloning Theorem implies that, if
we \textit{don't} already know a preparation procedure for $\left\vert
\psi\right\rangle $, then we can't learn one just by measuring $\left\vert
\psi\right\rangle $. \ (And conversely, the inability to learn a preparation
procedure implies the No-Cloning Theorem. \ If we could copy $\left\vert
\psi\right\rangle $\ perfectly, then we could keep repeating to make as many
copies as we wanted, then use quantum state tomography on the copies to learn
the amplitudes to arbitrary accuracy.)} \ Intuitively, it's not hard to see
why: a measurement of (say) a qubit $\left\vert \psi\right\rangle
=\alpha\left\vert 0\right\rangle +\beta\left\vert 1\right\rangle $ reveals
only a single, probabilistic bit of information about the continuous
parameters $\alpha$\ and $\beta$; the rest of the information vanishes
forever. \ The proof of the No-Cloning Theorem (in its simplest version) is as
easy as observing that the \textquotedblleft cloning map,\textquotedblright%
\begin{align*}
\left(  \alpha\left\vert 0\right\rangle +\beta\left\vert 1\right\rangle
\right)  \left\vert 0\right\rangle  &  \longrightarrow\left(  \alpha\left\vert
0\right\rangle +\beta\left\vert 1\right\rangle \right)  \left(  \alpha
\left\vert 0\right\rangle +\beta\left\vert 1\right\rangle \right) \\
&  =\alpha^{2}\left\vert 0\right\rangle \left\vert 0\right\rangle +\alpha
\beta\left\vert 0\right\rangle \left\vert 1\right\rangle +\alpha
\beta\left\vert 1\right\rangle \left\vert 0\right\rangle +\beta^{2}\left\vert
1\right\rangle \left\vert 1\right\rangle ,
\end{align*}
acts nonlinearly on the amplitudes, but in quantum mechanics, unitary
evolution must be linear.

Yet despite its mathematical triviality, the No-Cloning Theorem has the deep
consequence that quantum states have a certain \textquotedblleft
privacy\textquotedblright: unlike classical states, they can't be copied
promiscuously around the universe. \ One way to gain intuition for the
No-Cloning Theorem is to consider some striking cryptographic protocols that
rely on it. \ In \textit{quantum key distribution} \cite{bb84}---something
that's already (to a small extent) been commercialized and deployed---a
sender, Alice, transmits a secret key to a receiver, Bob, by encoding the key
in a collection of qubits. \ The crucial point is that if an eavesdropper,
Eve, tries to learn the key by measuring the qubits, then the very fact that
she measured the qubits will be detectable by Alice and Bob---so Alice and Bob
can simply keep trying until the channel is safe. \ \textit{Quantum money},
proposed decades ago by Wiesner \cite{wiesner} and developed further in recent
years \cite{aar:qcopy,achristiano,knots}, would exploit the No-Cloning Theorem
even more directly, to create cash that can't be counterfeited according to
the laws of physics, but that can nevertheless be verified as
genuine.\footnote{Unfortunately, unlike quantum key distribution, quantum
money is not yet practical, because of the difficulty of protecting quantum
states against decoherence for long periods of time.} \ A closely related
proposal, \textit{quantum software copy-protection} (see
\cite{aar:qcopy,achristiano}), would exploit the No-Cloning Theorem in a still
more dramatic way: to create quantum states $\left\vert \psi_{f}\right\rangle
$\ that can be used to evaluate some function $f$, but that can't feasibly be
used to create more states with which $f$ can be evaluated. \ Research on
quantum copy-protection has shown that, at least in a few special cases (and
maybe more broadly), it's possible to create a physical object that

\begin{enumerate}
\item[(a)] interacts with the outside world in an interesting and nontrivial
way, yet

\item[(b)] effectively hides from the outside world the information needed to
predict how the object will behave in future interactions.
\end{enumerate}

When put that way, the \textit{possible} relevance of the No-Cloning Theorem
to free-will discussions seems obvious! \ And indeed, a few bloggers and
others\footnote{See, for example,
www.daylightatheism.org/2006/04/on-free-will-iii.html} have previously
speculated about a connection. \ Interestingly, their motivation for doing so
has usually been to \textit{defend compatibilism} (see Section \ref{COMPAT}).
\ In other words, they've invoked the No-Cloning Theorem to explain why,
despite the mechanistic nature of physical law, human decisions will
nevertheless remain unpredictable \textit{in practice}. \ In a discussion of
this issue, one commenter\footnote{Sadly, I no longer have the reference.}
opined that, while the No-Cloning Theorem \textit{does} put limits on physical
predictability, human brains will also remain unpredictable for countless more
prosaic\ reasons that have nothing to do with quantum mechanics. \ Thus, he
said, invoking the No-Cloning Theorem in free-will discussions is
\textquotedblleft like hiring the world's most high-powered lawyer to get you
out of a parking ticket.\textquotedblright

Personally, though, I think the world's most high-powered lawyer might
ultimately be needed here! \ For the example of the Singulatarians (see
Section \ref{SINGULAR}) shows why, in these discussions, it doesn't suffice to
offer \textquotedblleft merely\ practical\textquotedblright\ reasons why
copying a brain state is hard. \ Every practical problem can easily be
countered by a speculative technological answer---one that assumes a future
filled with brain-scanning nanorobots and the like. \ If we want a proposed
obstacle to copying to survive unlimited technological imagination, then the
obstacle had better be grounded in the laws of physics.

So as I see it, the real question is: once we disallow quantum mechanics, does
there remain any \textit{classical} source of \textquotedblleft fundamental
unclonability\textquotedblright\ in physical law? \ Some might suggest, for
example, the impossibility of keeping a system perfectly isolated from its
environment during the cloning process, or the impossibility of measuring
continuous particle positions to infinite precision. \ But either of these
would require very nontrivial arguments (and if one wanted to invoke
continuity, one would also have to address the apparent breakdown of
continuity at the Planck scale). \ There are also formal analogues of the
No-Cloning Theorem in various classical settings, but none of them seem able
to do the work required here.\footnote{For example, No-Cloning holds for
classical probability distributions: there's no procedure that takes an input
bit $b$\ that equals $1$ with unknown probability $p$, and produces two output
bits that are both $1$ with probability $p$ independently. \ But this
observation lacks the import of the quantum No-Cloning Theorem, because
regardless of what one wanted to do with the bit $b$, one might as well have
\textit{measured} it immediately, thereby \textquotedblleft
collapsing\textquotedblright\ it to a deterministic bit---which \textit{can}
of course be copied.
\par
Also, seeking to clarify foundational issues in quantum mechanics, Spekkens
\cite{spekkens}\ constructed an \textquotedblleft epistemic toy
theory\textquotedblright\ that's purely classical, but where an analogue of
the No-Cloning Theorem holds. \ However, the toy theory involves
\textquotedblleft magic boxes\textquotedblright\ that we have no reason to
think can be physically realized.} \ As far as current physics can say,
\textit{if} copying a bounded physical system is fundamentally impossible,
then the reason for the impossibility seems ultimately traceable to quantum mechanics.

Let me end this section by discussing \textit{quantum teleportation}, since
nothing more suggestively illustrates the \textquotedblleft philosophical
work\textquotedblright\ that the No-Cloning Theorem can potentially do.
\ Recall, from Section \ref{UPLOADING}, the \textquotedblleft
paradox\textquotedblright\ raised by teleportation machines. \ Namely, after a
perfect copy of you has been reconstituted on Mars from the information in
radio signals, what should be done with the \textquotedblleft
original\textquotedblright\ copy of you back on Earth? \ Should it be euthanized?

Like other such paradoxes, this one need not trouble us if (because of the
No-Cloning Theorem, or whatever other reason) we drop the assumption that such
copying is possible---at least with great enough fidelity for the copy on Mars
to be a \textquotedblleft second instantiation of you.\textquotedblright%
\ \ However, what makes the situation more interesting is that there
\textit{is} a famous protocol, discovered by Bennett et al.\ \cite{bbcjpw},
for \textquotedblleft teleporting\textquotedblright\ an arbitrary quantum
state $\left\vert \psi\right\rangle $\ by sending classical information only.
\ (This protocol also requires quantum entanglement, in the form of
\textit{Bell pairs }$\frac{\left\vert 00\right\rangle +\left\vert
11\right\rangle }{\sqrt{2}}$,\textit{\ }shared between the sender and receiver
in advance, and one Bell pair gets \textquotedblleft used up\textquotedblright%
\ for every qubit teleported.)

Now, a crucial feature of the teleportation protocol is that, in order to
determine which classical bits to send, the sender needs to perform a
measurement on her quantum state $\left\vert \psi\right\rangle $\ (together
with her half of the Bell pair) that \textit{destroys} $\left\vert
\psi\right\rangle $. \ In other words, in quantum teleportation, the
destruction of the original copy is not an extra decision that one needs to
make; rather, it happens as an inevitable byproduct of the protocol itself!
\ Indeed this must be so, since otherwise quantum teleportation could be used
to violate the No-Cloning Theorem.

\subsection{The Freebit Picture\label{FREEBITS}}

At this point one might interject: theoretical arguments about the No-Cloning
Theorem are well and good, but even if accepted, they still don't provide any
concrete picture of how Knightian uncertainty could be relevant to human decision-making.

So let me sketch a possible picture (the only one I can think of, consistent
with current physics), which I call the \textquotedblleft freebit
picture.\textquotedblright\ \ At the Big Bang, the universe had some
particular quantum state $\left\vert \Psi\right\rangle $. \ If known,
$\left\vert \Psi\right\rangle $\ would of course determine the universe's
future history, modulo the probabilities arising from quantum measurement.
\ However, \textit{because} $\left\vert \Psi\right\rangle $\ is the state of
the whole universe (including us), we might refuse to take a \textquotedblleft
God's-eye view\textquotedblright\ of $\left\vert \Psi\right\rangle $, and
insist on thinking about $\left\vert \Psi\right\rangle $\ differently than
about an ordinary state that we prepare for ourselves in the lab. \ In
particular, we might regard at least some (not all) of the qubits that
constitute $\left\vert \Psi\right\rangle $\ as what I'll call
\textit{freebits}. \ A freebit is simply a qubit for which the most complete
physical description possible involves Knightian uncertainty.\ \ While the
details aren't so important, I give a brief mathematical account of freebits
in Appendix \ref{FREESTATES}. \ For now, suffice it to say that a
\textit{freestate} is a convex set of quantum mixed states, and a freebit is a
$2$-level quantum system in a freestate.

Thus, by the \textit{freebit picture}, I mean the picture of the world
according to which

\begin{enumerate}
\item[(i)] due to Knightian uncertainty about the universe's initial quantum
state $\left\vert \Psi\right\rangle $, at least some of the qubits found in
nature are regarded as freebits, and

\item[(ii)] the presence of these freebits makes predicting certain future
events---possibly including some human decisions---physically impossible, even
probabilistically and even with arbitrarily-advanced future technology.
\end{enumerate}

Section \ref{CHAOS} will say more about the \textquotedblleft
biological\textquotedblright\ aspects of the freebit picture: that is, the
actual chains of causation that could in principle connect freebits to (say) a
neuron firing or not firing. \ In the rest of this section, I'll discuss some
physical and conceptual questions about freebits themselves.

Firstly, why is it important that freebits be qubits, rather than
\textit{classical} bits subject to Knightian uncertainty? \ The answer is that
only for qubits can we appeal to the No-Cloning Theorem. \ Even if the value
of a classical bit $b$\ can't determined by measurements on the entire rest of
the universe, a superintelligent predictor could always learn $b$ by
non-invasively measuring $b$ \textit{itself}.\ \ But the same is not true for
a qubit.

Secondly, isn't Knightian uncertainty in the eye of the beholder? \ That is,
why couldn't one observer regard a given qubit as a \textquotedblleft
freebit,\textquotedblright\ while a different observer, with more information,
described the same qubit by ordinary quantum mixed state $\rho$? \ The answer
is that our criterion for what counts a freebit is extremely stringent.
\ Given a $2$-level quantum system $S$, \textit{if a superintelligent demon
could reliably learn the reduced density matrix }$\rho$\textit{\ of }%
$S$\textit{, via arbitrary measurements on anything in the universe (including
}$S$\textit{\ itself), then }$S$\textit{ is not a freebit.} \ Thus, to qualify
as a freebit, $S$ must be a \textquotedblleft freely moving
part\textquotedblright\ of the universe's quantum state $\left\vert
\Psi\right\rangle $: it must not be possible (even in principle) to trace
$S$'s causal history back to any physical process that generated $S$ according
to a known probabilistic ensemble. \ Instead, our Knightian uncertainty about
$S$ must (so to speak) go \textquotedblleft all the way
back,\textquotedblright\ and be traceable to uncertainty about the initial
state of the universe.

To illustrate the point:\ suppose we detect a beam of photons with varying
polarizations. \ For the most part, the polarizations look uniformly random
(i.e., like qubits in the maximally mixed state). \ But there is a slight bias
toward the vertical axis, and the bias is slowly changing over time, in a way
not fully accounted for by our model of the photons. \ So far, we can't rule
out the possibility that freebits might be involved. \ However, suppose we
later learn that the photons are coming from a laser in another room, and that
the polarization bias is due to drift in the laser that can be characterized
and mostly corrected. \ Then the scope for freebits is correspondingly reduced.

Someone might interject: \textquotedblleft but \textit{why} was there drift in
the laser? \ Couldn't freebits have been responsible for the drift
itself?\textquotedblright\ \ The difficulty is that,\ \textit{even if so}, we
still couldn't use those freebits to argue for Knightian uncertainty in the
laser's \textit{output}. \ For between the output photons, and whatever
freebits might have caused the laser to be configured as it was, there stands
a classical, macroscopic intermediary: the laser itself. \ If a demon had
wanted to predict the polarization drift in the output photons, the demon
could simply have traced the photons back to the laser, then
\textit{non-invasively measured the laser's classical degrees of
freedom}---cutting off the causal chain there and ignoring any further
antecedents. \ In general, given some quantum measurement outcome $Q$ that
we're trying to predict, if there exists a classical observable $C$ that could
have been non-invasively measured long before $Q$, and that if measured, would
have let the demon probabilistically predict $Q$ to arbitrary accuracy (in the
same sense that radioactive decay is probabilistically predictable), then I'll
call $C$ a \textit{past macroscopic determinant (PMD)} for $Q$.

In the freebit picture, we're exclusively interested in the quantum
states---if there are any!---that \textit{can't} be grounded in PMDs, but can
only be traced all the way back\ to the early universe, with no macroscopic
intermediary along the way that \textquotedblleft screen off\textquotedblright%
\ the early universe's causal effects. \ The reason is simple: because such
states, if they exist, are the only ones that our superintelligent demon, able
to measure all the macroscopic observables in the universe, would
\textit{still} have Knightian uncertainty about. \ In other words, such states
are the only possible freebits.

Of course this immediately raises a question:

\begin{quotation}
\noindent\textbf{(*) In the actual universe, \textit{are} there any quantum
states that can't be grounded in PMDs?}
\end{quotation}

A central contention of this essay is that pure thought doesn't suffice to
answer question (*):\ here we've reached the limits of where conceptual
analysis can take us. \ There are possible universes consistent with the rules
of quantum mechanics where the requisite states exist, and other such
universes where they don't exist,\ and deciding which kind of universe
\textit{we} inhabit seems to require scientific knowledge that we don't have.

Some people, while agreeing that logic and quantum mechanics don't suffice to
settle question (*), would nevertheless say we can settle it using simple
facts about astronomy. \ At least near the surface of the earth, they'd ask,
what quantum states could there possibly be that \textit{didn't} originate in
PMDs? \ Most of the photons impinging on the earth come from the sun, whose
physics is exceedingly well-understood. \ Of the subatomic particles that
could conceivably \textquotedblleft tip the scales\textquotedblright\ of (say)
a human neural event, causing it to turn out one way rather than another,
others might have causal pathways that lead back to other astronomical bodies
(such as supernovae), the earth's core, etc. \ But it seems hard to imagine
how any of the latter possibilities \textit{wouldn't} serve as PMDs: that is,
how they wouldn't effectively \textquotedblleft screen off\textquotedblright%
\ any Knightian uncertainty from the early universe.

To show that the above argument is inconclusive, one need only mention the
cosmic microwave background (CMB) radiation. \ CMB photons pervade the
universe, famously accounting for a few percent of the static in old-fashioned
TV sets. \ Furthermore, many CMB photons are believed to reach the earth
having maintained quantum coherence ever since being emitted at the so-called
\textit{time of last scattering}, roughly $380,000$ years after the Big Bang.
\ Finally, unlike (say) neutrinos or dark matter, CMB photons readily interact
with matter. \ In short, we're continually bathed with at least one type of
radiation that seems to satisfy most of the freebit picture's requirements!

On the other hand, no sooner is the CMB suggested for this role than we
encounter two serious objections. \ The first is that the time of last
scattering, when the CMB photons were emitted, is separated from the Big Bang
itself by $380,000$ years. \ So if we wanted to postulate CMB photons as
freebit carriers, then we'd also need a story about why the hot early universe
should \textit{not} be considered a PMD, and about how a qubit might have
\textquotedblleft made it intact\textquotedblright\ from the Big Bang---or at
any rate, from as far back as current physical theory can take us---to the
time of last scattering. \ The second objection asks us to imagine someone
\textit{shielded} from the CMB: for example, someone in a deep underground
mine. \ Would such a person be \textquotedblleft bereft of Knightian
freedom,\textquotedblright\ at least while he or she remained in the mine?

Because of these objections, I find that, while the CMB might be one
\textit{piece} of a causal chain conveying a qubit to us from the early
universe (without getting screened off by a PMD), it can't possibly provide
the full story. \ It seems to me that convincingly answering question (*)
would require something like a census of the possible causal chains from the
early universe to ourselves that are allowed by particle physics and
cosmology. \ I don't know whether the requisite census is beyond present-day
science, but it's certainly beyond \textit{me}! \ Note that, if it could be
shown that \textit{all} qubits today can be traced back to PMDs, and that the
answer to (*) is negative,\ then the freebit picture would be falsified.

\subsection{Amplification and the Brain\label{CHAOS}}

We haven't yet touched on an obvious question: \textit{once freebits have made
their way into (say) a brain, by whatever means, how could they then tip the
scales of a decision? \ }But it's not hard to suggest plausible answers to
this question, without having to assume anything particularly exotic about
either physics or neurobiology. \ Instead, one can appeal to the well-known
phenomenon of chaos (i.e., sensitive dependence on initial conditions) in
dynamical systems.

The idea that chaos in brain activity might somehow underlie free will\ is an
old one. \ However, that idea has traditionally been rejected, on the sensible
ground that a classical chaotic system is still perfectly deterministic! \ Our
inability to measure the initial conditions to unlimited precision, and our
consequent inability to predict very far into the future, seem at best like
practical limitations.

Thus, a revised idea has held that the role of chaos for free will might be to
take quantum fluctuations---which, as we know, are \textit{not} deterministic
(see Section \ref{QMHV})---and amplify those fluctuations to macroscopic
scale. \ However, this revised idea has also been rejected, on the (again
sensible) ground that, even if true, it would only make the brain a
\textit{probabilistic} dynamical system, which still seems \textquotedblleft
mechanistic\textquotedblright\ in any meaningful sense.

The freebit picture makes a further revision: namely, it postulates that
chaotic dynamics in the brain can have the effect of amplifying
\textit{freebits} (i.e., microscopic Knightian uncertainty) to macroscopic
scale. \ If nothing else, this overcomes the elementary objections above.
\ Yes, the resulting picture might still be wrong, but not for some simple
\textit{a priori} reason---and to me, that represents progress!

It's long been recognized that neural processes relevant to cognition can be
sensitive to microscopic\ fluctuations. \ An important example is the opening
and closing of a neuron's sodium-ion channels,\ which normally determines
whether and for how long a neuron fires. \ This process is modeled
probabilistically (in particular, as a Markov chain) by the standard
\textit{Hodgkin-Huxley equations} \cite{hodgkinhuxley} of neuroscience. \ Of
course, one then has to ask: is the apparent randomness in the behavior of the
sodium-ion channels ultimately traceable back to quantum mechanics (and if so,
by what causal routes)? \ Or does the \textquotedblleft
randomness\textquotedblright\ merely reflect our ignorance of relevant
classical details?

Balaguer \cite{balaguer} put the above question to various neuroscientists,
and was told that either that the answer is unknown or that it's outside
neuroscience's scope. \ For example, he quotes Sebastian Seung as saying:
\textquotedblleft The question of whether [synaptic transmission and spike
firing] are `truly random' processes in the brain isn't really a neuroscience
question. \ It's more of a physics question, having to do with statistical
mechanics and quantum mechanics.\textquotedblright\ \ He also quotes Christof
Koch as saying: \textquotedblleft At this point, we do not know to what extent
the random, i.e. stochastic, neuronal processes we observe are due to quantum
fluctuations (\`{a} la Heisenberg) that are magnified by chemical and
biological mechanisms or to what extent they just depend on classical physics
(i.e. thermodynamics) and statistical fluctuations in the underlying
molecules.\textquotedblright

In his paper \textquotedblleft A scientific perspective on human
choice\textquotedblright\ \cite{sompolinsky}, the neuroscientist Haim
Sompolinsky offers a detailed review of what's known about the brain's
sensitivity to microscopic fluctuations. \ Though skeptical about any role for
such fluctuations in cognition, he writes:

\begin{quotation}
\noindent In sum, given the present state of our understanding of brain
processes and given the standard interpretation of quantum mechanics, we
cannot rule out the possibility that the brain is truly an indeterministic
system; that because of quantum indeterminism, there are certain circumstances
where choices produced by brain processes are not fully determined by the
antecedent brain process and the forces acting on it. \ If this is so, then
the first prerequisite [i.e., indeterminism]\ of \textquotedblleft
metaphysical free will\textquotedblright\ ... may be consistent with the
scientific understanding of the brain.\footnote{However, Sompolinsky then goes
on to reject \textquotedblleft metaphysical free will\textquotedblright\ as
incompatible with a scientific worldview: if, he says, there were laws
relevant to brain function beyond the known laws of physics and chemistry,
then those laws would themselves be incorporated into science, leaving us back
where we started. \ I would agree if it weren't for the logical possibility of
\textquotedblleft Knightian laws,\textquotedblright\ which explained to us why
we couldn't even predict the probability distributions for certain events.}
\end{quotation}

To make the issue concrete: suppose that, with godlike knowledge of the
quantum state $\left\vert \Psi\right\rangle $ of the entire universe, you
wanted to change a particular decision made by a particular human being---and
do so \textit{without} changing anything else, except insofar as the other
changes flowed from the changed decision itself. \ Then the question that
interests us is: \textit{what sorts of changes to }$\left\vert \Psi
\right\rangle $\textit{\ would or wouldn't suffice to achieve your aim?} \ For
example, would it suffice to change the energy of a single photon impinging on
the subject's brain? \ Such a photon might get absorbed by an electron,
thereby slightly altering the trajectories of a few molecules near one
sodium-ion channel in one neuron, thereby initiating a chain of events that
ultimately causes the ion channel to open, which causes the neuron to fire,
which causes other neurons to fire, etc. \ If that sort of causal chain is
plausible---which, of course, is an empirical question---then at least as far
as neuroscience is concerned, the freebit picture would seem to have the raw
material that it needs.

Some people might shrug at this, and regard our story of \textquotedblleft the
photon that broke the camel's back\textquotedblright\ as so self-evidently
plausible that the question isn't even scientifically interesting! \ So it's
important to understand that there are two details of the story that matter
enormously, if we want the freebit picture to be viable. \ The first detail
concerns the \textit{amount of time} needed for a microscopic change in a
quantum state to produce a macroscopic change in brain activity. \ Are we
talking seconds? hours? days?\footnote{Or one could ask: if we model the brain
as a chaotic dynamical system, what's the Lyapunov exponent?} \ The second
detail concerns \textit{localization}. \ We'd like our change to the state of
a single photon to be \textquotedblleft surgical\textquotedblright\ in its
effects: it should change a person's neural firing patterns enough to alter
that person's actions, but any other macroscopic effects of the changed photon
state should be mediated through the change to the brain state. \ The reason
for this requirement is simply that, if it fails, then a superintelligent
predictor might \textquotedblleft non-invasively\textquotedblright\ learn
about the photon by measuring its \textit{other} macroscopic effects, and
ignoring its effects on the brain state.

To summarize our questions:

\begin{quotation}
\noindent\textit{What are the causal pathways by which a microscopic
fluctuation can get chaotically amplified, in human or other animal brains?
\ What are the characteristic timescales for those pathways? \ What
\textquotedblleft side effects\textquotedblright\ do the pathways produce,
separate from their effects on cognition?}\footnote{Another obvious question
is whether brains \textit{differ} in any interesting way from other
complicated dynamical systems like lava lamps or the Earth's atmosphere, in
terms of their response to microscopic fluctuations. \ This question will be
taken up in Sections \ref{WEATHER} and \ref{GERBIL}.}
\end{quotation}

In Section \ref{FALSIFY}, I'll use these questions---and the freebit picture's
dependence on their answers---to argue that the picture makes falsifiable
predictions. \ For now, I'll simply say that these questions strike me as wide
open to investigation, and not only in principle. \ That is, I can easily
imagine that in (say) fifty years, neuroscience, molecular biology, and
physics will be able to say more about these questions than they can today.
\ And crucially, the questions strike me as scientifically interesting
regardless of one's philosophical predilections. \ Indeed, one could reject
the freebit picture completely, and still see progress on these questions as a
\textquotedblleft rising tide that lifts all boats\textquotedblright\ in the
scientific understanding of free will. \ The freebit picture seems unusual
only in \textit{forcing} us to address these questions.

\subsection{Against Homunculi\label{HOMUNCULUS}}

A final clarification is in order about the freebit picture. \ One might worry
that freebits are playing the role of a homunculus: the \textquotedblleft
secret core\textquotedblright\ of a person; a smaller person inside who
directs the brain like a manager, puppeteer, or pilot; Ryle's ghost in the
machine.\ \ But in philosophy and cognitive science, the notion of a
homunculus has been rightly laughed off the stage. \ Like the theory that a
clock can only work if there's a smaller clock inside, the homunculus theory
blithely offers a black box where a \textit{mechanism} is needed, and it leads
to an obvious infinite regress: who's in charge of the homunculus?

Furthermore, if this were really the claim at issue, one would want to know:
why do humans (and other animals) \textit{have} such complicated brains in the
first place? \ Why shouldn't our skulls be empty, save for tiny
\textquotedblleft freebit antennae\textquotedblright\ for picking up signals
from the Big Bang?

But whatever other problems the freebit picture has, I think it's innocent of
the charge of homunculism. \ On the freebit picture---as, I'd argue, on
\textit{any} sane understanding of the world---the physical activity of the
brain retains its starring role in cognition. \ To whatever extent your
\textquotedblleft true self\textquotedblright\ has \textit{any} definite
location in spacetime, that location is in your brain. \ To whatever extent
your behavior is predictable, that predictability ultimately derives from the
predictability of your brain. \ And to whatever extent your choices have an
author, \textit{you }are their author, not some homunculus secretly calling
the shots.

But if this is so, one might ask, then \textit{what role could possibly be
left for freebits}, besides the banal role of an unwanted noise source,
randomly jostling neural firing patterns this way and that? \ Perhaps the
freebit picture's central counterintuitive claim is that freebits \textit{can}
play a more robust role than that of glorified random-number generator,
without usurping the brain's causal supremacy. \ Or more generally: \textit{an
organized, complex system can include a source of \textquotedblleft pure
Knightian unpredictability,\textquotedblright\ which foils probabilistic
forecasts made by outside observers, yet need not play any role in explaining
the system's organization or complexity.} \ While I confess that this claim is
strange, I fail to see any \textit{logical} difficulty with it, nor do I see
any way to escape it if the freebit picture is accepted.

To summarize, on the freebit picture, freebits are simply part of the
explanation for how a brain can reach decisions that are not probabilistically
predictable by outside observers, and that are therefore \textquotedblleft
free\textquotedblright\ in the sense that interests us. \ As such, on this
picture freebits are just one ingredient among many in the physical substrate
of the mind. \ I'd no more consider them the \textquotedblleft true
essence\textquotedblright\ of a person than the myelin coating that speeds
transmission between neurons.

\section{Freedom from the Inside Out\label{FIO}}

The discussion in Section \ref{HOMUNCULUS} might\ remind us about the
importance of stepping back. \ Setting aside any other
concerns,\ \textit{isn't it anti-scientific insanity to imagine that our
choices today could correlate nontrivially with the universe's microstate at
the Big Bang? \ Why shouldn't this idea just be thrown out immediately?}

In this section, I'll discuss an unusual perspective on time, causation, and
boundary conditions: one that, \textit{if} adopted, makes the idea of such a
correlation seem not particularly crazy at all. \ The interesting point is
that, despite its strangeness, this perspective seems perfectly compatible
with everything we know about the laws of physics. \ The perspective is not
new; it was previously suggested by Carl Hoefer\ \cite{hoefer} and
independently by Cristi Stoica\ \cite{stoica,stoica:qm}. \ (As Hoefer
discusses, centuries before either of them, Kant appears to have made related
observations in his \textit{Critique of Practical Reason}, while trying to
reconcile moral responsibility with free will! \ However, Kant's way of
putting things strikes me as obscure and open to other interpretations.)
\ Adopting Hoefer's terminology, I'll call the perspective \textquotedblleft
Freedom from the Inside Out,\textquotedblright\ or FIO for short.

The FIO perspective starts from the familiar fact that the known equations of
physics are \textit{time-reversible}: any valid solution to the equations is
still a valid solution if we replace $t$\ by $-t$.\footnote{One also needs to
interchange left with right and particles with antiparticles, but that doesn't
affect the substance of the argument.} \ Hence, there seems to be no
particular reason to imagine time as \textquotedblleft
flowing\textquotedblright\ from past to future. \ Instead, we might as well
adopt what philosophers call the \textquotedblleft Block
Universe\textquotedblright\ picture, where the whole $4$-dimensional spacetime
manifold is laid out at once as a frozen block.\footnote{Of course, if the
physical laws were probabilistic, then we'd have a probability distribution
over possible blocks. \ This doesn't change anything in the ensuing
discussion.} \ Time is simply one more coordinate parameterizing that
manifold, along with the spatial coordinates $x,y,z$. \ The equations
\textit{do} treat the $t$\ coordinate differently from the other three, but
not in any way that seems to justify the intuitive notion of $t$
\textquotedblleft flowing\textquotedblright\ in a particular direction, any
more than the $x,y,z$\ coordinates \textquotedblleft flow\textquotedblright%
\ in a particular direction. \ Of course, with the discovery of special
relativity, we learned that the choice of $t$\ coordinate\ is no more unique
than the choice of $x,y,z$\ coordinates; indeed, an event in a faraway galaxy
that you judge as years in the future, might well be judged as years in the
past by someone walking past you on the street. \ To many philosophers, this
seems to make the argument for a Block-Universe picture even stronger than it
had been in Newtonian physics.\footnote{As Einstein himself famously wrote to
Michele Besso's family in 1955: \textquotedblleft Now Besso has departed from
this strange world a little ahead of me. \ That means nothing. \ People like
us, who believe in physics, know that the distinction between past, present
and future is only a stubbornly persistent illusion.\textquotedblright}

The Block-Universe picture is sometimes described as \textquotedblleft leaving
no room for free will\textquotedblright---but that misses the much more
important point, that the picture also leaves no room for \textit{causation}!
\ If we adopt this mentality consistently,\ then \textit{it's every bit as
meaningless to say that \textquotedblleft Jack and Jill went up the hill
because of the prior microstate of the universe,\textquotedblright\ as it is
to say that\ \textquotedblleft Jack and Jill went up the hill because they
wanted to.\textquotedblright} \ Indeed, we might as well say that Jack and
Jill went up the hill because of the \textit{future} microstate of the
universe! \ Or rather, the concept of \textquotedblleft
because\textquotedblright\ plays no role in this picture: there are the
differential equations of physics, and the actual history of spacetime is one
particular solution to those equations, and that's that.

Now, the idea of \textquotedblleft Freedom from the Inside
Out\textquotedblright\ is simply to embrace the Block-Universe picture of
physics, then turn it on its head. \ We say: if this frozen spacetime block
has no \textquotedblleft intrinsic\textquotedblright\ causal arrows anyway,
then \textit{why not annotate it with causal arrows ourselves}, in whichever
ways seem most compelling and consistent to us? \ And at least \textit{a
priori}, who's to say that some of the causal arrows we draw can't point
\textquotedblleft backwards\textquotedblright\ in time---for example, from
human mental states and decisions to \textquotedblleft
earlier\textquotedblright\ microstates consistent with them? \ Thus, we might
decide that yesterday your brain absorbed photons in certain quantum states
\textit{because} today you were going to eat tuna casserole, and running the
differential equations backward from the tuna casserole produced the photons.
\ In strict Block-Universe terms, this seems \textit{absolutely} no worse than
saying that you ate tuna casserole today because of the state of the universe yesterday.

I'll let Hoefer explain further:

\begin{quotation}
\noindent The idea of freedom from the inside out is this: we are perfectly
justified in viewing our own actions \textit{not} as determined by the past,
\textit{nor} as determined by the future, but rather as simply determined (to
the extent that this word sensibly applies) \textit{by ourselves, by our own
wills}. \ In other words, they need not be viewed as \textit{caused} or
\textit{explained} by the physical states of other, vast regions of the block
universe. \ Instead, we can view our own actions, \textit{qua} physical
events, as primary explainers, determining---in a very partial way---physical
events outside ourselves to the past and future of our actions, in the block.
\ We adopt the perspective that the determination or explanation that matters
is from the \textit{inside} (of the block universe, where we live)
\textit{outward}, rather than from the \textit{outside} (e.g. the state of
things on a time slice 1 billion years ago) \textit{in}. \ And we adopt the
perspective of downward causation, thinking of our choices and intentions as
primary explainers of our physical actions, rather than letting microstates of
the world usurp this role. \ We are free to adopt these perspectives because,
quite simply, physics---including [a] postulated, perfected deterministic
physics---is perfectly compatible with them. \cite[p. 207-208, emphases in
original]{hoefer}
\end{quotation}

Some readers will immediately object as follows:

\begin{quotation}
\noindent Yes, \textquotedblleft causality\textquotedblright\ in the Block
Universe\ might indeed be a subtle, emergent concept. \ But doesn't the FIO
picture take that observation to a ludicrous extreme, by turning causality
into a free-for-all? \ For example, why couldn't an FIO believer declare that
the dinosaurs went extinct $65$ million years ago \textit{because} if they
hadn't, today I might not have decided to go bowling?
\end{quotation}

The reply to this objection is interesting. \ To explain it, we first need to
ask: if the notion of \textquotedblleft causality\textquotedblright\ appears
nowhere in fundamental physics, then why does it \textit{look} like past
events constantly cause future ones, and never the reverse? \ Since the late
$19^{th}$\ century, physicists have had a profound answer to that question, or
at least a reduction of the question to a different question.

The answer goes something like this: causality is an \textquotedblleft
emergent phenomenon,\textquotedblright\ associated with the Second Law of
Thermodynamics. \ Though no one really knows why, the universe was in a vastly
\textquotedblleft improbable,\textquotedblright\ low-entropy state at the Big
Bang---which means, for well-understood statistical reasons, that the further
in time we move \textit{away} from the Big Bang, the greater the entropy we'll
find. \ Now, the creation of reliable memories and records is essentially
always associated with an increase in entropy (some would argue by
definition). \ And in order for us, as observers, to speak sensibly about
\textquotedblleft$A$ causing $B$,\textquotedblright\ we must be able to create
records of $A$ happening \textit{and then} $B$ happening. \ But by the above,
this will essentially never be possible unless $A$ is closer in time than $B$
to the Big Bang.

We're now ready to see how the FIO picture evades the unacceptable conclusion
of a causal free-for-all. \ It does so by \textit{agreeing} with the usual
account of causality based on entropy increase, in all situations where the
usual account is relevant. \ While Hoefer\ \cite{hoefer} and
Stoica\ \cite{stoica,stoica:qm} are not explicit about this point, I would say
that on the FIO picture, it can only make sense to draw causal arrows
\textquotedblleft backwards in time,\textquotedblright\ in those rare
situations where entropy is \textit{not} increasing with time.

What are those situations? \ To a first approximation, they're the situations
where physical systems are allowed to evolve reversibly, free from contact
with their external environments. \ In practice, such perfectly-isolated
systems will almost always be microscopic, and the reversible equation
relevant to them will just be the Schr\"{o}dinger equation. \ One example of
such a system would be a photon of cosmic background radiation, which was
emitted in the early universe and has been propagating undisturbed ever since.
\ But these are precisely the sorts of systems that I'd consider candidates
for \textquotedblleft freebits\textquotedblright! \ As far as I can see, if we
want the FIO picture not to lead to absurdity, then we can entertain the
possibility of backwards-in-time causal arrows \textit{only} for these
systems. \ For this is where the \textquotedblleft normal\textquotedblright%
\ way of drawing causal arrows breaks down, and we have nothing else to guide us.

\subsection{The Harmonization Problem\label{HARMONIZATION}}

There's another potential problem with the FIO perspective---Hoefer
\cite{hoefer} calls it the \textquotedblleft Harmonization
Problem\textquotedblright---that is so glaring that it needs to be dealt with
immediately. \ The problem is this: once we let certain causal arrows point
backward in time, say from events today to the microstate at the Big Bang,
we've set up what a computer scientist would recognize as a giant
\textit{constraint satisfaction problem}. \ Finding a \textquotedblleft
solution\textquotedblright\ to the differential equations of physics is no
longer just a matter of starting from the initial conditions and evolving them
forward in time. \ Instead, we can now also have constraints involving
\textit{later} times---for example, that a person makes a particular
choice---from which we're supposed to propagate backward. \ But if this is so,
then \textit{why should a globally-consistent solution even exist, in
general?} \ In particular, what happens if there are \textit{cycles} in the
network of causal arrows? \ In such a case, we could easily face the classic
grandfather paradox of time travel to the past: for example, if an event $A$
causes another event $B$ in its future, but $B$ also causes
$\operatorname*{not}\left(  A\right)  $\ in its past---which in turn causes $\operatorname*{not}\left(  B\right)  $, which in turn causes $A$. \ Furthermore, even if
globally-consistent solutions \textit{do} exist, one is tempted to ask what
\textquotedblleft algorithm\textquotedblright\ Nature uses to build a solution
up. \ Should we imagine that Nature starts at the initial conditions and
propagates them forward, but then \textquotedblleft
backtracks\textquotedblright\ if a contradiction is found, adjusting the
initial conditions slightly in an attempt to avoid the contradiction? \ What
if it gets into an infinite loop this way?

With hindsight, we can see the discussion from Section \ref{UPLOADING} about
brain-uploading, teleportation, Newcomb's Paradox, and so on as highlighting
the same concerns about the consistency of spacetime---indeed, as using
speculative future technologies to make those concerns more vivid. \ Suppose,
for example, that your brain has been scanned, and a complete physical
description of it (sufficient to predict all your future actions) has been
stored on a computer on Pluto. \ And suppose you then make a decision.\ \ Then
from an FIO perspective, the question is: can we take your decision\ to
explain, not only the state of your brain at the moment of your choice, and
not only various microstates in your past lightcone\footnote{Given a spacetime
point $x$, the past lightcone of $x$ is the set of points from which $x$ can
receive a signal, and the future lightcone of $x$ is the set of points to
which it can send one.} (insofar as they need to be compatible with your
choice), but \textit{also what's stored in the computer memory billions of
kilometers away}? \ Should we say that you and the computer now make your
decisions in synchrony, whatever violence that might seem to do to the
locality of spacetime? \ Or should we say that the very act of copying your
brain state removed your freedom where previously you had it, or proved that
you were never free in the first place?

Fortunately, it turns out that in the freebit picture, \textit{none of these
problems arise.} \ The freebit picture might be rejected for other reasons,
but it can't be charged with logical inconsistency or with leading to closed
timelike curves. \ The basic observation is simply that we have to distinguish
between what I'll call \textit{macrofacts} and \textit{microfacts}. \ A
macrofact is a fact about a \textquotedblleft decohered, classical,
macroscopic\textquotedblright\ property of a physical system $S$ at a
particular time. \ More precisely, a macrofact is a fact $F$\ that could, in
principle, be learned by an external measuring device without disturbing the
system $S$ in any significant way. \ Note that, for $F$ to count as a
macrofact, it's not necessary that anyone has ever known $F$ or will ever know
$F$: only that $F$ \textit{could have been} known, if a suitable measuring
device had been around at the right place and the right time. \ So for
example, there is a macrofact about whether or not a Stegosaurus kicked a
particular rock 150 million years ago, even if no human will ever see the
rock, and even if the relevant information can no longer be reconstructed,
even in principle, from the quantum state of the entire solar system. \ There
are also macrofacts constantly being created in the interiors of stars and in
interstellar space.

By contrast, a microfact is a fact about an undecohered quantum state $\rho$:
a fact that couldn't have been learned, even in principle, without the
potential for altering $\rho$\ (if the measurement were performed in the
\textquotedblleft wrong\textquotedblright\ basis). \ For example, the
polarization of some particular photon of the cosmic microwave background
radiation is a microfact. \ A microfact might or might not concern a freebit,
but the quantum state of a freebit is always a microfact.

Within the freebit picture, the solution to the \textquotedblleft
Harmonization Problem\textquotedblright\ is now simply to impose the following
two rules.

\begin{enumerate}
\item[(1)] Backwards-in-time causal arrows can point only to microfacts, never
to macrofacts.

\item[(2)] No microfact can do \textquotedblleft double duty\textquotedblright%
: if it is caused by a fact to its future, then it is not itself the cause of
anything, nor is it caused by anything else.
\end{enumerate}

Together, these rules readily imply that no cycles can arise, and more
generally, that the \textquotedblleft causality graph\textquotedblright\ never
produces a contradiction. \ For whenever they hold, the causality graph with
be a \textit{directed acyclic graph} (a dag), with all arrows pointing forward
in time, except for some \textquotedblleft dangling\textquotedblright\ arrows
pointing backward in time that never lead anywhere else.

Rule (1) is basically imposed by fiat: it just says that, for all the events
we actually observe, we must seek their causes only to their past, never to
their future. \ This rule might someday be subject to change (for example, if
closed timelike curves were discovered), but for now, it seems like a pretty
indispensable part of a scientific worldview. \ By contrast, rule (2) can be
justified by appealing to the No-Cloning Theorem. \ If a microfact $f$\ is
directly caused by a macrofact $F$ to its future, then at some point a
\textit{measurement} must have occurred (or more generally, some decoherence
event that we can \textit{call} a \textquotedblleft
measurement\textquotedblright), in order to amplify $f$ to macroscopic scale
and correlate it with $F$. \ In the freebit picture, we think of the
correlation with $F$ as completely fixing $f$, which explains why $f$ can't
also be caused by anything other than $F$. \ But why can't $f$, itself, cause
some macrofact $F^{\prime}$\ (which, by rule (1), would need to be to $f$'s
future)? \ Here there are two cases: $F^{\prime}=F$\ or $F^{\prime}\neq F$.
\ If $F^{\prime}=F$, then we've simply created a \textquotedblleft
harmless\textquotedblright\ $2$-element cycle, where $f$ and $F$ cause each
other. \ It's purely by convention that we disallow such cycles, and declare
that $F$\ causes $f$ rather than the reverse. \ On the other hand, if
$F^{\prime}\neq F$, then we have two independent measurements of $f$ to $f$'s
future, violating the No-Cloning Theorem.\footnote{For the same reason, we
also have the rule that a microfact cannot cause two macrofacts to its future
via disjoint causal pathways. \ The only reason this rule wasn't mentioned
earlier is that it plays no role in eliminating cycles.} \ Note that this
argument wouldn't have worked if $f$ had been a macrofact, since macroscopic
information \textit{can} be measured many times independently.

Subtle questions arise when we ask about the possibility of microfacts causing
other microfacts. \ Rules (1) and (2) allow that sort of causation, as long as
it takes place forward in time---or, if backward in time, that it consists of
a single \textquotedblleft dangling\textquotedblright\ arrow only. \ If we
wanted, without causing any harm we could allow long chains (and even dags) of
microfacts causing other microfacts backward in time, possibly originating at
some macrofact to their future. \ We would need to be careful, though, that
none of those microfacts ever caused any facts to their future, since that
would create the possibility of cycles. \ A simpler option is just to declare
the entire concept of causality irrelevant to the microworld. \ On that view,
whenever a microfact $f$ \textquotedblleft causes\textquotedblright\ another
microfact $f^{\prime}$, unitarity makes it just as legitimate to say that
$f^{\prime}$\ causes $f$, or that neither causes the other. \ Because of the
reversibility of microscopic laws, the temporal order of $f$ and $f^{\prime}%
$\ is irrelevant: if $U\left\vert \psi\right\rangle =\left\vert \varphi
\right\rangle $, then $U^{\dagger}\left\vert \varphi\right\rangle =\left\vert
\psi\right\rangle $. \ This view would regard causality as inextricably bound
up with the thermodynamic arrow of time, and therefore with \textit{ir}%
reversible processes, and therefore with macrofacts.

\subsection{Microfacts and Macrofacts\label{MICROMACRO}}

An obvious objection to the distinction between microfacts and macrofacts is
that it's poorly-defined. \ The position of a rock might be \textquotedblleft
obviously\textquotedblright\ a macrofact, and the polarization of a photon
\textquotedblleft obviously\textquotedblright\ a microfact, but there is a
continuous transition between the two. \ Exactly how decohered and classical
does a fact have to be, before it counts as a \textquotedblleft
macrofact\textquotedblright\ for our purposes? \ This, of course, is
reminiscent of the traditional objection to Bohr's Copenhagen Interpretation
of quantum mechanics, and in particular, to its unexplained yet crucial
distinction between the \textquotedblleft microworld\textquotedblright\ of
quantum mechanics and the \textquotedblleft macroworld\textquotedblright\ of
classical observations.

Here, my response is basically to admit ignorance. \ The freebit picture is
not particularly sensitive to the precise \textit{way} we distinguish between
microfacts and macrofacts. \ But if the picture is to make sense, it does
require that there \textit{exist} a consistent way to make this distinction.
\ (Or at least, it requires that there exist a consistent way to
\textit{quantify} the macro-ness of a fact $f$. \ The degree of macro-ness of
$f$ might then correspond to the \textquotedblleft effort of
will\textquotedblright\ needed to affect $f$ retrocausally, with the effort
becoming essentially infinite for ordinary macroscopic facts!)

One obvious way to enforce a macro/micro distinction would be via a
\textit{dynamical collapse theory}, such as those of Ghirardi, Rimini, and
Weber \cite{grw} or Penrose \cite{penrose:shadows}. \ In these theories, all
quantum states periodically undergo \textquotedblleft spontaneous
collapse\textquotedblright\ to some classical basis, at a rate that grows with
their mass or some other parameter, and in such a way that the probability of
spontaneous collapse is close to $0$ for all quantum systems that have so far
been studied, but close to $1$ for ordinary \textquotedblleft
classical\textquotedblright\ systems. \ Unfortunately, the known
dynamical-collapse theories tend to suffer from technical problems, such as
small violations of conservation of energy, and of course there is as yet no
experimental evidence for them. \ More fundamentally, I personally cannot
believe that Nature would solve the problem of the \textquotedblleft
transition between microfacts and macrofacts\textquotedblright\ in such a
seemingly ad hoc way, a way that does so much violence to the clean rules of
linear quantum mechanics.

As I'll discuss further in Section \ref{MWI}, my own hope is that a principled
distinction between microfacts and macrofacts could ultimately emerge from
cosmology. \ In particular, I'm partial to the idea that, in a deSitter
cosmology like our own, \textit{a \textquotedblleft
macrofact\textquotedblright\ is simply any fact of which the news\ is already
propagating outward at the speed of light}, so that the information can never,
even in principle, be gathered together again in order to \textquotedblleft
uncause\textquotedblright\ the fact. \ A microfact would then be any fact for
which this propagation \textit{hasn't} yet happened. \ The advantage of this
distinction is that it doesn't involve the slightest change to the principles
of quantum mechanics. \ The disadvantage is that the distinction is
\textquotedblleft teleological\textquotedblright: the line between microfacts
and macrofacts is defined by what \textit{could} happen arbitrarily far in the
future. \ In particular, this distinction implies that if, hypothetically, we
could surround the solar system by a perfect reflecting boundary, then within
the solar system, there would no longer be \textit{any} macrofacts! \ It also
implies that there can be no macrofacts in an \textit{anti}-deSitter (adS)
cosmology, which \textit{does} have such a reflecting boundary. \ Finally, it
suggests that there can probably be few if any macrofacts inside the event
horizon of a black hole. \ For even if the information in the black hole
interior \textit{eventually} emerges, in coded form, in the Hawking radiation,
the Hawking evaporation process is so slow ($\thicksim10^{67}$\ years for a
solar-mass black hole) that there would seem to be plenty of time for an
observer outside the hole to gather most of the radiation, and thereby prevent
the information from spreading further.\footnote{To be more precise here, one
would presumably need to know the detailed \textit{mapping} between the qubits
of Hawking radiation and the degrees of freedom inside the hole, which in turn
would require a quantum theory of gravity.}

Because I can't see a better alternative---and also, because I rather like the
idea of cosmology playing a role in the foundations of quantum mechanics!---my
current inclination is to bite the bullet,\ and accept these and other
implications of the macro/micro distinction I've suggested. \ But there's
enormous scope here for better ideas (or, of course, new developments in
physics) to change my thinking.

\section{Further Objections\label{OBJECTIONS}}

In this section, I'll present five objections to the freebit picture, together
with my responses. \ Some of these objections are obvious, and are generic to
\textit{any} analysis of \textquotedblleft freedom\textquotedblright\ that
puts significant stock in the actual unpredictability of human choices.
\ Others are less obvious, and are more specific to the freebit picture.

\subsection{The Advertiser Objection\label{AD}}

The first objection is simply that human beings are depressingly predictable
in practice:\ if they weren't, then they wouldn't be so easily manipulable!
\ So, does surveying the sorry history of humankind---in which most people,
most of the time, did exactly the boring, narrowly self-interested things one
might have expected them to do---furnish any evidence at all for the
\textit{existence} of freebits?

\textbf{Response.} \ I already addressed a related objection in Section
\ref{LIBET}, but this one seems so important that it's worth returning to it
from a different angle.

It's obvious that humans are at least partly predictable---and sometimes
\textit{extremely} predictable, depending on what one is trying to predict.
\ For example, it's vanishingly unlikely that tomorrow, the CEO of General
Motors will show up to work naked, or that Noam Chomsky will announce his
support for American militarism. \ On the other hand, that doesn't mean we
know how to program a computer to \textit{simulate} these people, anticipating
every major or minor decision they make throughout their lives! \ It seems
crucial to maintain the distinction between the partial predictability that
even the most vocal free-will advocate would concede, and the
\textquotedblleft physics-like\textquotedblright\ predictability of a comet.
\ To illustrate, imagine a machine that correctly predicted \textit{most}
decisions of \textit{most} people: what they'll order for dinner, which movies
they'll select, which socks they'll take out of the drawer, and so on. \ (In a
few domains, this goal is already being approximated by recommendation
systems, such as those of Amazon and Netflix.) \ But imagine that the machine
was regularly blindsided by the most \textit{interesting, consequential,
life-altering} decisions. \ In that case, I suspect most people's intuitions
about their own \textquotedblleft freedom\textquotedblright\ would be shaken
slightly, but would basically remain intact.\ \ By analogy, for most computer
programs that arise in practice, it's easy to decide whether they halt, but
that hardly decreases the importance of the \textit{general} unsolvability of
the halting problem. \ Perhaps, as Kane \cite{fourviews}\ speculates, we truly
exercise freedom only for a relatively small number of \textquotedblleft
self-forming actions\textquotedblright\ (SFAs)---that is, actions that help to
\textit{define} who we are---and the rest of the time are essentially
\textquotedblleft running on autopilot.\textquotedblright\ \ Perhaps these
SFAs are common in childhood and early adulthood, but become rare later in
life, as we get set in our ways and correspondingly more predictable.

Having said this, I concede that the intuition in favor of humans'
predictability is a powerful one. \ Indeed, even supposing humans \textit{did}
have the capacity for Knightian freedom, one could argue that that capacity
can't be particularly important, if almost all humans choose to
\textquotedblleft run on autopilot\textquotedblright\ almost all of the time!

However, against the undeniable fact of humans so often being\ manipulable
like lab rats, there's a \textit{second} undeniable fact, which should be
placed on the other side of the intuitive ledger. \ This second fact is the
conspicuous failure of investors, pundits, intelligence analysts, and so on
\textit{actually to predict}, with any reliability, what individuals or even
entire populations will do. \ Again and again the best forecasters are
blindsided (though it must be admitted that \textit{after the fact},
forecasters typically excel at explaining the inevitability of whatever people
decided to do!).

\subsection{The Weather Objection\label{WEATHER}}

What's special about the human brain? \ If we want to describe it as having
\textquotedblleft Knightian freedom,\textquotedblright\ then why not countless
\textit{other} complicated dynamical systems, such as the Earth's weather, or
a lava lamp?

\textbf{Response.} \ For systems like the weather or a lava lamp, I think a
plausible answer can actually be given. \ These systems are indeed
unpredictable, but the unpredictability seems much more probabilistic than
Knightian in character. \ To put it another way, the famous \textquotedblleft
butterfly effect\textquotedblright\ seems likely to be an artifact of
\textit{deterministic} models of those systems; one expects it to get washed
out as soon as we model the systems' microscopic components probabilistically.
\ To illustrate, imagine that the positions and momenta of all the molecules
in the Earth's atmosphere had been measured to roughly the maximum precision
allowed by quantum mechanics; and that, on the basis of those measurements, a
supercomputer had predicted a $23\%$ probability of a thunderstorm in Shanghai
at a specific date next year. \ Now suppose we changed the initial conditions
by adding a single butterfly. \ In classical, deterministic physics, that
could certainly change whether the storm happens, due to chaotic effects---but
that isn't the relevant question. \ The question is: how much does adding the
butterfly change the \textit{probability} of the storm? \ The answer seems
likely to be \textquotedblleft hardly at all.\textquotedblright\ \ After all,
the original $23\%$ was already obtained by averaging over a huge number of
possible histories. \ So unless the butterfly somehow changes the
\textit{general features} of the statistical ensemble, its effects should be
washed out by the unmeasured randomness in the millions of \textit{other}
butterflies (and whatever else is in the atmosphere).

Yet with brains, the situation seems plausibly different. \ For brains seem
\textquotedblleft balanced on a knife-edge\textquotedblright\ between order
and chaos: were they as orderly as a pendulum, they couldn't support
interesting behavior; were they as chaotic as the weather, they couldn't
support rationality. \ More concretely, a brain is composed of neurons, each
of which (in the crudest idealization) has a firing rate dependent on whether
or not the sum of signals from incoming neurons exceeds some threshold. \ As
such, one expects there to be many molecular-level changes one could make to a
brain's state that don't affect the overall firing pattern at all, but a few
changes---for example, those that push a critical neuron \textquotedblleft
just over the edge\textquotedblright\ to firing or not firing---that affect
the overall firing pattern drastically. \ So for a brain---unlike for the
weather---a single freebit \textit{could} plausibly influence the probability
of some macroscopic outcome, even if we model all of the system's constituents quantum-mechanically.

A closely-related difference between brains and the weather is that, while
both are presumably chaotic systems able to amplify tiny effects, only in the
case of brains are the amplifications likely to have \textquotedblleft
irreversible\textquotedblright\ consequences. \ Even if a butterfly flapping
its wings can cause a thunderstorm halfway around the world, a butterfly
almost certainly won't change the average \textit{frequency} of
thunderstorms---at least, not without changing something other than the
weather as an intermediary. \ To change the frequency of thunderstorms, one
needs to change the trajectory of the earth's \textit{climate}, something less
associated with butterflies than with macroscopic \textquotedblleft
forcings\textquotedblright\ (for example, increasing the level of carbon
dioxide in the atmosphere, or hitting the earth with an asteroid). \ With
brains, by contrast, it seems perfectly plausible that a single neuron's
firing or not firing could affect the rest of a person's life (for example, if
it caused the person to make a \textit{very} bad decision).

\subsection{The Gerbil Objection\label{GERBIL}}

Even if it's accepted that brains are very different from lava lamps or the
weather considered purely as dynamical systems, one could attempt a different
\textit{reductio ad absurdum} of the freebit picture, by constructing a
physical system that \textit{behaves} almost exactly like a brain, yet whose
Knightian uncertainty is \textquotedblleft decoupled\textquotedblright\ from
its intelligence in what seems like an absurd way. \ Thus, consider the
following thought experiment:\ on one side of a room, we have a digital
computer, whose internal operations are completely deterministic. \ The
computer is running an AI program that easily passes the Turing test:\ many
humans, let's say, have maintained long Internet correspondences with the AI
on diverse subjects, and not one ever suspected its consciousness or humanity.
\ On the other side of the room, we have a gerbil in a box. \ The gerbil
scurries in its box, in a way that we can imagine to be subject to at least
some Knightian uncertainty. \ Meanwhile, an array of motion sensors regularly
captures information about the gerbil's movements and transmits it across the
room to the computer, which uses the information as a source of random bits
for the AI. \ Being extremely sophisticated, of course the AI doesn't make
\textit{all} its decisions randomly. \ But if it can't decide between two
roughly-equal alternatives, then it sometimes uses the gerbil movements to
break a tie, much as an indecisive human might flip a coin in a similar situation.

The problem should now be obvious. \ By assumption, we have a system that
\textit{acts} with human-level intelligence (i.e., it passes the Turing test),
and that's \textit{also} subject to Knightian uncertainty, arising from
amplified quantum fluctuations in a mammalian nervous system. \ So if we
believe that humans have a \textquotedblleft capacity for
freedom\textquotedblright\ because of those two qualities, then we seem
\textit{obligated} to believe that the AI/gerbil hybrid has that capacity as
well. \ Yet if we simply disconnect the computer from the gerbil box, then the
AI loses its \textquotedblleft capacity for freedom\textquotedblright! \ For
then the AI's responses, though they might still \textit{seem} intelligent,
could be \textquotedblleft unmasked as mechanistic\textquotedblright\ by
anyone who possessed a copy of the AI's program. \ Indeed, even if we replaced
the gerbil box by an \textquotedblleft ordinary\textquotedblright%
\ quantum-mechanical random number generator, the AI's responses would still
be \textit{probabilistically} predictable; they would no longer involve
Knightian uncertainty.

Thus, a believer in the freebit picture seems forced to an insane conclusion:
that the gerbil, though presumably oblivious to its role, is like a magic
amulet that gives the AI a \textquotedblleft capacity for
freedom\textquotedblright\ it wouldn't have had otherwise. \ Indeed, the
gerbil seems uncomfortably close to the soul-giving potions of countless
children's stories (stories that always end with the main character realizing
that she \textit{already had a soul}, even without the potion!). \ Yet, if we
reject this sort of thinking in the gerbil-box scenario, then why shouldn't we
also reject it for the human brain? \ With the brain, it's true, the
\textquotedblleft Knightian-indeterminism-providing gerbil\textquotedblright%
\ has been moved physically closer to the locus of thought: now it scurries
around in the synaptic junctions and neurons, rather than in a box on the
other side of the room. \ But why should proximity make a difference?
\ \textit{Wherever} we put the gerbil, it just scurries around aimlessly!
\ Maybe the scurrying is probabilistic, maybe it's \textquotedblleft
Knightian,\textquotedblright\ but either way, it clearly plays no more
\textit{explanatory} role in intelligent decision-making than the writing on
the Golem's forehead.

In summary, it seems the only way to rescue the freebit picture is to paper
over an immense causal gap---between the brain's \textquotedblleft Knightian
noise\textquotedblright\ and its cognitive information processing---with
superstition and magic.

\textbf{Response.} \ Of all the arguments directed specifically against the freebit picture, this
one strikes me as the most serious, which is why I tried to present it
in a way that would show its intuitive force.

On reflection, however, there \textit{is} at least one potentially-important
difference between the AI/gerbil system and the brain. \ In the AI/gerbil
system, the \textquotedblleft intelligence\textquotedblright\ and
\textquotedblleft Knightian noise\textquotedblright\ components were
\textit{cleanly separable} from one another. \ That is, the computer
\textit{could} easily be disconnected from the gerbil box, and reattached to a
\textit{different} gerbil box, or an ordinary random-number generator, or
nothing at all. \ And after this was done, there's a clear sense in which the
AI would still be running \textquotedblleft the exact same
program\textquotedblright: only the \textquotedblleft indeterminism-generating
peripheral\textquotedblright\ would have been swapped out. \ For this reason,
it seems best to think of the gerbil as simply \textit{yet another part of the
AI's external environment}---along (for example) with all the questions sent
to the AI over the Internet, which could \textit{also} be used as a source of
Knightian indeterminism.

With the brain, by contrast, it's not nearly so obvious that the
\textquotedblleft Knightian indeterminism source\textquotedblright%
\ \textit{can} be physically swapped out for a different one, without
destroying or radically altering the brain's \textit{cognitive} functions as
well. \ Yes, we can imagine futuristic nanorobots swarming through a brain,
recording all the \textquotedblleft macroscopically measurable
information\textquotedblright\ about the connections between neurons and the
strengths of synapses, then building a new brain that was \textquotedblleft
macroscopically identical\textquotedblright\ to the first, differing only in
its patterns of microscopic noise. \ But how would we know whether the robots
had recorded \textit{enough} information about the original brain? \ What if,
in addition to synaptic strengths, there was also cognitively-relevant
information at a smaller scale? \ Then we'd need more advanced nanorobots,
able to distinguish even smaller features. \ Ultimately, we could imagine
robots able to record \textit{all} \textquotedblleft
classical\textquotedblright\ or even \textquotedblleft
quasi-classical\textquotedblright\ features. \ But by definition, the robots
would then be measuring features that were \textit{somewhat}
quantum-mechanical, and therefore inevitably changing those features.

Of course, this is hardly a conclusive argument, since maybe there's a gap of
many orders of magnitude between (a) the smallest possible scale of cognitive
relevance, and (b) the scale where the No-Cloning Theorem becomes relevant.
\ In that case, at least in principle, the nanorobots really \textit{could}
complete their scan of the brain's \textquotedblleft cognitive
layer\textquotedblright\ without risking the slightest damage to it; only the
(easily-replaced?) \textquotedblleft Knightian indeterminism
layer\textquotedblright\ would be altered by the nanorobots' presence.
\ Whether this is possible is an empirical question for neuroscience.

However, the discussion of brain-scanning raises a broader point: that,
against the gerbil-box scenario, we need to weigh some other, older thought
experiments where many people's intuitions go the other way. \ Suppose the
nanorobots \textit{do} eventually complete their scan of all the
\textquotedblleft macroscopic, cognitively-relevant\textquotedblright%
\ information in \textit{your} brain, and suppose they then transfer the
information to a digital computer, which proceeds to run a macroscopic-scale
simulation of your brain. \ Would that simulation \textit{be} you? \ If your
\textquotedblleft original\textquotedblright\ brain were destroyed in this
process, or simply anesthetized, would you expect to wake up\ as the digital
version? \ (Arguably, this is not even a philosophical question, just a
straightforward empirical question asking you to predict a future observation!)

Now, suppose you believe there's some conceivable digital doppelg\"{a}nger
that would \textit{not} \textquotedblleft be you,\textquotedblright\ and you
also believe that the difference between you and it resides somewhere in the
physical world. Then since (by assumption) the doppelg\"{a}nger is
functionally indistinguishable from you, it would seem to follow that the
difference between you and it \textit{must} reside in \textquotedblleft
functionally-irrelevant\textquotedblright\ degrees of freedom, such as
microscopic ones. \ Either that, or else the boundary between the
\textquotedblleft functional\textquotedblright\ and \textquotedblleft
non-functional\textquotedblright\ degrees of freedom is not even sharp enough
for the doppelg\"{a}ngers to be created in the first place.

My conclusion is that \textit{either} you can be uploaded, copied, simulated,
backed up, and so forth, leading to all the puzzles of personal identity
discussed in Section \ref{UPLOADING}, or else you \textit{can't} bear the same
sort of \textquotedblleft uninteresting\textquotedblright\ relationship to the
\textquotedblleft non-functional\textquotedblright\ degrees of freedom in your
brain that the AI bore to the gerbil box.

To be clear, nothing I've said even hints at any \textit{sufficient condition}
on a physical system, in order for the system to have attributes such as free
will or consciousness. \ That is, even if human brains were subject to
Knightian noise at the microscopic scale, and even if the sources of such
noise could \textit{not} be cleanly separated from the cognitive functions,
human beings might still fail to be \textquotedblleft truly\textquotedblright%
\ conscious\ or free---whatever those things mean!---for other reasons. \ At
most, I'm investigating plausible \textit{necessary} conditions for Knightian
freedom as defined in this essay (and hence, in my personal perspective, for
\textquotedblleft free will\textquotedblright\ also).

\subsection{The Initial-State Objection\label{INITIAL}}

The fourth objection holds that the notion of \textquotedblleft
freebits\textquotedblright\ from the early universe\ nontrivially influencing
present-day events is not merely strange, but inconsistent with known physics.
\ More concretely, Stenger \cite{stenger} has claimed that it \textit{follows}
from known physics that the initial state at the Big Bang was essentially
random, and can't have encoded any \textquotedblleft
interesting\textquotedblright\ information.\footnote{Here Stenger's concern
was not free will or human predictability, but rather ruling out the
possibility (discussed by some theologians) that God could have arranged the
Big Bang with foreknowledge about life on Earth.} \ His argument is basically
that the temperature at the Big Bang was enormous; and in quantum field theory
(neglecting gravity), such extreme temperatures seem manifestly incompatible
with any nontrivial structure.

\textbf{Response.} \ Regardless of whether Stenger's \textit{conclusion}
holds, today there are strong indications, from cosmology and quantum gravity,
that something has to be wrong with the above \textit{argument} for a thermal
initial state.

First, by this argument, the universe's entropy should have been maximal at
the Big Bang, but the Second Law tells us that the entropy was minimal!
\ Stenger \cite{stenger}\ notices the obvious problem, and tries to solve it
by arguing that the entropy at the Big Bang really \textit{was} maximal, given
the tiny size of the observable universe at that time. \ On that view, the
reason why entropy can increase as we move away from the Big Bang is simply
that the universe is expanding, and with it the dimension of the state space.
\ But others, such as Carroll \cite{carroll:eternity} and Penrose
\cite{penrose:road}, have pointed out severe problems with that answer. \ For
one thing, if the dimension of the state space can increase, then we give up
on \textit{reversibility}, a central feature of quantum mechanics. \ For
another, this answer has the unpalatable implication that the entropy should
turn around and decrease in a hypothetical Big Crunch. \ The alternative,
which Carroll and Penrose favor, is to hold that \textit{despite} the enormous
temperature at the Big Bang, the universe's state then was every bit as
\textquotedblleft special\textquotedblright\ and low-entropy as the Second Law
demands, but its specialness\ must have resided in \textit{gravitational}
degrees of freedom that we don't yet fully understand.

The second indication that the \textquotedblleft thermal Big
Bang\textquotedblright\ picture is incomplete is that \textit{quantum field
theory has misled us in a similar way before}. \ In 1975, Hawking
\cite{hawking} famously used quantum field theory in curved spacetime to
calculate the temperature of black holes, and to prove the existence of the
Hawking radiation by which black holes eventually lose their mass and
disappear. \ However, Hawking's calculation seemed to imply that the radiation
was thermal---so that in particular, it couldn't encode any nontrivial
information about objects that fell into the black hole. \ This led to the
\textit{black-hole information-loss paradox}, since quantum-mechanical
reversibility forbids the quantum states of the infalling objects simply to
\textquotedblleft disappear\textquotedblright\ in this way. \ Today, largely
because of the \textit{AdS/CFT correspondence} (see \cite{deboer}), there's a
near-consensus among experts that the quantum states of infalling objects
\textit{don't} disappear as Hawking's calculation suggested. \ Instead, at
least from the perspective of someone outside the black hole, the infalling
information should \textquotedblleft stick around\textquotedblright\ on or
near the event horizon, in not-quite-understood quantum-gravitational degrees
of freedom, before finally emerging in garbled form in the Hawking radiation.
\ And if quantum field theory says otherwise, that's because quantum field
theory is only a limiting case of a quantum theory of gravity.

The AdS/CFT correspondence is just one realization of the \textit{holographic
principle}, which has emerged over the last two decades as a central feature
of quantum theories of gravity. \ It's now known that many physical theories
have both a $D$-dimensional\ \textquotedblleft bulk\textquotedblright%
\ formulation\ and a $\left(  D-1\right)  $-dimensional\ \textquotedblleft
boundary\textquotedblright\ formulation.\ \ In general, these two formulations
\textit{look} completely different: states that are smooth and regular in one
formulation might look random and garbled in the other; questions that are
trivial to answer in one formulation might seem hopeless in the
other.\footnote{As the simplest example, the boundary formulation makes it
obvious that the total entropy in a region should be upper-bounded by its
\textit{surface area}, rather than its volume. \ In the bulk formulation, that
property is strange and unexpected.} \ Nevertheless, there exists an
isomorphism between states and observables, by which the boundary formulation
\textquotedblleft holographically encodes\textquotedblright\ everything that
happens in the bulk formulation. \ As a classic example, if Alice jumps into a
black hole, then she might perceive herself falling smoothly toward the
singularity. \ Meanwhile Bob, far from the black hole, might describe exactly
the same physical situation using a \textquotedblleft dual
description\textquotedblright\ in which Alice never makes it past the event
horizon, and instead her quantum information gets \textquotedblleft
pancaked\textquotedblright\ across the horizon in a horrendously complicated
way. \ In other words, not only is not absurd to suppose that a
\textquotedblleft disorganized mess of entropy\textquotedblright\ on the
boundary of a region could be \textquotedblleft isomorphic\textquotedblright%
\ to a richly-structured state in the region's interior, but there are now
detailed examples where that's exactly what happens.\footnote{Admittedly, the
known examples involve isomorphisms between two theories with different
numbers of spatial dimensions but both with a time dimension. \ There don't
seem to be any nontrivial examples where the boundary theory lives on an
initial spacelike or null hypersurface of the bulk theory. \ (One could, of
course, produce a trivial example, by simply defining the \textquotedblleft
boundary theory\textquotedblright\ to consist of the initial conditions of the
bulk theory, with no time evolution! \ By \textquotedblleft
nontrivial,\textquotedblright\ I mean something more interesting than that.)}

The bottom line is that, when discussing extreme situations like the Big Bang,
it's \textit{not okay} to ignore quantum-gravitational degrees of freedom
simply because we don't yet know how to model them. \ And including those
degrees of freedom seems to lead straight back to the unsurprising conclusion
that \textit{no one knows} what sorts of correlations might have been present
in the universe's initial microstate.

\subsection{The Wigner's-Friend Objection\label{MWI}}

A final objection comes from the Many-Worlds Interpretation of quantum
mechanics---or more precisely, from taking seriously the possibility that a
conscious observer could be measured in a coherent superposition of states.

Let's start with the following thought experiment, called \textit{Wigner's
friend }(after Eugene Wigner, who wrote about it in 1962 \cite{wigner:friend}%
). \ An intelligent agent $A$ is placed in a coherent superposition of two
different mental states, like so:%
\[
\left\vert \Phi\right\rangle =\frac{\left\vert 1\right\rangle \left\vert
A_{1}\right\rangle +\left\vert 2\right\rangle \left\vert A_{2}\right\rangle
}{\sqrt{2}}.
\]
We imagine that these two states correspond to two different questions that
the agent could be asked: for example, $\left\vert 1\right\rangle \left\vert
A_{1}\right\rangle $\ represents $A$ being asked its favorite color, while
$\left\vert 2\right\rangle \left\vert A_{2}\right\rangle $\ denotes $A$ being
asked its favorite ice cream flavor. \ Then, crucially, a second agent $B$
comes along and measures $\left\vert \Phi\right\rangle $\ in the basis%
\begin{equation}
\left\{  \frac{\left\vert 1\right\rangle \left\vert A_{1}\right\rangle
+\left\vert 2\right\rangle \left\vert A_{2}\right\rangle }{\sqrt{2}}%
,\frac{\left\vert 1\right\rangle \left\vert A_{1}\right\rangle -\left\vert
2\right\rangle \left\vert A_{2}\right\rangle }{\sqrt{2}}\right\}  ,
\label{basis}%
\end{equation}
in order to verify that $A$ really \textit{was} in a superposition of two
mental states, not just in one state or the other.

Now, there are many puzzling questions one can ask about this scenario: most
obviously, \textquotedblleft what is it like\textquotedblright\ for $A$ to be
manipulated in this way? \ If $A$ perceives itself in a definite
state---either $A_{1}$ \textit{or} $A_{2}$, but not both---then will $B$'s
later measurement of $A$ in the basis (\ref{basis}) appear to $A$ as a
violation of the laws of physics?

However, let's pass over this well-trodden ground, and ask a different
question more specific to the freebit picture. \ According to that picture,
$A$'s \textquotedblleft free decision\textquotedblright\ of how to answer
whichever question it was asked should be correlated with one or more freebits
$w$. \ But if we write out the combined state of the superposed $A$ and the
freebits,%
\[
\frac{\left\vert 1\right\rangle \left\vert A_{1}\right\rangle +\left\vert
2\right\rangle \left\vert A_{2}\right\rangle }{\sqrt{2}}\otimes\left\vert
w\right\rangle ,
\]
then a problem becomes apparent. \ Namely, the same freebits $w$ need to do
\textquotedblleft double duty,\textquotedblright\ correlating with $A$'s
decision in both the $\left\vert 1\right\rangle $\ branch and the $\left\vert
2\right\rangle $\ branch! \ In other words, even supposing microscopic details
of the environment could somehow \textquotedblleft explain\textquotedblright%
\ what happens in one branch, how could the \textit{same} details explain both
branches? \ As $A_{1}$\ contemplated its favorite color, would it find itself
oddly constrained by $A_{2}$'s simultaneous contemplations of its favorite ice
cream flavor, or vice versa?

One might think we could solve this problem by stipulating that $w$\ is split
into two collections of freebits, $\left\vert w\right\rangle =\left\vert
w_{1}\right\rangle \otimes\left\vert w_{2}\right\rangle $, with $\left\vert
w_{1}\right\rangle $\ corresponding to $A_{1}$'s\ response and $\left\vert
w_{2}\right\rangle $\ corresponding to $A_{2}$'s. \ But this solution quickly
runs into an exponential explosion: if we considered a state like%
\[
\frac{1}{2^{500}}\sum_{x\in\left\{  0,1\right\}  ^{1000}}\left\vert
x\right\rangle \left\vert A_{x}\right\rangle ,
\]
we would find we needed $2^{1000}$\ freebits to allow each $A_{x}$\ to make a
yes/no decision independently of all the other $A_{x}$'s.

Another \textquotedblleft obvious\textquotedblright\ way out would be if the
freebits were entangled with $A$:%
\[
\left\vert \Phi^{\prime}\right\rangle =\frac{\left\vert 1\right\rangle
\left\vert A_{1}\right\rangle \left\vert w_{1}\right\rangle +\left\vert
2\right\rangle \left\vert A_{2}\right\rangle \left\vert w_{2}\right\rangle
}{\sqrt{2}}.
\]
The problem is that there seems to be no way to \textit{produce} such
entanglement, without violating quantum mechanics. \ If $w$ is supposed to
represent microscopic details of $A$'s environment, ultimately traceable to
the early universe, then it would be extremely mysterious to find $A$\ and $w$
entangled. \ Indeed, in a Wigner's-friend experiment, such entanglement would
show up as a \textit{fundamental decoherence source}: something which was
\textit{not} traceable to any \textquotedblleft leakage\textquotedblright\ of
quantum information from $A$, yet which somehow prevented $B$ from observing
quantum interference between the $\left\vert 1\right\rangle \left\vert
A_{1}\right\rangle $\ and\ $\left\vert 2\right\rangle \left\vert
A_{2}\right\rangle $\ branches, when $B$ measured in the basis (\ref{basis}).
\ If, on the other hand, $B$ \textit{did} ultimately trace the entanglement
between $A$ and $w$\ to a \textquotedblleft leakage\textquotedblright\ of
information from $A$, then $w$\ would have been revealed to have never
contained freebits at all! \ For in that case, $w$\ would \textquotedblleft
merely\textquotedblright\ be the result of unitary evolution coupling $A$ to
its environment---so it could presumably be predicted by $B$, who could even
verify its hypothesis by measuring the state $\left\vert \Phi^{\prime
}\right\rangle $\ in the basis%
\[
\left\{  \frac{\left\vert 1\right\rangle \left\vert A_{1}\right\rangle
\left\vert w_{1}\right\rangle +\left\vert 2\right\rangle \left\vert
A_{2}\right\rangle \left\vert w_{2}\right\rangle }{\sqrt{2}},\frac{\left\vert
1\right\rangle \left\vert A_{1}\right\rangle \left\vert w_{1}\right\rangle
-\left\vert 2\right\rangle \left\vert A_{2}\right\rangle \left\vert
w_{2}\right\rangle }{\sqrt{2}}\right\}  .
\]

\textbf{Response.} As in Section \ref{MICROMACRO}---where asking about the
definition of \textquotedblleft microfacts\textquotedblright\ and
\textquotedblleft macrofacts\textquotedblright\ led to a closely-related
issue---my response to this important objection is to \textit{bite the
bullet}. \ That is, I accept the existence of a deep incompatibility between
the freebit picture and the physical feasibility of the Wigner's-friend
experiment. \ To state it differently: \textit{if} the freebit picture is
correct, \textit{and} the Wigner's-friend experiment can be carried out, then
I think we're forced to conclude that---at least for the duration of the
experiment---\textit{the subject no longer has the \textquotedblleft capacity
for Knightian freedom,\textquotedblright} and is now a \textquotedblleft
mechanistic,\textquotedblright\ externally-characterized physical system
similar to a large quantum computer.

I realize that the position above sounds crazy, but it becomes less so once
one starts thinking about what would \textit{actually} be involved in
performing a Wigner's-friend experiment on a human subject. \ Because of the
immense couplings between a biological brain and its external
environment\ (see Tegmark \cite{tegmark:qmbrain} for example), the experiment
is likely impossible with any entity we would currently recognize as
\textquotedblleft human.\textquotedblright\ \ Instead, as a first step (!),
one would presumably need to solve the problem of \textit{brain-uploading}:
that is, transferring a human brain into a digital substrate. \ Only then
could one even \textit{attempt} the second part, of transferring the
now-digitized brain onto a quantum computer, and preparing and measuring it in
a superposition of mental states. \ I submit that, while the resulting entity
\textit{might} be \textquotedblleft freely-willed,\textquotedblright%
\ \textquotedblleft conscious,\textquotedblright\ etc., it certainly wouldn't
be uncontroversially so, nor can we form any intuition by reasoning by
analogy with ourselves (even if we were inclined to try).

Notice in particular that, if the agent $A$ could be manipulated in
superposition, then as a direct byproduct of those manipulations, $A$ would
presumably undergo \textit{the same mental processes over and over, forwards
in time as well as backwards in time}. \ For example, the \textquotedblleft
obvious\textquotedblright\ way for $B$ to measure $A$ in the basis
(\ref{basis}), would simply be for $B$ to \textquotedblleft
uncompute\textquotedblright\ whatever unitary transformation $U$ had placed
$A$ in the superposed state $\left\vert \Phi\right\rangle $\ in the first
place. \ Presumably, the process would then be repeated many times, as $B$
accumulated more statistical evidence for the quantum interference pattern.
\ So, during the uncomputation steps, would $A$\ \textquotedblleft experience
time running backwards\textquotedblright? \ Would the inverse map $U^{\dag}$
feel different than $U$? \ Or would all the applications of $U$ and $U^{\dag}%
$\ be \textquotedblleft experienced simultaneously,\textquotedblright\ being
functionally indistinguishable from one another? \ I hope I'm not alone in
feeling a sense of vertigo about these questions! \ To me, it's at least a
plausible speculation that $A$ doesn't experience \textit{anything}, and that
the reasons why it doesn't are related to $B$'s very ability to manipulate $A$
in these ways.

More broadly, the view I've taken here on superposed agents\ strikes me as
almost a\textit{ consequence} of the view I took earlier, on agents whose
mental states can be perfectly measured, copied, simulated, and predicted by
other agents. \ For there's a close connection between being able to
\textit{measure} the exact state $\left\vert S\right\rangle $ of a physical
system, and being able to detect quantum interference in a superposition of
the form%
\[
\left\vert \psi\right\rangle =\frac{\left\vert S_{1}\right\rangle +\left\vert
S_{2}\right\rangle }{\sqrt{2}},
\]
consisting of two \textquotedblleft slight variants\textquotedblright\ of
$\left\vert S\right\rangle $. \ If we know $\left\vert S\right\rangle $, then
among other things, we can load a copy of $\left\vert S\right\rangle $\ onto a
quantum computer, and thereby prepare and measure a superposition like
$\left\vert \psi\right\rangle $---provided, of course, that one counts the
quantum computer's \textit{encodings} as $\left\vert S_{1}\right\rangle $\ and
$\left\vert S_{2}\right\rangle $\ as \textquotedblleft just as good as the
real thing.\textquotedblright\ \ Conversely, the ability to detect
interference between $\left\vert S_{1}\right\rangle $\ and $\left\vert
S_{2}\right\rangle $\ presupposes that we know, and can control, \textit{all}
the degrees of freedom that make them different: quantum mechanics tells us
that, if any degrees of freedom are left unaccounted for, then we will simply
see a probabilistic mixture of $\left\vert S_{1}\right\rangle $\ and
$\left\vert S_{2}\right\rangle $, not a superposition.

But if this is so, one might ask, then \textit{what makes humans any
different?} \ According to the most literal reading of quantum mechanics'
unitary evolution rule---which some call the Many-Worlds
Interpretation---don't we \textit{all} exist in superpositions of enormous
numbers of branches, and isn't our inability to measure the interference
between those branches merely a \textquotedblleft practical\textquotedblright%
\ problem, caused by rapid decoherence? \ Here I reiterate the speculation put
forward in Section \ref{MICROMACRO}: that the decoherence of a state
$\left\vert \psi\right\rangle $ should be considered \textquotedblleft
fundamental\textquotedblright\ and \textquotedblleft
irreversible,\textquotedblright\ precisely when $\left\vert \psi\right\rangle
$ becomes entangled with degrees of freedom that are receding toward our
deSitter horizon at the speed of light, and that can no longer be collected
together even in principle. \ That sort of decoherence could be avoided, at
least in principle, by a fault-tolerant quantum computer, as in the
Wigner's-friend thought experiment above. \ But it plausibly \textit{can't} be
avoided by any entity that we would currently recognize as \textquotedblleft
human.\textquotedblright

One could also ask: \textit{if} the freebit picture were accepted, what would
be the implications for the foundations of quantum mechanics? \ For example,
would the Many-Worlds Interpretation then have to be rejected?
\ Interestingly, I think the answer is no. \ Since I haven't suggested any
change to the formal rules of quantum mechanics, \textit{any} interpretations
that accept those rules---including Many-Worlds, Bohmian mechanics, and
various Bayesian/subjectivist interpretations---would in some sense
\textquotedblleft remain on the table\textquotedblright\ (to whatever extent
they were on the table before!). \ As far as we're concerned in this essay, if
one wants to believe that different branches of the wavefunction, in which
one's life followed a different course than what one observes,
\textquotedblleft really exist,\textquotedblright\ that's fine; if one wants
to deny the reality of those branches, that's fine as well. \ Indeed, if the
freebit picture were correct, then \textit{Nature would have conspired so that
we had no hope, even in principle, of distinguishing the various
interpretations by experiment}.

Admittedly, one might think it's obvious that the interpretations can't be
distinguished by experiment, with or without the freebit picture. \ Isn't that
why they're \textit{called} \textquotedblleft
interpretations\textquotedblright? \ But as Deutsch \cite{deutsch:universal}%
\ already pointed out in 1985, scenarios like Wigner's friend seriously
challenge the idea that the interpretations are empirically equivalent. \ For
example, if one could perform a quantum interference experiment on
\textit{one's own mental state}, then couldn't one directly experience what it
was like for the different components of the wavefunction describing that
state to evolve unitarily?\footnote{A crucial caveat is that, after the
interference experiment was over, one would retain no reliable memories or
publishable records about \textquotedblleft what it was like\textquotedblright%
! \ For the very fact of such an experiment implies that one's memories are
being created and destroyed at will. \ Without the destruction of memories, we
can't get interference. \ After the experiment is finished, one might have
\textit{something} in one's memory, but what it is could have been
probabilistically predicted even before the experiment began, and can in no
way depend on \textquotedblleft what it was like\textquotedblright\ in the
middle. \ Still, \textit{at least while the experiment was underway}, maybe
one would know which interpretation of quantum mechanics was correct!} \ And
wouldn't that, essentially, vindicate a Many-Worlds-like perspective, while
ruling out the subjectivist views that refuse to countenance the reality of
\textquotedblleft parallel copies of oneself\textquotedblright? \ By denying
that the subject of a Wigner's-friend experiment has the \textquotedblleft
capacity for Knightian freedom,\textquotedblright\ the freebit picture
suggests that maybe there's nothing that it's like to \textit{be} such a
subject---and hence, that debates about quantum interpretation can freely
continue forever, with not even the in-principle possibility of an empirical resolution.

\section{Comparison to Penrose's Views\label{PENROSE}}

Probably the most original thinker to have speculated about physics and mind
in the last half-century has been Roger Penrose. \ In his books, including
\textit{The Emperor's New Mind} \cite{penrose}\ and \textit{Shadows of the
Mind} \cite{penrose:shadows},\footnote{A later book, \textit{The Road to
Reality} \cite{penrose:road}, says little directly about mind, but is my
favorite. \ I think it makes Penrose's strongest case for a gap in our
understanding of quantum mechanics, thermodynamics, and cosmology that radical
new ideas will be needed to fill.} Penrose has advanced three related ideas,
all of them extremely controversial:

\begin{enumerate}
\item[(1)] That arguments related to G\"{o}del's Incompleteness Theorem imply
that the physical action of the human brain cannot be algorithmic (i.e., that
it must violate the Church-Turing Thesis).

\item[(2)] That there must be an \textquotedblleft objective\textquotedblright%
\ physical process that collapses quantum states and produces the definite
world we experience, and that the best place to look for such a process is in
the interface of quantum mechanics, general relativity, and cosmology (the
\textquotedblleft specialness\textquotedblright\ of the universe's initial
state providing the only known source of time-asymmetry in physics, not
counting quantum measurement).

\item[(3)] That objective collapse events, possibly taking place in the
cytoskeletons of the brain's neurons (and subsequently amplified by
\textquotedblleft conventional\textquotedblright\ brain activity), provide the
best candidate for the source of noncomputability demanded by (1).
\end{enumerate}

\noindent An obvious question is how Penrose's views relate to the ones
discussed here. \ Some people might see the freebit picture as
\textquotedblleft Penrose lite.\textquotedblright\ \ For it embraces Penrose's
core belief in a relationship between the mysteries of mind and those of
modern physics, and even follows Penrose in focusing on certain
\textit{aspects} of physics, such as the \textquotedblleft
specialness\textquotedblright\ of the initial state. \ On the other hand, the
account here rejects almost all of Penrose's further speculations: for
example, about noncomputable dynamics in quantum gravity and the special role
of the cytoskeleton in exploiting those dynamics.\footnote{Along another axis,
though, some people might see the freebit picture as \textit{more} radical, in
that it suggests the impossibility of \textit{any} non-tautological
explanation for certain events and decisions, even an explanation invoking
oracles for Turing-uncomputable problems.} \ Let me now elaborate on seven
differences between my account and Penrose's.

\begin{enumerate}
\item[(1)] \textbf{I make no attempt to \textquotedblleft explain
consciousness.\textquotedblright} \ Indeed, that very goal seems misguided to
me, at least if \textquotedblleft consciousness\textquotedblright\ is meant in
the phenomenal sense rather than the neuroscientists' more restricted
senses.\footnote{Just like \textquotedblleft free will,\textquotedblright\ the
word \textquotedblleft consciousness\textquotedblright\ has been the victim of
ferocious verbal overloading, having been claimed for everything from
\textquotedblleft that which disappears under anesthesia,\textquotedblright%
\ to \textquotedblleft that which a subject can give verbal reports
about,\textquotedblright\ to \textquotedblleft the brain's executive control
system\textquotedblright!\ \ Worse, \textquotedblleft
consciousness\textquotedblright\ has the property that, even if one specifies
\textit{exactly} what one means by it, readers are nevertheless certain to
judge anything one says against their own preferred meanings. \ For this
reason, just as I ultimately decided to talk about \textquotedblleft
freedom\textquotedblright\ (or \textquotedblleft Knightian
freedom\textquotedblright) rather than \textquotedblleft free
will\textquotedblright\ in this essay, so I'd much rather use less fraught
terms for executive control, verbal reportability, and so on, and restrict the
word \textquotedblleft consciousness\textquotedblright\ to mean
\textquotedblleft the otherwise-undefinable thing that people have tried to
get at for centuries with the word `consciousness,' supposing that thing
exists.\textquotedblright} \ For as countless philosophers have pointed out
over the years (see McGinn \cite{mcginn} for example), \textit{all} scientific
explanations seem equally compatible with a \textquotedblleft zombie
world,\textquotedblright\ which fulfills the right causal relations but where
no one \textquotedblleft really\textquotedblright\ experiences anything.
\ More concretely, even if Penrose were right that the human brain had
\textquotedblleft super-Turing\textquotedblright\ computational powers---and I
see no reason to think he is---I've never understood how that would help with
what Chalmers \cite{chalmers} calls the \textquotedblleft hard
problem\textquotedblright\ of consciousness. \ For example, could a Turing
machine equipped with an oracle for the halting problem perceive the
\textquotedblleft redness of red\textquotedblright\ any better than a Turing
machine without such an oracle?

Given how much Penrose says about consciousness, I find it strange that he
says almost nothing about the related mystery of \textit{free will}. \ My
central claim, in this essay, is that there exists an \textquotedblleft
empirical counterpart\textquotedblright\ of free will (what I call
\textquotedblleft Knightian freedom\textquotedblright), whose investigation
really \textit{does} lead naturally to questions about physics and cosmology,
in a way that I don't know to happen for any of the usual empirical
counterparts of consciousness.

\item[(2)] \textbf{I make no appeal to the Platonic perceptual abilities of
human mathematicians.} \ Penrose's arguments rely on human mathematicians'
supposed power to \textquotedblleft see\textquotedblright\ the consistency of
(for example) the Peano axioms of arithmetic (rather than simply
\textit{assuming} or \textit{asserting} that consistency, as a computer
program engaging in formal reasoning might do). \ As far as I can tell, to
whatever extent this \textquotedblleft power\textquotedblright\ exists at all,
it's just a particularly abstruse type of \textit{qualia} or
\textit{subjective experience}, as empirically inaccessible as any other type.
\ In other words, instead of talking about the consistency of Peano
arithmetic, I believe Penrose might as well have fallen back on the standard
arguments about how a robot could never \textquotedblleft
really\textquotedblright\ enjoy fresh strawberries, but at most \textit{claim}
to enjoy them.

In both cases, the reply seems obvious: \textit{how do you know} that the
robot doesn't really enjoy the strawberries that it claims to enjoy, or see
the consistency of arithmetic that it claims to see? \ And how do you know
other people \textit{do} enjoy or see those things? \ In any case, none of the
arguments in this essay turn on these sorts of considerations. \ If any
important difference is to be claimed between a digital computer and a human
brain, then I insist that the difference correspond to something empirical:
for example, that computer memories can be reliably measured and copied
without disturbing them, while brain states quite possibly can't. \ The
difference must \textit{not} rely, even implicitly, on a question-begging
appeal to the author's or reader's own subjective experience.

\item[(3)] \textbf{I make no appeal to G\"{o}del's Incompleteness Theorem.}
\ Let me summarize Penrose's (and earlier, Lucas's \cite{lucas})
\textquotedblleft G\"{o}delian argument for human
specialness.\textquotedblright\ \ Consider any finitely describable machine
$M$ for deciding the truth or falsehood of mathematical statements, which
never errs but which might sometimes output \textquotedblleft I don't
know.\textquotedblright\ \ Then by using the code of $M$, it's possible to
construct a mathematical statement $S_{M}$---one example is a formal encoding
of \textquotedblleft$M$ will never affirm this statement\textquotedblright%
---that we humans, \textquotedblleft looking in from the
outside,\textquotedblright\ can clearly see is true, yet that $M$ itself can't
affirm without getting trapped in a contradiction.

The difficulties with this argument\ have been explained at length elsewhere
\cite{aar:qcsd,dennett,russellnorvig}; some standard replies to it are given
in a footnote.\footnote{Why is it permissible to assume that $M$ never errs,
if no \textit{human} mathematician (or even, arguably, the entire mathematical
community) has ever achieved that standard of infallibility?
\par
Even if $M$ never \textit{did} affirm $S_{M}$, or never erred more generally,
how could we ever \textit{know} that? \ Indeed, much like with consciousness
itself, even if one person had the mysterious Platonic ability to
\textquotedblleft see\textquotedblright\ $M$'s soundness, how could that
person ever convince a skeptical third party?
\par
Finally, if we believed that the human brain was itself finitely describable,
then why couldn't we construct a similar mathematical statement (e.g.,
\textquotedblleft Penrose will never affirm this statement\textquotedblright),
which \textit{Penrose} couldn't affirm without contradicting himself, even
though a different human, or indeed an AI program, could easily affirm it?}

Here, I'll simply say that I think the Penrose-Lucas argument establishes
\textit{some} valid conclusion, but the conclusion is much weaker than what
Penrose wants, and it can also be established much more straightforwardly,
without G\"{o}delian considerations. \ \noindent The valid conclusion is that,
\textit{if you know the code of an AI, }then regardless of how intelligent the
AI seems to be, you can \textquotedblleft unmask\textquotedblright\ it as an
automaton,\ blindly following instructions.\textit{ \ }To do so, however, you
don't need to trap the AI in a self-referential paradox: it's enough to verify
that the AI's responses are precisely the ones predicted (or probabilistically
predicted) by the code that you possess! \ Both with the Penrose-Lucas
argument and with this simpler argument, it seems to me that the real issue is
not whether the AI follows a program, but rather, whether it follows a program
that's \textit{knowable by other physical agents}. \ That's why this essay
focusses from the outset on the latter issue.

\item[(4)] \textbf{I don't suggest any barrier to a suitably-programmed
digital computer passing the Turing Test.} \ Of course, if the freebit picture
were correct, then there \textit{would} be a barrier to duplicating the mental
state and predicting the responses of a \textit{specific} human. \ Even here,
though, it's possible that, through non-invasive measurements, one could learn
enough to create a digital \textquotedblleft mockup\textquotedblright\ of a
given person that would fool that person's closest friends and relatives,
possibly for decades. \ (For this purpose, it might not matter if the mockup's
responses eventually diverged badly from the original person's!) \ And such a
mockup would certainly succeed at the \textquotedblleft
weaker\textquotedblright\ task of passing the Turing Test---i.e., of fooling
interrogators into thinking it was human, at least until its code was
revealed. \ If these sorts of mockups \textit{couldn't} be built, then it
would have to be for reasons well beyond anything explored in this essay.

\item[(5)] \textbf{I don't imagine anything particularly exotic about the
biology of the brain.} \ In \textit{The Emperor's New Mind} \cite{penrose},
Penrose speculates that the brain might act as what today we would call an
adiabatic quantum computer: a device that generates highly-entangled quantum
states, and might be able to solve certain optimization problems faster than
any classical algorithm. \ (In this case, presumably, the entanglement would
be \textit{between neurons}.) \ In \textit{Shadows} \cite{penrose:shadows},
Penrose goes further, presenting a proposal of himself and Stuart Hameroff
that ascribes a central role to \textit{microtubules}, a component of the
neuronal cytoskeleton. \ In this proposal, the microtubules would basically be
\textquotedblleft antennae\textquotedblright\ sensitive to
yet-to-be-discovered quantum-gravity effects. \ Since Penrose \textit{also}
conjectures that a quantum theory of gravity would include Turing-uncomputable
processes (see point 7 below), the microtubules would therefore let human
beings surpass the capabilities of Turing machines.

Unfortunately, subsequent research hasn't been kind to these ideas.
\ Calculations of decoherence rates leave almost no room for the possibility
of quantum coherence being maintained in the hot, wet environment of the brain
for anything like the timescales relevant to cognition, or for long-range
entanglement between neurons (see Tegmark \cite{tegmark:qmbrain} for example).
\ As for microtubules, they are common structural elements in cells---not only in
neurons---and no clear evidence has emerged that they are particularly
sensitive to the quantum nature of spacetime. \ And this is setting aside the
question of the evolutionary pathways by which the \textquotedblleft
quantum-gravitational antennae\textquotedblright\ could have arisen.

The freebit perspective requires none of this: at least from a biological
perspective, its picture of the human brain is simply that of conventional
neuroscience. \ Namely, the human brain is a massively-parallel,
highly-connected \textit{classical} computing organ, whose design was shaped
by millions of years of natural selection. \ Neurons perform a role vaguely
analogous to that of \textit{gates} in an electronic circuit (though neurons
are far more complicated in detail), while synaptic strengths serve as a
readable and writable memory. \ If we restrict to issues of principle, then
perhaps the most salient \textit{difference} between the brain and today's
electronic computers is that the brain is a \textquotedblleft digital/analog
hybrid.\textquotedblright\ \ This means, for example, that we have no
practical way to measure the brain's exact \textquotedblleft computational
state\textquotedblright\ at a given time, copy the state, or restore the brain
to a previous state; and it is not even obvious whether these things can be
done in principle. \ It also means that the brain's detailed activity might be
sensitive to microscopic fluctuations (for example, in the sodium-ion
channels) that get chaotically amplified; this amplification might even occur
over timescales relevant to human decision-making (say, $30$ seconds). \ Of
course, if those fluctuations were quantum-mechanical in origin---and at a
small enough scale, they \textit{would} be---then they couldn't be measured
even in principle without altering them.

From the standpoint of neuroscience, the last parts of the preceding paragraph
are certainly not established, but neither does there seem to be any good
evidence against them.\ \ I regard them as plausible guesses, and hope future
work will confirm or falsify them. \ To the view above, the freebit picture
adds only a single further speculation---a speculation that, moreover, I think
\textit{does not even encroach on neuroscience's \textquotedblleft
turf}.\textquotedblright\ \ This is simply that, if we consider the quantum
states $\rho$\ relevant to the microscopic fluctuations, then those states are
subject to at least some Knightian uncertainty (i.e., they are
\textquotedblleft freestates\textquotedblright\ as defined in Appendix
\ref{FREESTATES}); and furthermore, at least some of the Knightian uncertainty
could ultimately be traced, if we wished, back to our ignorance of the
detailed microstate of the early universe. \ This might or might not be true,
but it seems to me that it's not a question for neuroscience at all, but for
physics and cosmology (see Section \ref{FALSIFY}). \ What the freebit picture
\textquotedblleft needs\textquotedblright\ from neuroscience, then, is
extremely modest---certainly compared to what Penrose's picture needs!

\item[(6)] \textbf{I don't propose an \textquotedblleft objective
reduction\textquotedblright\ process that would modify quantum mechanics.}
\ Penrose speculates that, when the components of a quantum state achieve a
large enough energy difference that they induce \textquotedblleft appreciably
different\textquotedblright\ configurations of the spacetime metric (roughly,
configurations that differ by one Planck length or more), new
quantum-gravitational laws beyond of unitary quantum mechanics should come
into effect, and cause an \textquotedblleft objective
reduction\textquotedblright\ of the quantum state. \ This hypothetical process
would underlie what we perceive as a measurement or collapse. \ Penrose has
given arguments that his reduction process, if it existed, would have escaped
detection by current quantum interference experiments, but \textit{could}
conceivably be detected or ruled out by experiments in the foreseeable future
\cite{mspb}. \ Penrose's is far from the only \textquotedblleft objective
reduction\textquotedblright\ model on the table: for example, there's a
well-known earlier model due to Ghirardi, Rimini, and Weber (GRW) \cite{grw},
but that one was purely \textquotedblleft phenomenological,\textquotedblright%
\ rather than being tied to gravity or some other known part of physics.

If an objective reduction process were ever discovered, then it would provide
a ready-made distinction between microfacts and macrofacts (see Section
\ref{MICROMACRO}), of exactly the sort the freebit picture needs. \ Despite
this, I'm profoundly skeptical that any of the existing objective reduction
models are close to the truth. \ The reasons for my skepticism are, first,
that the models seem too ugly and ad hoc (GRW's more so than Penrose's); and
second, that the AdS/CFT correspondence now provides evidence that quantum
mechanics can emerge unscathed even from the combination with gravity.
\ That's why, in Sections \ref{MICROMACRO}\ and \ref{MWI}, I speculated that
the distinction between microfacts and macrofacts might ultimately be defined
in terms of deSitter space cosmology, with a macrofact being any fact
\textquotedblleft already irreversibly headed toward the deSitter
horizon.\textquotedblright

\item[(7)] \textbf{I don't propose that quantum gravity leads to
Turing-uncomputable processes.\ \ }One of Penrose's most striking claims is
that the laws of physics should involve \textit{uncomputability}: that is,
transitions between physical states that cannot in principle be simulated by a
Turing machine, even given unlimited time. \ Penrose arrives at this
conclusion via his G\"{o}del argument (see point 3); he then faces the
formidable challenge of where to \textit{locate} the necessary uncomputability
in anything plausibly related to physics. \ Note that this is
\textit{separate} from the challenge (discussed in point 5) of how to make the
human brain sensitive to the uncomputable phenomena, supposing they exist!
\ In \textit{Shadows} \cite{penrose:shadows}, Penrose seems to admit that this
is a weak link in his argument. \ As evidence\ for uncomputability, the best
he can offer is a theorem of Markov that the $4$-manifold homeomorphism
problem is undecidable (indeed, equivalent to the halting problem)
\cite{markov:4manif}, and a speculation of Geroch and Hartle
\cite{gerochhartle}\ that maybe that fact has something to do with quantum
gravity, since some attempted formulations of quantum gravity involve sums
over $4$-manifolds.

Personally, I see no theoretical or empirical reason to think that the laws of
physics should let us solve Turing-uncomputable problems---either with our
brains, or with any other physical system. \ Indeed, I would go further: in
\cite{aar:np}, I summarized the evidence that the laws of physics seem to
\textquotedblleft conspire\textquotedblright\ to prevent us from solving
$\mathsf{NP}$-complete problems (like the Traveling Salesman Problem) in
polynomial time.\ \ But the $\mathsf{NP}$-complete problems, being solvable in
\textquotedblleft merely\textquotedblright\ exponential time, are child's play
compared to Turing-uncomputable problems like the halting problem! \ For this
reason, I regard it as a serious drawback of Penrose's proposals that they
demand uncomputability in the dynamical laws, and as an advantage of the
freebit picture that it suggests nothing of the kind. \ Admittedly, the
freebit picture does require that there be no\ complete,
computationally-simple description of the \textit{initial conditions}%
.\footnote{More precisely, the initial state, when encoded in some natural way
as a binary string, must have non-negligibly large \textquotedblleft
sophistication\textquotedblright: see Appendix \ref{KOLMOG}.} \ But it seems
to me that nothing in established physics should have led us to expect that
such a description would exist \textit{anyway}!\footnote{By contrast, when it
comes to dynamical behavior, we have centuries of experience discovering laws
that can indeed be simulated on a computer, given the initial conditions as
input, and no experience discovering laws that can't be so simulated.} \ The
freebit picture is silent on whether detailed properties of the initial state
can be actually be \textit{used} to solve otherwise-intractable computational
problems, such as $\mathsf{NP}$-complete problems, in a reliable way. \ But
the picture certainly gives no reason to \textit{think} this is possible, and
I see no evidence for its possibility from any other source.
\end{enumerate}

\section{\textquotedblleft Application\textquotedblright\ to Boltzmann
Brains\label{BOLTZMANN}}

In this section, I'll explain how the freebit perspective, if adopted, seems
to resolve the notorious \textquotedblleft Boltzmann brain
problem\textquotedblright\ of cosmology. \ No doubt some people will feel that
the cure is even worse than the disease! \ But even if one thinks that, the
mere fact of a connection between freebits and Boltzmann brains seems worth
spelling out.

First, what is the Boltzmann brain problem? \ Suppose that---as now seems all
but certain \cite{perlmutter}---our universe will \textit{not} undergo a Big
Crunch, but will simply continue to expand forever, its entropy increasing
according to the Second Law. \ Then eventually, after the last black holes
have decayed into Hawking radiation, the universe will reach the state known
as \textit{thermal equilibrium}: basically an entropy-maximizing soup\ of
low-energy photons, flying around in an otherwise cold and empty vacuum. \ The
difficulty is that, even in thermal equilibrium, there's still a tiny but
nonzero probability that \textit{any given (finite) configuration} will arise
randomly: for example, via a chance conglomeration of photons, which could
give rise to other particles via virtual processes. \ In general, we expect a
configuration with total entropy $S$ to arise at a particular time and place
with probability of order $\thicksim1/\exp\left(  S\right)  $. \ But eternity
being a long time, even such exponentially-unlikely fluctuations should not
only occur, but occur\textit{ infinitely often}, for the same reason why all
of Shakespeare's plays presumably appear, in coded form, infinitely often in
the decimal expansion of $\pi$.\footnote{This would follow from the
conjecture, as yet unproved, that $\pi$\ is a \textquotedblleft base-$10$
normal number\textquotedblright: that is, that just like for a random
sequence, every possible sequence of $k$ consecutive digits appears in $\pi$'s
decimal expansion with asymptotic frequency $1/10^{k}$.} \

So in particular, we would eventually expect (say) beings physically identical
to you, who'd survive just long enough to have whatever mental experiences
you're now having, then disappear back into the void. \ These hypothetical
observers are known in the trade as\ \textit{Boltzmann brains} (see
\cite{dks}), after Ludwig Boltzmann, who speculated about related matters in
the late $19^{th}$\ century. \ So, how do you know that \textit{you} aren't a
Boltzmann brain?

But the problem is worse. \ Since in an eternal universe, you would have
infinitely many Boltzmann-brain doppelg\"{a}ngers, any observer with your
memories and experiences seems infinitely \textit{more} likely to be a
Boltzmann brain, than to have arisen via the \textquotedblleft
normal\textquotedblright\ processes of Darwinian evolution and so on starting
from a Big Bang! \ Silly as it sounds, this has been a major problem plaguing
recent cosmological proposals, since they keep wanting to assign enormous
probability measure to Boltzmann brains (see \cite{carroll:eternity}).

But now suppose you believed the freebit picture, and also believed that
possessing Knightian freedom is a necessary condition for counting as an
observer.\ \ Then I claim that the Boltzmann brain problem would immediately
go away. \ The reason is that, in the freebit picture, Knightian
freedom\ \textit{implies} a certain sort of correlation with the universe's
initial state at the Big Bang---so that lack of complete knowledge of the
initial state corresponds to lack of complete predictability (even
probabilistic predictability) of one's actions by an outside observer. \ But a
Boltzmann brain wouldn't have that sort of correlation with the initial state.
\ By the time thermal equilibrium is reached, the universe will (by
definition) have \textquotedblleft forgotten\textquotedblright\ all details of
its initial state, and any freebits will have long ago been \textquotedblleft
used up.\textquotedblright\ \ In other words, there's no way to make a
Boltzmann brain think one thought rather than another by toggling freebits.
\ So, on this account,\ Boltzmann brains wouldn't be \textquotedblleft
free,\textquotedblright\ even during their brief moments of existence. \ This,
perhaps, invites the further speculation that there's nothing that it's like
to be a Boltzmann brain.

\subsection{What Happens When We Run Out of Freebits?\label{RUNOUT}}

The above discussion of Boltzmann brains leads to a more general observation
about the freebit picture. \ Suppose that

\begin{enumerate}
\item[(1)] the freebit picture holds,

\item[(2)] the observable universe has a finite extent (as it does, assuming a
positive cosmological constant), and

\item[(3)] the holographic principle\ (see Section \ref{INITIAL}) holds.
\end{enumerate}

Then the number of freebits accessible to any one observer must be
finite---simply because the number of bits of \textit{any} kind is then
upper-bounded by the observable universe's finite holographic entropy. \ (For
details, see Bousso \cite{bousso:vac} or Lloyd \cite{lloyd}, both of whom
estimate the observable universe's information capacity as roughly
$\thicksim10^{122}$\ bits.)

But the nature of freebits is that they get permanently \textquotedblleft used
up\textquotedblright\ whenever they are amplified to macroscopic scale. \ So
under the stated assumptions, we conclude that only a finite number of
\textquotedblleft free decisions\textquotedblright\ can possibly be made,
before the observable universe runs out of freebits! \ In my view, this should
not be too alarming. \ After all, even \textit{without} the notion of
freebits, the Second Law of Thermodynamics (combined with the holographic
principle and the positive cosmological constant) already told us that the
observable universe can witness at most $\thicksim10^{122}$\ \textquotedblleft
interesting events,\textquotedblright\ of any kind, before it settles into
thermal equilibrium. \ For more on the theme of freebits as a finite resource,
see Appendix \ref{KOLMOG}.

\section{Indexicality and Freebits\label{INDEXFREE}}

The Boltzmann brain problem is just one example of what philosophers call an
\textit{indexical} puzzle: a puzzle involving the \textquotedblleft
first-person facts\textquotedblright\ of who, what, where, and when
\textit{you} are, which seems to persist even after all the \textquotedblleft
third-person facts\textquotedblright\ about the physical world have been
specified. \ Indexical puzzles, and the lack of consensus on how to handle
them, underlie many notorious debates in science and philosophy---including
the debates surrounding the anthropic principle and \textquotedblleft
fine-tuning\textquotedblright\ of cosmological parameters, the multiverse and
string theory landscape,\ the Doomsday argument, and the Fermi paradox of
where all the extraterrestrial civilizations are. \ I won't even try to
summarize the vast literature on these problems (see\ Bostrom \cite{bostrom}
for an engaging introduction). \ Still, it might be helpful to go through a
few examples, just to illustrate what we mean by indexical puzzles.

\begin{itemize}
\item When doing Bayesian statistics, it's common to use a \textit{reference
class}: roughly speaking, the set of observers from which you consider
yourself to have been \textquotedblleft sampled.\textquotedblright\ \ For an
uncontroversial example, suppose you want to estimate the probability that you
have some genetic disease, in order to decide (say) whether it's worth getting
tested. \ In reality, you either have the disease or you don't. \ Yet it seems
perfectly unobjectionable to estimate the \textit{probability} that you have
the disease, by imagining that you were \textquotedblleft chosen
randomly\textquotedblright\ from among all people with the same race, sex, and
so forth, then looking up the relevant statistics. \ However, things quickly
become puzzling when we ask how \textit{large} a reference class can be
invoked. \ Can you consider yourself to have been \textquotedblleft sampled
uniformly at random\textquotedblright\ from the set of all humans who ever
lived or ever will live? \ If so, then why not also include early hominids,
chimpanzees, dolphins, extraterrestrials, or sentient artificial
intelligences? \ Many would simply reply that you're \textit{not} a dolphin or
any of those other things, so there's no point worrying about the hypothetical
possibility of having been one. \ The problem is that you're \textit{also} not
some other person of the same race and sex---you're \textit{you}---but for
medical and actuarial purposes, we clearly \textit{do} reason as if you
\textquotedblleft could have been\textquotedblright\ someone different. \ So
why can your reference class include those other people but not dolphins?

\item Suppose you're an astronomer who's trying to use Bayesian statistics to
estimate the probability of one cosmological model versus another one,
conditioned on the latest data about the cosmic background radiation and so
forth. \ Of course, as discussed in Section \ref{KNIGHTIAN}, any such
calculation requires a specification of \textit{prior probabilities}. \ The
question is: should your prior include the assumption that, all else being
equal, we're twice as likely to find ourselves in a universe that's twice as
large (and thus presumably has twice as many civilizations, in expectation)?
\ If so, then how do we escape the absurd-seeming conclusion that we're
\textit{certain} to live in an infinite universe, if such a universe is
possible at all---since we expect there to be infinitely many more observers
in an infinite universe than in a finite one? \ Surely we can't deduce the
size of the universe without leaving our armchairs, like a medieval
scholastic? \ The trouble is that, as Bostrom \cite{bostrom} points out,
\textit{not} adjusting the prior probabilities for the expected number of
observers leads to its own paradoxes. \ As a fanciful example, suppose we're
trying to decide between two theories, which on physical grounds are equally
likely. \ Theory $A$ predicts the universe will contain a single civilization
of two-legged creatures, while theory $B$ predicts the universe will contain a
single civilization of two-legged creatures, \textit{as well as} a trillion
equally-advanced civilizations of nine-legged creatures. \ Observing ourselves
to be two-legged, can we conclude that theory $A$ is overwhelmingly more
likely---since if theory $B$ were correct, then we would almost certainly have
been pondering the question on nine legs? \ A straightforward application of
Bayes' rule seems to imply that the answer should be yes---\textit{unless} we
perform the adjustment for the number of civilizations that led to the first paradox!

\item Pursuing thoughts like the above quickly leads to the notorious
\textit{Doomsday argument}. \ According to that argument, the likelihood that
human civilization will kill itself off in the near future is much larger than
one would na\"{\i}vely think---where \textquotedblleft
na\"{\i}vely\textquotedblright\ means \textquotedblleft before taking
indexical considerations into account.\textquotedblright\ \ The logic is
simple: suppose human civilization will continue for billions of years longer,
colonizing the galaxy and so forth. \ In that case, our own position near the
very beginning of human history would seem absurdly improbable---the more so
when one takes into account that such a long-lived, spacefaring civilization
would probably have a much larger population than exists today (just as
\textit{we} have a much larger population than existed hundreds of years ago).
\ If we're Bayesians, and willing to speak at all about ourselves as drawn from an ensemble (as most people
are, in the medical example), that presumably means we should revise downward
the probability that the \textquotedblleft spacefaring\textquotedblright%
\ scenario is correct, and revise upward the probability of scenarios that
give us a more \textquotedblleft average\textquotedblright\ position in human
history. \ But because of exponential population growth, one expects the
latter to be heavily weighted toward scenarios where civilization kills itself
off in the very near future. \ Many commentators have tried to dismiss this
argument as flat-out erroneous (see Leslie \cite{leslie}\ or Bostrom
\cite{bostrom}\ for common objections to the argument and responses to
them).\footnote{Many people point out that cavemen could have made exactly the
same argument, and would have been wrong. \ This is true but irrelevant: the
whole point of the Doomsday argument is that \textit{most} people who make it
will be right!
\par
Another common way to escape the argument's conclusion is to postulate the
existence of large numbers of extraterrestrial civilizations, which are there
regardless of what humans do or don't do. \ If the extraterrestrials are
included in our reference class, they can then \textquotedblleft
swamp\textquotedblright\ the effect of the number of future humans in the
Bayesian calculation.} \ However, while the \textquotedblleft modern,
Bayesian\textquotedblright\ version of the Doomsday argument might indeed be
wrong, it's not wrong because of some trivial statistical oversight. \ Rather,
the argument might be wrong because it embodies an interestingly false way of
thinking about indexical uncertainty.
\end{itemize}

Perplexing though they might be,\ what do any of these indexical\ puzzles have
to do with \textit{our} subject, free will? \ After all, presumably no one
thinks that we have the \textquotedblleft free will\textquotedblright\ to
choose where and when we're born! \ Yet, while I've never seen this connection
spelled out before, it seems to me that indexical puzzles like those above
\textit{do} have some bearing on the free will debate. \ For the indexical
puzzles make it apparent that, even if we assume the laws of physics are
completely mechanistic, there remain large aspects of our experience that
those laws fail to determine, even probabilistically. \ Nothing in the laws
picks out one particular chunk of suitably organized matter from the immensity
of time and space, and says, \textquotedblleft here, \textit{this} chunk is
you; its experiences are your experiences.\textquotedblright\ \ Nor does
anything in the laws give us even a probability distribution over the possible
such chunks. \ Despite its obvious importance even for empirical questions,
our uncertainty about who we are and who we \textquotedblleft could have
been\textquotedblright\ (i.e., our reference class) bears all the marks of
Knightian uncertainty. \ Yet once we've admitted indexical uncertainty into
our worldview,\ it becomes less clear why we should reject the sort of
uncertainty that the freebit picture needs! \ If whether you find yourself
born in $8^{th}$-century China, $21^{st}$-century California, or Planet Zorg
is a variable subject to Knightian uncertainty, then why not what you'll have
for dessert tonight?

More concretely, suppose that there are numerous planets nearly identical to
Earth, down to the same books being written, people with the same names
and same DNA being born, etc. \ If the universe is spatially infinite---which
cosmologists consider a serious possibility\footnote{Astronomers can only see
as far as light has reached since the Big Bang. \ If a positive spatial
curvature was ever detected on cosmological scales, it would strongly suggest
that the universe \textquotedblleft wraps around\textquotedblright---much like
hypothetical ancients might have deduced that the earth was round by measuring
the curvature of a small patch. \ So far, though, except for local
perturbations, the universe appears perfectly flat to within the limits of
measurement, suggesting that it is either infinite or else extends far beyond
our cosmological horizon. \ On the other hand, it is logically possible that
the universe could be \textit{topologically} closed (and hence finite),
despite having zero spatial curvature. \ Also, assuming a positive
cosmological constant, sufficiently far parts of the universe would be forever
out of causal contact with us---leading to philosophical debate about whether
those parts should figure into scientific explanations, or even be considered
to \textquotedblleft exist.\textquotedblright}---then there's no need to
imagine this scenario: for simple probabilistic reasons, it's almost certainly
true! \ Even if the universe is spatially finite, the probability of such a
\textquotedblleft twin Earth\textquotedblright\ would approach $1$ as the
number of stars, galaxies, and so on went to infinity. \ Naturally, we'd
expect any two of these \textquotedblleft twin Earths\textquotedblright\ to be
separated by exponentially large distances---so that, because of the dark
energy pushing the galaxies apart, we would \textit{not} expect a given
\textquotedblleft Earth-twin\textquotedblright\ ever to be in communication
with any of its counterparts.

Assume for simplicity that there are at most two of these Earth-twins, call
them $A$ and $B$. \ (That assumption will have no effect on our conclusions.)
\ Let's suppose that, because of (say) a chaotically amplified quantum
fluctuation, these two Earths are about to diverge significantly in their
histories for the first time. \ Let's further suppose that \textit{you}---or
rather, beings on $A$ and $B$ respectively who look, talk, and act like
you---are the proximate cause of this divergence. \ On Earth $A$, the quantum
fluctuation triggers a chain of neural firing events in \textquotedblleft
your\textquotedblright\ brain that ultimately causes \textquotedblleft
you\textquotedblright\ to take a job in a new city. \ On Earth $B$, a
different quantum fluctuation triggers a chain of neural firings that causes
\textquotedblleft you\textquotedblright\ to stay where you are.

We now ask: from the perspective of a superintelligence that knows everything
above, what's the total probability $p$\ that \textquotedblleft
you\textquotedblright\ take the new job? \ Is it simply $\frac{1}{2}$, since
the two actual histories should be weighted equally? \ What if Earth $B$ had a
greater probability than Earth $A$\ of having formed in the first place---or
did under one cosmological theory but not another? \ And why do we need to
average over the two Earths at all? \ Maybe \textquotedblleft you$_{A}%
$\textquotedblright\ is the \textquotedblleft real\textquotedblright\ you, and
taking the new job is a defining property of who you are, much as Shakespeare
\textquotedblleft wouldn't be Shakespeare\textquotedblright\ had he not
written his plays. \ So maybe you$_{B}$ isn't even part of your reference
class: it's just a faraway doppelg\"{a}nger you'll never meet, who looks and
acts like you (at least up to a certain point in your life) but \textit{isn't}
you. \ So maybe $p=1$. \ Then again, maybe you$_{B}$\ is the \textquotedblleft
real\textquotedblright\ you and $p=0$. \ Ultimately, not even a
superintelligence could calculate $p$\ without knowing something about
\textit{what it means to be \textquotedblleft you,\textquotedblright} a topic
about which the laws of physics are understandably silent.

Now, someone who accepted the freebit picture would say that the
superintelligence's inability to calculate $p$ is no accident. \ For whatever
quantum fluctuation separated Earth $A$ from Earth $B$ could perfectly well
have been a freebit. \ In that case, before you made the decision, the right
representation of your physical state would have been a Knightian combination
of you$_{A}$\ and you$_{B}$. \ (See Appendix \ref{FREESTATES}\ for details of
how these Knightian combinations fit in with the ordinary density matrix
formalism of quantum mechanics.) \ After you make the decision, the ambiguity
is resolved in favor of you$_{A}$\ or you$_{B}$. \ Of course, you$_{A}$ might
then turn out to be a Knightian combination of two further entities,
you$_{A_{1}}$\ and you$_{A_{2}}$, and so on.

For me, the appeal of this view is that it \textquotedblleft cancels two
philosophical mysteries against each other\textquotedblright: free will and
indexical uncertainty. \ As I said in Section \ref{FWFREEDOM}, free
will seems to me to \textit{require} some source of Knightian uncertainty in the
physical world. \ Meanwhile, indexical uncertainty is a type of Knightian
uncertainty that's been considered troublesome and unwanted---though attempts
to replace it with probabilistic uncertainty have led to no end of apparent
paradoxes, from the Doomsday argument to Bostrom's
observer-counting\ paradoxes to the Boltzmann brain problem. \ So it seems
natural to try to fit the free-will peg into the indexical-uncertainty hole.

\section{Is The Freebit Picture Falsifiable?\label{FALSIFY}}

An obvious question about the freebit picture is whether it leads to any new,
falsifiable predictions. \ At one level, I was careful not to
\textquotedblleft commit\textquotedblright\ myself to such predictions in
this essay! \ My goal was to clarify some conceptual issues about the physical
predictability of the brain. \ Whenever I ran up against an unanswered
scientific question---for example, about the role of amplified quantum events
in brain activity---I freely confessed my ignorance.

On the other hand, it's natural to ask: \textit{are there empirical conditions
that the universe has to satisfy, in order for the freebit perspective to have
even a \textbf{chance} of being related to \textquotedblleft free
will\textquotedblright\ as most people understand the concept?}

I submit that the answer to the above question is yes. \ To start with the
most straightforward\ predictions: first, it's necessary that psychology will
never become physics. \ If human beings could be predicted as accurately as
comets, then the freebit picture would be falsified. \ For in such a world, it
would have \textit{turned out} that, whatever it is we called
\textquotedblleft free will\textquotedblright\ (or even \textquotedblleft the
illusion of free will\textquotedblright) in ordinary life, that property was
not associated with any fundamental unpredictability of our choices.

It's also necessary that a quantum-gravitational description of the early
universe will \textit{not} reveal it to have a simply-describable pure or
mixed state. \ Or at least, it's necessary that indexical uncertainty---that
is, uncertainty about our own location in the universe or multiverse---will
forever prevent us from reducing arbitrary questions about our own future to
well-posed mathematical problems. \ (Such a math problem might ask, for
example, for the probability distribution $\mathcal{D}$\ that results when the
known evolution equations of physics are applied to the known initial state at
the Big Bang, marginalized to the vicinity of the earth, and conditioned on
some relevant subset of branches of the quantum-mechanical wavefunction in
which we happen to find ourselves.)

However, the above \textquotedblleft predictions\textquotedblright\ have an
unsatisfying, \textquotedblleft god-of-the-gaps\textquotedblright\ character:
they're simply predictions that certain scientific problems will never be
completely solved! \ Can't we do better, and give \textit{positive}
predictions? \ Perhaps surprisingly, I think we can.

The first prediction was already discussed in Section \ref{CHAOS}. \ In order
for the freebit picture to work, \textit{it's necessary that quantum
uncertainty---for example, in the opening and closing of sodium-ion
channels---can not only get chaotically amplified by brain activity, but can
do so \textquotedblleft surgically\textquotedblright\ and on \textquotedblleft
reasonable\textquotedblright\ timescales.} \ In other words, the elapsed time
between (a) the amplification of a quantum event and (b) the neural firings
influenced by that amplification, must not be so long that the idea of a
connection between the two retreats into absurdity. \ (Ten seconds would
presumably be fine, whereas a year would presumably not be.) \ Closely related
to that requirement, the quantum event must not affect countless
\textit{other} classical features of the world, separately from its effects on
the brain activity. \ For if it did, then a prediction machine could in
principle measure those other classical features to forecast the brain
activity, with no need for potentially-invasive measurements on either the
original quantum state or the brain itself.

It's tempting to compare the empirical situation for the freebit picture to
that for supersymmetry in physics. \ Both of these frameworks are very hard to
falsify---since no matter what energy scale has been explored, or how far into
the future neural events have been successfully predicted, a diehard proponent
could always hold out for the superparticles or freebits making their effects
known at the \textit{next} higher energy, or the next longer timescale. \ Yet
despite this property, supersymmetry and freebits are both \textquotedblleft
falsifiable in degrees.\textquotedblright\ \ In other words, if the
superparticles can be chased up to a sufficiently high energy, then
\textit{even if present, they would no longer do most of the work they were
originally invented to do}. \ The same is true of freebits, if the time
between amplification and decision is long enough.

\begin{figure}[ptb]%
\centering
\includegraphics[
height=2.0003in,
width=3in
]%
{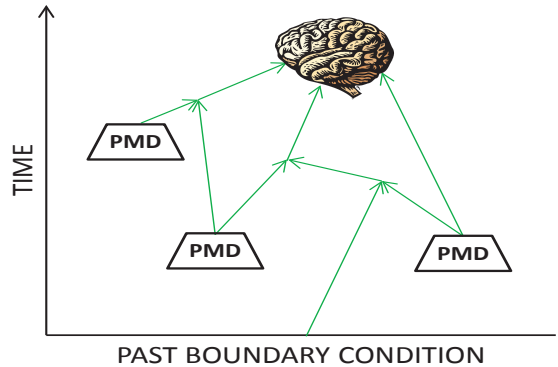}%
\caption{Tracing the causal antecedents of a human decision backwards in time,
stopping \textit{either} at past macroscopic determinants (PMDs) or at the
initial boundary condition}%
\label{chainsfig}%
\end{figure}

Moreover, there's also a \textit{second} empirical prediction of the freebit
picture, one that doesn't involve the notion of a \textquotedblleft reasonable
timescale.\textquotedblright\ \ Recall, from Section \ref{FREEBITS}, the
concept of a past macroscopic determinant (PMD): a set of \textquotedblleft
classical\textquotedblright\ facts (for example, the configuration of a laser)
to the causal past of a quantum state $\rho$\ that, if known, completely
determine $\rho$. \ Now consider an omniscient demon, who wants to influence
your decision-making process by changing the quantum state of a single photon
impinging on your brain. \ Imagine that there are indeed photons that would
serve the demon for this purpose. \ However, now imagine that \textit{all such
photons can be \textquotedblleft grounded\textquotedblright\ in PMDs}. \ That
is, imagine that the photons' quantum states \textit{cannot} be altered,
maintaining a spacetime history consistent with the laws of physics, without
\textit{also} altering classical degrees of freedom in the photons' causal
past. \ In that case, the freebit picture would once again fail. \ For
\textit{if} a prediction machine had simply had the foresight to measure the
PMDs, then (by assumption) it could also calculate the quantum states $\rho$,
and therefore the probability of your reaching one decision rather than
another. \ Indeed, not only could the machine probabilistically predict your
actions; it could even provide a complete quantum-mechanical account of where
the probabilities came from. \ Given such a machine, your choices would remain
\textquotedblleft unpredictable\textquotedblright\ only in the sense that a
radioactive atom is unpredictable, a sense that doesn't interest us (see
Section \ref{KNIGHTIANPHYS}). \ The conclusion is that, for the freebit
picture to work, it's necessary that some of the relevant quantum states
\textit{can't} be grounded\ in PMDs, but only traced back to the early
universe (see Figure \ref{chainsfig}).%
%TCIMACRO{\FRAME{ftbpFU}{3in}{2.0003in}{0pt}{\Qcb{Tracing the causal
%antecedents of a human decision backwards in time, stopping \QTR{it}{either}
%at past macroscopic determinants (PMDs) or at the initial boundary
%condition}}{\Qlb{chainsfig}}{chains.eps}%
%maintain-aspect-ratio TRUE;  display "USEDEF";  valid_file "F";  width 3in;
%height 2.0003in;  depth 0pt;  original-width 0pt;  original-height 0pt;
%cropleft "0";  croptop "1";  cropright "1";  cropbottom "0"; }} }%
%BeginExpansion
%EndExpansion

\section{Conclusions\label{CONC}}

At one level, all I did in this essay was to invoke what David Deutsch, in his
book \textit{The Beginning of Infinity} \cite{deutsch:infinity}, called the
\textquotedblleft momentous dichotomy\textquotedblright:

\begin{quotation}
\noindent\textit{Either a given technology is possible, or else there must be
some reason (say, of physics or logic) why it isn't possible.}
\end{quotation}

Granted, the above statement is a near-tautology. \ But as Deutsch points out,
the implications of applying the tautology consistently can be enormous. \ One
illustrative application, \textit{slightly} less contentious than free will,
involves my own field of quantum computing. \ The idea there is to apply the
principles of quantum mechanics to build a new type of computer, one that
could solve certain problems (such as factoring integers) exponentially faster
than we know how to solve them with any existing computer. \ Quantum computing
research is being avidly pursued around the world, but there are also a few
vocal skeptics of such research. \ Many (though not all) of the skeptics
appear to subscribe to the following three positions:

\begin{enumerate}
\item[(a)] A useful quantum computer is an almost self-evident absurdity:
noise, decoherence, and so forth must conspire to prevent such a machine from
working, just as the laws of physics always conspire to prevent
perpetual-motion machines from working.

\item[(b)] No addition or revision is needed to quantum mechanics: the
physical framework that underlies quantum computing, and that describes the
state of an isolated physical system as a vector in an exponentially-large
Hilbert space (leading to an apparent exponential \textit{slowdown} when
simulating quantum mechanics with a conventional computer).

\item[(c)] Reconciling (a) and (b) is not even a particularly interesting
problem. \ At any rate, the burden is on the quantum computing
\textit{proponents} to sort these matters out! \ Skeptics can be satisfied
that \textit{something} must prevent quantum computers from working, and leave
it at that.
\end{enumerate}

What Deutsch's dichotomy suggests is that such blas\'{e} incuriosity, in the
face of a glaring conflict between ideas, is \textit{itself} the absurd
position. \ Such a position only pretends to be the \textquotedblleft
conservative\textquotedblright\ one: secretly it is radical, in that it
rejects the whole idea of science advancing by identifying apparent conflicts
and then trying to resolve them.

In this essay, I applied Deutsch's momentous dichotomy to a different question:

\begin{quotation}
\noindent Could there exist a machine, consistent with the laws of physics,
that \textquotedblleft non-invasively cloned\textquotedblright\ all the
information in a particular human brain that was relevant to behavior---so
that the human could emerge from the machine unharmed, but would thereafter be
fully probabilistically predictable given his or her future sense-inputs, in
much the same sense that a radioactive atom is probabilistically predictable?
\end{quotation}

My central thesis is simply that \textit{there is no \textquotedblleft safe,
conservative answer\textquotedblright\ to this question.}\ \ Of course, one
can debate what exactly the question means, and how we would know whether the
supposed cloning machine had succeeded. \ (See Appendix \ref{MEAN} for my best
attempt at formalizing the requirements.) \ But I contend that philosophical
analysis can only take us so far. \ The question also has an \textquotedblleft
empirical core\textquotedblright\ that could turn out one way or another,
depending on details of the brain's physical organization that are not yet
known. \ In particular, does the brain possess what one could call a
\textit{clean digital abstraction layer}: that is, a set of macroscopic
degrees of freedom that

\begin{enumerate}
\item[(1)] encode everything relevant to memory and cognition,

\item[(2)] can be accurately modeled as performing a classical digital
computation, and

\item[(3)] \textquotedblleft notice\textquotedblright\ the microscopic,
quantum-mechanical degrees of freedom at most as pure random-number sources,
generating noise according to prescribed probability distributions?
\end{enumerate}

Or is such a clean separation between the macroscopic and microscopic levels
unavailable---so that any attempt to clone\ a brain would either miss much of
the cognitively-relevant information, or else violate the No-Cloning Theorem?

In my opinion, \textit{neither} answer to the question should make us wholly
comfortable: if it does, then we haven't sufficiently thought through the
implications! \ Suppose, on the one hand, that the brain-cloning device is
possible. \ Then we immediately confront countless \textquotedblleft paradoxes
of personal identity\textquotedblright\ like those discussed in Section
\ref{UPLOADING}. \ Would you feel comfortable being (painlessly) killed,
provided that a perfect clone of your brain's digital abstraction layer
remained? \ Would it matter if the cloned data was moved to a new biological
body, or was only \textquotedblleft simulated\textquotedblright%
\ electronically? \ If the latter, what would count as an acceptable
simulation? \ Would it matter if the simulation was run backwards in time, or
in heavily-encrypted form, or in only one branch of a quantum computation?
\ Would you literally expect to \textquotedblleft wake up\textquotedblright%
\ as your clone? \ What if two clones were created: would you expect to wake
up as each one with 50\% probability? \ When applying Bayesian reasoning,
should you, all other things equal, judge yourself twice as likely to wake up
in a possible world with twice as many clones of yourself? \ The point is
that, in a world with the cloning device, these would no longer be
metaphysical conundrums, but in some sense, just \textit{straightforward
empirical questions} about what you should expect to observe! \ Yet even
people who agreed on every possible \textquotedblleft
third-person\textquotedblright\ fact about the physical world and its
contents, might answer such questions completely differently.

To be clear: it seems legitimate to me, given current knowledge, to conjecture
that there is no principled obstruction to perfect brain-cloning, and indeed
that this essay was misguided even to speculate about such an obstruction.
However, \textit{if} one thinks that by taking the \textquotedblleft
pro-cloning\textquotedblright\ route, one can sidestep the need for any
\textquotedblleft weird metaphysical commitments\textquotedblright\ of one's
own, then I'd say one is mistaken.

So suppose, on the other hand, that the perfect brain-cloning device is
\textit{not} possible. \ Here, exactly like in the case of quantum computing,
no truly inquisitive person will ever be satisfied by a bare
\textit{assertion} of the device's impossibility, or by a listing of practical
difficulties. \ Instead, such a person will demand to know: what
\textit{principle} explains why perfect brain-cloning can't succeed, not even
a million years from now? \ How do we reconcile its impossibility with
everything we know about the mechanistic nature of the brain? \ Indeed, how
should we think about the laws of physics \textit{in general}, so that the
impossibility of perfect brain-cloning would no longer seem surprising or inexplicable?

As soon as we try to answer these questions, I've argued that we're driven,
more-or-less inevitably, to the view that the brain's detailed evolution would
have to be buffeted around by chaotically-amplified \textquotedblleft
Knightian surprises,\textquotedblright\ which I called freebits.\ \ Before
their amplification, these freebits would need to live in quantum-mechanical
degrees of freedom, since otherwise a cloning machine could (in principle)
non-invasively copy them. \ Furthermore, our ignorance about the freebits
would ultimately need to be traceable back to ignorance about the microstate
of the early universe---again because otherwise, cloning would become possible
in principle, through vigilant monitoring of the freebits' sources.

Admittedly, all this sounds like a tall order! \ But, strange though it
sounds, I don't see that any of it is ruled out by current scientific
understanding---though conceivably it \textit{could} be ruled out in the
future. \ In any case, setting aside one's personal beliefs, it seems
worthwhile to understand that \textit{this} is the picture one seems forced
to, if one starts from the hypothesis that brain-cloning (with all its
metaphysical difficulties) should be fundamentally impossible, then tries to
make that hypothesis compatible with our knowledge of the laws of physics.

\subsection{Reason and Mysticism\label{MYSTICISM}}

At this point, I imagine some readers might press me: but what do I
\textit{really} think? \ Do I actually take seriously the notion of
quantum-mechanical\ \textquotedblleft freebits\textquotedblright\ from the
early universe playing a role in human decisions? \ The easiest response is
that, in laying out my understanding of the various alternatives---yes, brain
states might be perfectly clonable, but if we want to avoid the philosophical
weirdness that such cloning would entail, then we're led in such-and-such an
interesting direction, etc.---I \textit{already said} what I really think.
\ In pondering these riddles, I don't have any sort of special intuition, for
which the actual arguments and counterarguments that I can articulate serve as
window-dressing. \ The arguments exhaust my intuition.

I'll observe, however, that even uncontroversial facts can be made to sound
incredible when some of their consequences are spelled out; indeed, the
spelling out of such consequences has always been a mainstay of popular
science writing. \ To give some common examples:\ everything you can see in
the night sky was once compressed into a space smaller than an atom. \ The
entire story of your life, including the details of your death, is almost
certainly encoded (in fact, infinitely many times) somewhere in the decimal
expansion of $\pi$. \ If Alois Hitler and Klara P\"{o}lzl\ had moved an inch
differently while having intercourse, World War II would probably have been
prevented. \ When you lift a heavy bag of groceries, what you're really
feeling is the coupling of gravitons to a stress-energy tensor generated
mostly by the rapid movements of gluons. \ In each of these cases, the same
point \textit{could} be made more prosaically, and many would prefer that it
was. \ But when we state things vividly, at least we can't be accused of
trying to \textit{hide} the full implications of our abstract claims, for fear
of being laughed at were those implications understood.

Thus, suppose I'd merely argued, in this essay, that it's possible that humans
will never become as predictable (even probabilistically predictable) as
digital computers, because of chaotic amplification of unknowable microscopic
events, our ignorance of which can be traced as far backward in time as one
wishes. \ In that case, some people would likely agree and others would likely
disagree, with many averring (as I do) that the question remains open.
\ Hardly anyone, I think, would consider the speculation an absurdity or a
gross affront to reason. \ But if exactly the same idea is phrased in terms of
a \textquotedblleft quantum pixie-dust\textquotedblright\ left over from the
Big Bang, which gets into our brains and gives us the capacity for free
will---well then, \textit{of course} it sounds crazy! \ Yet the second
phrasing is nothing more than a dramatic rendering of the worldview that the
first phrasing implies.

Perhaps some readers will accuse me of mysticism. \ To this, I can only reply
that the view I flirt with in this essay seems like \textquotedblleft
mysticism\textquotedblright\ of an unusually tame sort: one that embraces the
mechanistic and universal nature of the laws of physics, as they've played out
for 13.7 billion years; that can accept even the \textquotedblleft collapse of
the wavefunction\textquotedblright\ as an effect brought about by ordinary
unitary evolution; that's consumed by doubts; that welcomes corrections and
improvement; that declines to plumb the cosmos\ for self-help tips or moral
strictures about our sex lives; and that sees science---not introspection, not
ancient texts, not \textquotedblleft science\textquotedblright\ redefined to
mean something different, but \textit{just science in the ordinary sense}---as
our best hope for making progress on the ancient questions of existence.

To any \textquotedblleft mystical\textquotedblright\ readers, who want human
beings to be as free as possible from the mechanistic\ chains\ of
cause and effect, I say: \textit{this picture represents the absolute maximum
that I can see how to offer you, if I confine myself to speculations that I
can imagine making contact with our current scientific understanding of the
world.} \ Perhaps it's less than you want; on the other hand, it does seem
like more than the usual compatibilist account offers! \ To any
\textquotedblleft rationalist\textquotedblright\ readers, who cheer when
consciousness,\ free will,\ or similarly woolly notions get steamrolled by the
advance of science, I say: you can feel vindicated, if you like, that despite
searching (almost literally) to the ends of the universe, I wasn't able to
offer the \textquotedblleft mystics\textquotedblright\ anything more than I
was! \ And even what I \textit{do} offer might be ruled out by future discoveries.

Indeed, the freebit picture's falsifiability is perhaps the single most
important point about it. \ Consider the following questions:\ On what
timescales can microscopic fluctuations in biological systems get amplified,
and change the probabilities of macroscopic outcomes? \ What other side
effects do those fluctuations have? \ Is the brain interestingly different in
its noise sensitivity than the weather, or other complicated dynamical
systems? \ Can the brain's microscopic fluctuations be fully understood
probabilistically, or are they subject to Knightian uncertainty? \ That is,
can the fluctuations all be grounded in past macroscopic determinants, or are
some ultimately cosmological in origin? \ Can we have a complete theory of
cosmological initial conditions? \ Few things would make me happier than if
progress on these questions led to the discovery that the freebit picture was
wrong.\ \ For then at least we would have learned something.

\section{Acknowledgments}

I thank Yakir Aharonov, David Albert, Julian Barbour, Silas Barta, Wolfgang Beirl, Alex Byrne,
Sean Carroll, David Chalmers, Alessandro Chiesa, Andy Drucker, Owain Evans,
Andrew Hodges, Sabine Hossenfelder, Guy Kindler, Seth Lloyd, John Preskill,
Huw Price, Haim Sompolinsky, Cristi Stoica, Jaan Tallinn, David Wallace, and
others I've no doubt forgotten for helpful discussions about the subject of
this essay; and especially Dana Moshkovitz Aaronson, David Aaronson, Steve
Aaronson, Cris Moore, Jacopo Tagliabue, and Ronald de Wolf\ for their comments on earlier
drafts. \ Above all, I thank S. Barry Cooper and Andrew Hodges for
\textquotedblleft commissioning\textquotedblright\ this essay, and for their
near-infinite patience humoring my delays with it. \ It goes without saying
that none of the people mentioned necessarily endorse anything I say here
(indeed, some of them definitely don't!).

\bibliographystyle{plain}
\bibliography{thesis}

\section{Appendix: Defining \textquotedblleft Freedom\textquotedblright%
\label{MEAN}}

In this appendix, I'll use the notion of Knightian uncertainty (see Section
\ref{KNIGHTIANPHYS})\ to offer a possible mathematical formalization of
\textquotedblleft freedom\textquotedblright\ for use in free-will
discussions.\ \ Two caveats are immediately in order. \ The first is that my
formalization only tries to capture what I've called \textquotedblleft
Knightian freedom\textquotedblright---a strong sort of in-principle physical
unpredictability---and not \textquotedblleft metaphysical free
will.\textquotedblright\ \ For as discussed in Section \ref{FWFREEDOM}, I
don't see how \textit{any} definition grounded in the physical universe could
possibly capture the latter, to either the believers' or the deniers'
satisfaction. \ Also, as we'll see, formalizing \textquotedblleft Knightian
freedom\textquotedblright\ is \textit{already} a formidable challenge!

The second caveat is that, by necessity, my definition will be in terms of
more \textquotedblleft basic\textquotedblright\ concepts, which I'll need to
assume as unanalyzed primitives. \ Foremost among these is the concept of a
\textit{physical system}: something that occupies space; exchanges information
(and matter, energy, etc.) with other physical systems; and crucially, retains
an identity through time even as its internal state changes. \ Examples of
physical systems are black holes, the earth's atmosphere, human bodies, and
digital computers. \ Without some concept like this, it seems to me that we
can never specify \textit{whose} freedom we're talking about, or even which
physical events we're trying to predict.

Yet as philosophers know well, the concept of \textquotedblleft physical
system\textquotedblright\ already has plenty of traps for the unwary. \ As one
illustration, should we say that a human body remains \textquotedblleft the
same physical system\textquotedblright\ after its death? \ If so, then an
extremely reliable method to predict a human subject immediately suggests
itself: namely, first shoot the subject; then predict that the subject will
continue to lay on the ground doing nothing!

Now, it might be objected that this \textquotedblleft prediction
method\textquotedblright\ shouldn't count, since it \textit{changes the
subject's state} (to put it mildly),\ rather than just passively gathering
information about the subject. \ The trouble with that response is that
putting the subject in an fMRI machine, interviewing her, or even just having
her sign a consent form or walking past her on the street also change her
state! \ If we don't allow \textit{any} interventions that change the
subject's state from what it \textquotedblleft would have been
otherwise,\textquotedblright\ then prediction---at least with the
science-fiction accuracy we're imagining---seems hopeless, but in an
uninteresting way. \ So which interventions are allowed and which aren't?

I see no alternative but to take the \textit{set of allowed interventions} as
another unanalyzed primitive. \ When formulating the prediction task (and
hence, in this essay, when defining Knightian freedom), we simply declare that
certain interventions---such as interviewing the subject, putting her in an
fMRI machine, or perhaps even having nanorobots scan her brain state---are
allowed; while other interventions, such as killing her, are not allowed.

An important boundary case, much discussed by philosophers, is an intervention
that would destroy each neuron of the subject's brain one by one, replacing
the neurons by microchips claimed to be functionally equivalent. \ Is
\textit{that} allowed? \ Note that such an operation could certainly make it
easier to predict the subject---since from that point forward, the predictor
would only have to worry about simulating the microchips, not the messy
biological details of the original brain. \ Here I'll just observe that, if we
like, we can disallow such drastic interventions, without thereby taking any
position on the conundrum of what such \textquotedblleft
siliconization\textquotedblright\ would do to the subject's conscious
experience. \ Instead we can simply say that, while the subject might indeed
be perfectly predictable after the operation, that fact \textit{doesn't settle
the question at hand}, which was about the subject's predictability
\textit{before} the operation. \ For a large part of what we wanted to know
was to what extent the messy biological details \textit{do} matter, and we
can't answer that question by defining it out of consideration.

But one might object: if the \textquotedblleft messy biological
details\textquotedblright\ need to be left in place when trying to predict a
brain, what \textit{doesn't} need to be left in place? \ After all, brains are
not isolated physical systems: they constantly receive inputs from the sense
organs, from hormones in the bloodstream, etc. \ So when modeling a brain, do
we also need to model the entire environment in which that brain is
immersed---or at least, all aspects of the environment that might conceivably
affect behavior? \ If so, then prediction seems hopeless, but again, not for
any \textquotedblleft interesting\textquotedblright\ reasons: merely for the
boring reason that we can't possibly measure \textit{all} relevant aspects of
the subject's environment, being embedded in the environment ourselves.

Fortunately, I think there's a way around this difficulty, at the cost of one
more unanalyzed primitive. \ Given a physical system $S$, denote by $I\left(
S\right)  $ the set of \textit{screenable inputs} to $S$---by which I mean,
the inputs to $S$ that we judge that any would-be predictor of $S$ should also
be provided, in order to ensure a \textquotedblleft fair
contest.\textquotedblright\ \ For example, if $S$ is a human brain, then
$I\left(  S\right)  $\ would probably include (finite-precision digital
encodings of) the signals entering the brain through the optic, auditory, and
other sensory systems, the levels of various hormones in the blood, and other
measurable variables at the interface between the brain and its external
environment. \ On the other hand, $I\left(  S\right)  $\ probably
\textit{wouldn't} include, for example, the exact quantum state of every
photon impinging on the brain. \ For arguably, we have no idea how to screen
off all those microscopic \textquotedblleft inputs,\textquotedblright\ short
of siliconization or some equally drastic intervention.

Next, call a system $S$ \textit{input-monitorable} if there exists an allowed
intervention to $S$, the result of which is that, after the intervention, all
signals in $I\left(  S\right)  $\ get \textquotedblleft
carbon-copied\textquotedblright\ to the predictor's computer at the same time
as they enter $S$. \ For example, using some future technology, a brain might
be input-monitored by installing microchips that scan all the electrical
impulses in the optic and auditory nerves, the chemical concentrations in the
blood-brain barrier, etc., and that faithfully transmit that information to a
predictor in the next room via wireless link. \ Crucially, and in contrast to
siliconization, input-monitoring doesn't strike me as raising any profound
issues of consciousness or selfhood. \ That is, it seems fairly clear that an
input-monitored human would still be \textquotedblleft the same
human,\textquotedblright\ just hooked up to some funny devices!
\ Input-monitoring also differs from siliconization in that it seems much
closer to practical realization.

\textit{The definition of freedom\ that I'll suggest will only make sense for
input-monitorable physical systems.} \ If $S$\ is not input-monitorable, then
I'll simply hold that the problem of \textquotedblleft predicting $S$'s
behavior\textquotedblright\ isn't well-enough defined: $S$\ is so intertwined
with its environment that one can't say where predicting $S$ ends and
predicting its environment begins. \ One consequence is that, in this
framework, we can't even \textit{pose the question} of whether humans have
Knightian freedom, unless we agree (at least provisionally) that humans are
input-monitorable. \ Fortunately, as already suggested, I don't see any major
scientific obstacles to supposing that humans \textit{are} input-monitorable,
and I even think input-monitoring could plausibly be achieved in $50$ or $100$ years.

Admittedly, in discussing whether humans are input-monitorable, a lot depends
on our choice of screenable inputs $I\left(  S\right)  $. \ If $I\left(
S\right)  $\ is small, then input-monitoring $S$ might be easy, but predicting
$S$ after the monitoring is in place might be hard or impossible, simply
because of the predictor's ignorance about crucial features of $S$'s
environment. \ By contrast, if $I\left(  S\right)  $\ is large, then
input-monitoring $S$ might be hard or impossible, but supposing $S$ were
input-monitored, predicting $S$ might be easy. \ Since our main interest is
the inherent difficulty of prediction, our preference should always be for the
largest $I\left(  S\right)  $\ possible.

So, suppose $S$ is input-monitorable, and suppose we've arranged things so
that all the screenable inputs to $S$---what $S$ sees, what $S$ hears,
etc.---are transmitted in real-time to the predictor. \ We then face the
question: what aspects of $S$ are we trying to predict, and what does it mean
to predict those aspects?

Our next primitive concept will be that of $S$'s \textit{observable
behaviors}. \ For the earth's atmosphere, observable behaviors might include
snow and thunderstorms; for a human brain, they might include the signals sent
out by the motor cortex, or even just a high-level description of which words
will be spoken and which decisions taken. \ Fortunately, it seems to me that,
for any sufficiently complex system, the prediction problem is \textit{not}
terribly sensitive to which observable behaviors we focus on, provided those
behaviors belong to a large \textquotedblleft universality
class.\textquotedblright\ \ By analogy, in computability theory, it doesn't
matter whether we ask whether a given computer program will ever halt, or
whether the program will ever return to its initial state, or whether it will
ever print \textquotedblleft YES\textquotedblright\ to the console, or some
other question about the program's future behavior. \ For these problems are
all \textit{reducible} to each other: if we had a reliable method to predict
whether a program would halt, then we could also predict whether the program
would print \textquotedblleft YES\textquotedblright\ to the console, by
modifying the program so that it prints \textquotedblleft
YES\textquotedblright\ if and only if it halts. \ In the same way, if we had a
reliable method to predict a subject's hand movements in arbitrary situations,
I claim that we could also predict the subject's speech. \ For
\textquotedblleft arbitrary situations\textquotedblright\ include those where
we direct the subject to translate everything she says into sign language!
\ And thus, assuming we've built a \textquotedblleft hand-prediction
algorithm\textquotedblright\ that works in those situations, we must also have
built (or had the ability to build) a speech-prediction algorithm as well.

So suppose we fix some set $B$ of observable behaviors of $S$. \ What should
count as \textit{predicting} the behaviors? \ From the outset, we should admit
that $S$'s behavior might be inherently probabilistic---as, for example, if it
depended on amplified quantum events taking place inside $S$. \ So we should
be satisfied if we can predict $S$\ in \textquotedblleft
merely\textquotedblright\ the same sense that physicists can predict a
radioactive atom: namely, by giving a probability distribution over $S$'s
possible future behaviors.

Here difficulties arise, which are well-known in the fields of finance and
weather forecasting. \ How exactly do we test predictions that take the form
of probability distributions, if the predictions apply to events that might
not be repeatable? \ Also, what's to prevent someone from declaring success on
the basis of absurdly conservative \textquotedblleft
predictions\textquotedblright: for example, \textquotedblleft%
50/50\textquotedblright\ for every yes/no question? \ Briefly, I'd say that
\textit{if} the predictor's forecasts take the form of probability
distributions, then to whatever extent those forecasts are unimpressive
(\textquotedblleft50/50\textquotedblright), \textit{the burden is on the
predictor} to convince skeptics that the forecasts nevertheless encoded
everything that \textit{could} be predicted about $S$ via allowed
interventions. \ That is, the predictor needs to rule out the hypothesis that
the probabilities merely reflected ignorance about unmeasured but measurable
variables. \ In my view, this would ultimately require the predictor to give a
causal account of $S$'s behavior, which showed explicitly how the observed
outcome depended on quantum events---the only sort of events that we know to
be probabilistic on physical grounds (see Section \ref{QMHV}).

But it's not enough to let the predictions be probabilistic; we need to scale
back our ambitions still further. \ For even with a system $S$ as simple as a
radioactive atom, there's no hope of calculating the \textit{exact}
probability that (say) $S$ will decay within a certain time interval---if only
because of the error bars in the physical constants that enter into that
probability. \ But intuitively, this lack of precision doesn't make the atom
any less \textquotedblleft mechanistic.\textquotedblright\ Instead, it seems
to me that we should call a probabilistic system\ $S$\ \textquotedblleft
mechanistic\textquotedblright\ \textit{if---and only if---the differences
between our predicted probabilities for }$S$\textit{'s behavior and the
\textquotedblleft true\textquotedblright\ probabilities can be made as small
as desired by repeated experiments.}

Yet we are still not done. \ For what does the predictor $P$ \textit{already
know} about $S$, before $P$'s data-gathering process even starts? \ If $P$
were initialized with a \textquotedblleft magical copy\textquotedblright\ of
$S$, then of course predicting $S$ would be trivial.\footnote{I thank Ronald
de Wolf for this observation.} \ On the other hand, it also seems unreasonable
not to tell $P$ \textit{anything} about $S$: for example, if $P$ could
accurately predict $S$, but only if given the hint that $S$ is a human being,
that would still be rather impressive, and intuitively incompatible with $S$'s
\textquotedblleft freedom.\textquotedblright\ \ So, as our final primitive
concept, we assume a \textit{reference class} $C$ of possible physical
systems; $P$ is then told only that $S\in C$\ and needs to succeed under that
assumption. \ For example, $C$\ might be \textquotedblleft the class of all
members of the species \textit{Homo sapiens},\textquotedblright\ or even the
class of all systems macroscopically identifiable as some \textit{particular}
Homo sapien.\footnote{We could also formulate a stronger notion of a
\textquotedblleft universal predictor,\textquotedblright\ which has to work
for \textit{any} physical system $S$ (or equivalently, whose reference class
$C$ is the set of all physical systems). \ My own guess is that, \textit{if}
there exists a predictor for \textquotedblleft sufficiently
complex\textquotedblright\ systems like human brains, then there also exists a
universal predictor. \ But I won't attempt to argue for that here.}

This, finally, leads to my attempted definition of freedom. \ Before offering
it, let me stress that \textit{nothing in the essay depends much on the
details of the definition---and indeed, I'm more than willing to tinker with
those details.} \ So then what's the point of \textit{giving} a definition?
\ One reason is to convince skeptics that the concept of \textquotedblleft
Knightian freedom\textquotedblright\ \textit{can} be made precise, once one
has a suitable framework with which to discuss these issues. \ A second reason
is to illustrate just how much little-examined complexity lurks in the
commonsense notion of a physical system's being \textquotedblleft
predictable\textquotedblright---and to show how non-obvious the questions of
freedom and predictability actually become, once we start to unravel that complexity.

\begin{quotation}
\noindent Let $S$ be any input-monitorable physical system drawn from the
reference class $C$. \ Suppose that, as the result of allowed interventions,
$S$ is input-monitored by another physical system $P=P\left(  C\right)  $ (the
\textquotedblleft predictor\textquotedblright), starting at some
time\footnote{Here we don't presuppose that time is absolute or continuous.
\ Indeed, all we need is that $S$\ passes through a discrete series of
\textquotedblleft instants,\textquotedblright\ which can be ordered by
increasing values of $t$.} $0$.$\ $ Given times $0\leq t<u\leq\infty$, let
$I_{t,u}$\ encode all the information in the screenable inputs that
$S$\ receives between times $t$ and $u$, with $I_{t,\infty}$\ denoting the
information received from time $t$ onwards. \ Likewise, let $B_{t,u}$\ encode
all the information in $S$'s observable behaviors between times $t$ and $u$.
\ (While this is not essential, we can assume that $I_{t,u}$\ and $B_{t,u}$
both consist of finite sequences of bits whenever $u<\infty$.)

Let $\mathcal{D}\left(  B_{t,u}|I_{t,u}\right)  $\ be the \textquotedblleft
true\textquotedblright\ probability distribution\footnote{Or probability
measure over infinite sequences, in the case $u=\infty$.} over $B_{t,u}$
conditional on the inputs $I_{t,u}$, where \textquotedblleft
true\textquotedblright\ means the distribution that would be predicted by a
godlike intelligence who knew the exact physical state of $S$\ and its
external environment at time $t$. \ We assume that $\mathcal{D}\left(
B_{t,u}|I_{t,u}\right)  $\ satisfies the \textquotedblleft causal
property\textquotedblright: that $\mathcal{D}\left(  B_{t,v}|I_{t,u}\right)
=\mathcal{D}\left(  B_{t,v}|I_{t,v}\right)  $\ depends only on $I_{t,v}$\ for
all $v<u$.

Suppose that, from time $0$ to $t$, the predictor $P$ has been monitoring the
screenable inputs $I_{0,t}$ and observable behaviors $B_{0,t}$, and more
generally, interacting with $S$\ however it wants via allowed interventions
(for example, submitting questions to $S$ by manipulating $I_{0,t}$, and
observing the responses in $B_{0,t}$). \ Then, at time $t$, we ask $P$ to
output a description of a function $f$, which maps the future inputs
$I_{t,\infty}$\ to a distribution $\mathcal{E}\left(  B_{t,\infty}%
|I_{t,\infty}\right)  $ satisfying the causal property. \ Here $\mathcal{E}%
\left(  B_{t,\infty}|I_{t,\infty}\right)  $\ represents $P$'s\ best estimate
for the distribution $\mathcal{D}\left(  B_{t,\infty}|I_{t,\infty}\right)  $.
\ Note that the description of $f$ might be difficult to \textquotedblleft
unpack\textquotedblright\ computationally---for example, it might consist of a
complicated algorithm that outputs a description of $\mathcal{E}\left(
B_{t,u}|I_{t,u}\right)  $\ given as input $u\in\left(  t,\infty\right)  $\ and
$I_{t,u}$. \ All we require is that the description be
\textit{information-theoretically} complete, in the sense that one
\textit{could} extract $\mathcal{E}\left(  B_{t,u}|I_{t,u}\right)  $\ from it
given enough computation time.

Given $\varepsilon,\delta>0$, we call $P$\ a $\left(  t,\varepsilon
,\delta\right)  $\textit{-predictor} for the reference class $C$ if the
following holds. \ For all $S\in C$, with probability at least $1-\delta
$\ over any \textquotedblleft random\textquotedblright\ inputs in $I_{t,u}%
$\ (controlled neither by $S$ nor by $P$),\ we have%
\[
\left\Vert \mathcal{E}\left(  B_{t,\infty}|I_{t,\infty}\right)  -\mathcal{D}%
\left(  B_{t,\infty}|I_{t,\infty}\right)  \right\Vert <\varepsilon
\]
for the actual future inputs $I_{t,\infty}$\ (not necessarily for every
\textit{possible} $I_{t,\infty}$). \ Here $\left\Vert \mathcal{\cdot
}\right\Vert $\ denotes the variation distance.\footnote{Variation distance is
a standard measure of distance between two probability distributions, and is
defined by $\left\Vert \left\{  p_{x}\right\}  -\left\{  q_{x}\right\}
\right\Vert :=\frac{1}{2}\sum_{x}\left\vert p_{x}-q_{x}\right\vert $.}

We call $C$ \textit{mechanistic} if for all $\varepsilon,\delta>0$, there
exists a $t=t_{\varepsilon,\delta}$\ and a $P=P_{\varepsilon,\delta}$\ such
that $P$\ is a $\left(  t,\varepsilon,\delta\right)  $-predictor for $C$. \ We
call $C$ \textit{free} if $C$ is not mechanistic.
\end{quotation}

Two important sanity checks\ are the following:

\begin{itemize}
\item[(a)] According to the above definition, classes $C$ of physical systems
like thermostats, digital computers, and radioactive nuclei are indeed
mechanistic (given reasonable sets of screenable inputs, allowed
interventions, and observable behaviors). \ For example, suppose that $C$ is
the set of all possible configurations of a particular digital computer; the
allowed interventions include reading the entire contents of the disk drives
and memory\ and \textquotedblleft eavesdropping\textquotedblright\ on all the
input ports (all of which is known to be technologically doable without
destroying the computer); and the observable behaviors include everything sent
to the output ports. \ In that case, even with no further interaction, the
predictor can clearly emulate the computer arbitrarily far into the future.
\ Indeed, even if the computer $S\in C$\ has an internal quantum-mechanical
random number generator, the probability distribution $\mathcal{D}$\ over
\textit{possible} future behaviors can still be approximately extremely well.

\item[(b)] On the other hand, at least mathematically, one can construct
classes of systems $C$ that are free. \ Indeed, this is trivial to arrange, by
simply restricting the screenable inputs so that $S$'s future behavior is
determined by some input stream to which $P$ does not have access.
\end{itemize}

Like many involved definitions in theoretical computer science, cryptography,
economics, and other areas, my definition of freedom is \textquotedblleft
merely\textquotedblright\ an attempt to approximate an informal concept that
one had prior to formalization. \ And indeed, there are many changes one could
contemplate to the definition. \ To give just a few examples, instead of
requiring that $\mathcal{E}\left(  B_{t,\infty}|I_{t,\infty}\right)
$\ approximate $\mathcal{D}\left(  B_{t,\infty}|I_{t,\infty}\right)  $\ only
for the actual future inputs $I_{t,\infty}$, one could demand that it do so
for all \textit{possible} $I_{t,\infty}$. \ Or one could assume a distribution
over the future $I_{t,\infty}$, and require success on \textit{most} of them.
\ Or one could require success only for most $S\in C$, again assuming a
distribution over $S$'s. \ Or one could switch around the quantifiers---e.g.,
requiring a single predictor $P$\ that achieves greater and greater prediction
accuracy $\varepsilon>0$\ the longer it continues. \ Or one could drop the
requirement that $P$ forecast all of $B_{t,\infty}$, requiring only that it
forecast $B_{t,u}$\ for some large but finite $u$. \ It would be extremely
interesting to develop the mathematical theory of these different sorts of
prediction---something I reluctantly leave to future work. \ Wouldn't it be
priceless if, after millennia of debate, the resolution of the question
\textquotedblleft are humans free?\textquotedblright\ turned out to be
\textquotedblleft yes if you define `free' with the $\varepsilon,\delta
$\ quantifiers inside, but no if you put the quantifiers
outside?\textquotedblright

A central limitation of the definition, as it stands, is that it's qualitative
rather than quantitative, closer in spirit to computability theory than
complexity theory. \ More concretely, the definition of \textquotedblleft
mechanistic\textquotedblright\ only requires that there \textit{exist} a
finite time $t$ after which the predictor succeeds; it puts no limit on the
amount of time. \ But this raises a problem: what if the predictor could
succeed in learning to emulate a human subject, but only after observing the
subject's behavior for (say) $10^{100}$\ years? \ Does making the subject
immortal, in order to give the predictor enough time, belong to the set of
allowed interventions? \ Likewise, suppose that, after observing the subject's
behavior for $20$ years, the predictor becomes able to predict the subject's
future behavior probabilistically, but \textit{only for the next }%
$20$\textit{\ years}, not indefinitely? \ The definition doesn't consider this
sort of \textquotedblleft time-limited\textquotedblright\ prediction, even
though intuitively, it seems almost as hard to reconcile with free will as the
unlimited kind. \ On the other hand, the actual numbers matter: a predictor
that needed $20$ years of data-gathering, in order to learn enough to predict
the subject's behavior for the $5$ seconds immediately afterward, would seem
intuitively compatible with freedom. \ In any case, in this essay I mostly
ignore quantitative timing issues (except for brief discussions in Sections
\ref{LIBET}\ and \ref{FALSIFY}), and imagine for simplicity that we have a
predictor that after some finite time learns to predict the subject's
responses arbitrarily far into the future.

\section{Appendix: Prediction and Kolmogorov Complexity\label{KOLMOG}}

As mentioned in Section \ref{KNIGHTIAN}, some readers will take issue with the
entire concept of Knightian uncertainty---that is, with uncertainty that can't
even be properly quantified using probabilities. \ Among those readers, some
might be content to assert that there exists a \textquotedblleft true,
objective\textquotedblright\ prior probability distribution $\mathcal{D}$ over
all events in the physical world---and while we might not know any
prescription to \textit{calculate} $\mathcal{D}$\ that different agents can
agree on, we can be sure that agents are irrational to whatever extent their
own priors deviate from $\mathcal{D}$. \ However, more sophisticated readers
might try to \textit{derive} the existence of a roughly-universal prior, using
ideas from \textit{algorithmic information theory} (see Li and Vit\'{a}nyi
\cite{livitanyi} for an excellent introduction). \ In this appendix, I'd like
to sketch how the latter argument would go and offer a response to it.

Consider an infinite sequence of bits $b_{1},b_{2},b_{3},\ldots$, which might
be generated randomly, or by some hidden computable pattern, or by some
process with elements of both. \ (For example, maybe the bits are uniformly
random, except that every hundredth bit is the \textquotedblleft majority
vote\textquotedblright\ of the previous $99$ bits.) \ We can imagine, if we
like, that these bits represent a sequence of yes-or-no decisions made by a
human being. \ For each $n\geq1$, a superintelligent predictor is given
$b_{1},\ldots,b_{n-1}$, and asked to predict $b_{n}$. \ Then the idea of
algorithmic statistics is to give a \textit{single} rule, which can be proved
to predict $b_{n}$\ \textquotedblleft almost as well\textquotedblright\ as any
other computable rule, in the limit $n\rightarrow\infty$.

Here's how it works. \ Choose any Turing-universal programming language $L$,
which satisfies the technical condition of being \textit{prefix-free}: that
is, adding characters to the end of a valid program never yields another valid
program. \ Let $P$ be a program written in $L$, which runs for an infinite
time and has access to an unlimited supply of random bits, and which generates
an infinite sequence $B=\left(  b_{1},b_{2},\ldots\right)  $\ according to
some probability distribution $\mathcal{D}_{P}$. \ Let $\left\vert
P\right\vert $\ be the number of bits in $P$. \ Then for its initial guess as
to the behavior of $B$, our superintelligent predictor will use the so-called
\textit{universal prior} $\mathcal{U}$, in which each distribution
$\mathcal{D}_{P}$\ appears with probability $2^{-\left\vert P\right\vert }/C$,
for some normalizing constant $C=\sum_{P}2^{-\left\vert P\right\vert }\leq1$.
\ (The reason for the prefix-free condition was to ensure that the sum
$\sum_{P}2^{-\left\vert P\right\vert }$\ converges.) \ Then, as the bits
$b_{1},b_{2},\ldots$\ start appearing, the predictor repeatedly updates
$\mathcal{U}$\ using Bayes' rule, so that its estimate for $\Pr\left[
b_{n}=1\right]  $\ is always%
\[
\frac{\Pr_{\mathcal{U}}\left[  b_{1}\ldots b_{n-1}1\right]  }{\Pr
_{\mathcal{U}}\left[  b_{1}\ldots b_{n-1}\right]  }.
\]
Now suppose that the \textquotedblleft true\textquotedblright\ distribution
over $B$\ is $\mathcal{D}_{Q}$, for some particular program $Q$. \ Then I
claim that, in the limit $n\rightarrow\infty$, a predictor that starts with
$\mathcal{U}$\ as its prior will do just as well as if it had started with
$\mathcal{D}_{Q}$. \ The proof is simple: by definition, $\mathcal{U}$\ places
a \textquotedblleft constant fraction\textquotedblright\ of its probability
mass on $\mathcal{D}_{Q}$\ from the beginning (where the \textquotedblleft
constant,\textquotedblright\ $2^{-\left\vert Q\right\vert }/C$, admittedly
depends on $\left\vert Q\right\vert $). \ So for all $n$ and $b_{1}\ldots
b_{n}$,%
\[
\frac{\Pr_{\mathcal{U}}\left[  b_{1}\ldots b_{n}\right]  }{\Pr_{\mathcal{D}%
_{Q}}\left[  b_{1}\ldots b_{n}\right]  }=\frac{\Pr_{\mathcal{U}}\left[
b_{1}\right]  \Pr_{\mathcal{U}}\left[  b_{2}|b_{1}\right]  \cdots
\Pr_{\mathcal{U}}\left[  b_{n}|b_{1}\ldots b_{n-1}\right]  }{\Pr
_{\mathcal{D}_{Q}}\left[  b_{1}\right]  \Pr_{\mathcal{D}_{Q}}\left[
b_{2}|b_{1}\right]  \cdots\Pr_{\mathcal{D}_{Q}}\left[  b_{n}|b_{1}\ldots
b_{n-1}\right]  }\geq2^{-\left\vert Q\right\vert }.
\]
Hence%
\[%
%TCIMACRO{\dprod \limits_{n=1}^{\infty}}%
%BeginExpansion
{\displaystyle\prod\limits_{n=1}^{\infty}}
%EndExpansion
\frac{\Pr_{\mathcal{U}}\left[  b_{n}|b_{1}\ldots b_{n-1}\right]  }%
{\Pr_{\mathcal{D}_{Q}}\left[  b_{n}|b_{1}\ldots b_{n-1}\right]  }%
\geq2^{-\left\vert Q\right\vert }%
\]
as well. \ But for all $\varepsilon>0$, this means that $\mathcal{U}$\ can
assign a probability to the correct value of $b_{n}$\ less than $1-\varepsilon
$\ times the probability assigned by $\mathcal{D}_{Q}$, only for $O\left(
\left\vert Q\right\vert /\varepsilon\right)  $ values of $n$ or fewer.

Thus, an \textquotedblleft algorithmic Bayesian\textquotedblright\ might argue
that there are only two possibilities: either a physical system is predictable
by the universal prior $\mathcal{U}$, or else---to whatever extent it
isn't---in any meaningful sense the system behaves randomly. \ There's no
third possibility that we could identify with Knightian uncertainty or freebits.

One response to this argument---perhaps the response Penrose would
prefer---would be that we can easily defeat the so-called \textquotedblleft
universal predictor\textquotedblright\ $\mathcal{U}$, using a sequence of
bits\ $b_{1},b_{2},\ldots$\ that's deterministic but noncomputable. \ One way
to construct such a sequence is to \textquotedblleft diagonalize against
$\mathcal{U}$,\textquotedblright\ defining%
\[
b_{n}:=\left\{
\begin{array}
[c]{cc}%
0 & \text{if }\Pr_{\mathcal{U}}\left[  b_{1}\ldots b_{n-1}1\right]
>\Pr_{\mathcal{U}}\left[  b_{1}\ldots b_{n-1}0\right]  ,\\
1 & \text{otherwise}%
\end{array}
\right.
\]
for all $n\geq1$. \ Alternatively, we could let $b_{n}$ be the $n^{th}%
$\ binary digit of\ Chaitin's constant $\Omega$\ \cite{chaitin}
(basically,\ the probability that a randomly generated computer program
halts).\footnote{For completeness, let me prove that the universal predictor
$\mathcal{U}$\ fails to predict the digits of $\Omega=0.b_{1}b_{2}b_{3}\ldots
$. \ Recall that $\Omega$\ is \textit{algorithmically random}, in the sense
that for all $n$, the shortest program to generate $b_{1}\ldots b_{n}$\ has
length $n-O\left(  1\right)  $. \ Now, suppose by contradiction that
$\Pr_{\mathcal{U}}\left[  b_{1}\ldots b_{n}\right]  \geq L/2^{n}$, where $L\gg
n$. \ Let $A_{n}$ be a program that dovetails over all programs $Q$, in order
to generate better and better lower bounds on $\Pr_{\mathcal{U}}\left[
x\right]  $\ for all $n$-bit strings $x$\ (converging to the correct
probabilities in the infinite limit). \ Then we can specify $b_{1}\ldots
b_{n}$\ by saying: \textquotedblleft when $A_{n}$ is run, $b_{1}\ldots b_{n}%
$\ is the $j^{th}$\ string $x\in\left\{  0,1\right\}  ^{n}$\ such that $A_{n}%
$'s lower bound on $\Pr_{\mathcal{U}}\left[  x\right]  $\ exceeds $L/2^{n}%
$.\textquotedblright\ \ Since there are at most $2^{n}/L$ such strings $x$,
this description requires at most $n-\log_{2}L+\log_{2}n+O\left(  1\right)  $
bits. \ Furthermore, it clearly gives us a procedure to generate $b_{1}\ldots
b_{n}$. \ But if $L\gg n$, then this contradicts the fact that $b_{1}\ldots
b_{n}$\ has description length $n-O\left(  1\right)  $. \ Therefore%
\[
\Pr_{\mathcal{U}}\left[  b_{1}\ldots b_{n}\right]  =%
%TCIMACRO{\dprod \limits_{i=1}^{n}}%
%BeginExpansion
{\displaystyle\prod\limits_{i=1}^{n}}
%EndExpansion
\Pr_{\mathcal{U}}\left[  b_{i}|b_{1}\ldots b_{i-1}\right]  =O\left(  \frac
{n}{2^{n}}\right)  ,
\]
and $\mathcal{U}$\ hardly does better than chance.} \ In either case,
$\mathcal{U}$\ will falsely judge the $b_{n}$'s to be random even in the limit
$n\rightarrow\infty$. \ Note that an even more powerful predictor
$\mathcal{U}^{\prime}$, equipped with a suitable oracle, could predict either
of these sequences perfectly. \ But then we could construct new sequences
$b_{1}^{\prime},b_{2}^{\prime},\ldots$\ that were unpredictable even by
$\mathcal{U}^{\prime}$, and so on.

The response above is closely related to a notion called
\textit{sophistication} from algorithmic information theory (see
\cite{antunesfortnow,gtv,livitanyi}). \ Given a binary string $x$, recall that
the \textit{Kolmogorov complexity} $K\left(  x\right)  $\ is the length of the
shortest program, in some Turing-universal programming language, whose output
(given a blank input) is $x$. \ To illustrate, the Kolmogorov complexity of
the first $n$ bits of $\pi\approx11.00100100\ldots_{2}$\ is small ($\log
_{2}n+O\left(  \log\log n\right)  $), since one only has to provide $n$ (which
takes $\log_{2}n$\ bits), together with a program for computing $\pi$\ to a
given accuracy (which takes some small, fixed number of bits, independent of
$n$). \ By contrast, if $x$\ is an $n$-bit string chosen uniformly at random,
then $K\left(  x\right)  \approx n$\ with overwhelming probability, simply by
a counting argument. \ Now, based on those two examples, it's tempting to
conjecture that \textquotedblleft every string is either highly-patterned or
random\textquotedblright: that is, either

\begin{enumerate}
\item[(i)] $K\left(  x\right)  $\ is small, or else

\item[(ii)] $K\left(  x\right)  $\ is large, but only because of
\textquotedblleft boring, random, patternless entropy\textquotedblright\ in
$x$.
\end{enumerate}

\noindent Yet the above conjecture, when suitably formalized, turns out to be
false. \ Given a set of strings $S\subseteq\left\{  0,1\right\}  ^{n}$, let
$K\left(  S\right)  $\ be the length of the shortest program that lists the
elements of $S$. \ Then given an $n$-bit string $x$ and a small parameter $c$,
one can define the $c$\textit{-sophistication} of $x$, or
$\operatorname*{Soph}_{c}\left(  x\right)  $, to be the minimum of $K\left(
S\right)  $, over all sets $S\subseteq\left\{  0,1\right\}  ^{n}$\ such that
$x\in S$\ and%
\[
K\left(  S\right)  +\log_{2}\left\vert S\right\vert \leq K\left(  x\right)
+c.
\]
Intuitively, the sophistication of $x$\ is telling us, in a near-minimal
program for $x$, how many bits of the program need to be \textquotedblleft
interesting code\textquotedblright\ rather than \textquotedblleft
algorithmically random data.\textquotedblright\ \ Certainly
$\operatorname*{Soph}_{c}\left(  x\right)  $ is well-defined and at most
$K\left(  x\right)  $, since we can always just take $S$\ to be the singleton
set $\left\{  x\right\}  $. \ Because of this, highly-patterned strings are
unsophisticated. \ On the other hand, random strings are \textit{also}
unsophisticated, since for them, we can take $S$\ to be the entire set
$\left\{  0,1\right\}  ^{n}$. \ Nevertheless, it's possible to prove
\cite{antunesfortnow,gtv} that there exist highly sophisticated strings:
indeed, strings $x$\ such that $\operatorname*{Soph}_{c}\left(  x\right)  \geq
n-O\left(  \log n\right)  $. \ These strings could thus be said to inhabit a
third category between \textquotedblleft patterned\textquotedblright\ and
\textquotedblleft random.\textquotedblright\ \ Not surprisingly, the
construction of sophisticated strings makes essential use of uncomputable processes.

However, for reasons explained in Section \ref{PENROSE}, I'm exceedingly
reluctant to postulate uncomputable powers in the laws of physics (such as an
ability to generate the digits of $\Omega$). \ Instead, I would say that, if
there's scope for freedom, then it lies in the fact that even when a sequence
of bits $b_{1},b_{2},\ldots$ is computable, the universal predictor\ is only
guaranteed to work in the limit $n\rightarrow\infty$. \ Intuitively, once the
predictor has figured out the program $Q$ generating the $b_{n}$'s, it can
\textit{then} predict future $b_{n}$'s as well as such prediction is possible.
\ However, the number of serious mistakes that the predictor makes before
converging on the correct $Q$ could in general be as large as $Q$'s
bit-length. \ Worse yet, there's no finite time after which the predictor can
\textit{know} that it's converged on the correct $Q$. \ Rather, in principle
the predictor can always be surprised by a bit $b_{n}$\ that diverges from the
predictions of whatever hypothesis $Q$ it favored in the past, whereupon it
needs to find a new hypothesis $Q^{\prime}$, and so on.

Some readers might object that, in the real world, it's reasonable to assume
an upper bound on the number of bits needed to describe a given physical
process (for example, a human brain). \ In that case, the predictor would
indeed have an absolute upper bound on $\left\vert Q\right\vert $, and hence
on the number of times it would need to revise its hypothesis substantially.

I agree that such bounds on $\left\vert Q\right\vert $\ almost certainly
exist---indeed, they must exist, if we accept the holographic principle from
quantum gravity (see Section \ref{INITIAL}). \ For me, the issue is simply
that the relevant bounds seem too large to be of any practical interest.
\ Suppose, for example, that we believed $10^{14}$\ bits---or roughly one bit
per synapse---sufficed to encode everything of interest about a particular
human brain. \ While that strikes me as an underestimate, it still works out
to roughly $40,000$\ bits per second, assuming an $80$-year lifespan. \ In
other words, it seems that a person of normal longevity would have more than
enough bits to keep the universal predictor $\mathcal{U}$\ on its toes!

The above estimate leads to amusing thought: \textit{if} one lived forever,
then perhaps one's \textquotedblleft store of freedom\textquotedblright\ would
eventually get depleted, much like an $n$-bit computer program can surprise
$\mathcal{U}$\ at most $O\left(  n\right)  $\ times. \ (Arguably, this
depletion happens to some extent over our actual lifetimes, as we age and
become increasingly predictable and set in our ways.) \ From this perspective,
freedom could be considered merely a \textquotedblleft finite-$n$
effect\textquotedblright---but this would be one case where the value of $n$ matters!

\section{Appendix: Knightian Quantum States\label{FREESTATES}}

In Section \ref{NOCLONE}, I introduced the somewhat whimsically-named
\textit{freestate}: a representation of knowledge that combines probabilistic,
quantum-mechanical, and Knightian uncertainty, thereby generalizing density
matrices, which combine probabilistic and quantum uncertainty. \ (The
\textquotedblleft freebits\textquotedblright\ referred to throughout the essay
are then just $2$-level freestates.) \ While there might be other ways to
formalize the concept of freestates, in this appendix I'll give a particular
formalization that I prefer.

A good starting point is to combine probabilistic and Knightian uncertainty,
leaving aside quantum mechanics. \ For simplicity, consider a bit
$b\in\left\{  0,1\right\}  $. \ In the probabilistic case, we can specify our
knowledge of $b$ with a single real number, $p=\Pr\left[  b=1\right]
\in\left[  0,1\right]  $. \ In the Knightian case, however, we might have a
set of \textit{possible} probabilities: for example,%
\begin{equation}
p\in\left\{  0.1\right\}  \cup\left[  0.2,0.3\right]  \cup\left(
0,4.0.5\right)  . \tag{*}\label{star}%
\end{equation}
This seems rather complicated! \ Fortunately, we can make several
simplifications. \ Firstly, since we don't care about infinite precision, we
might as well take all the probability intervals to be closed. \ More
importantly, I believe we should assume \textit{convexity}: that is, if
$p<q$\ are both possible probabilities for some event $E$, then so is every
intermediate probability $r\in\left[  p,q\right]  $. \ My argument is simply
that Knightian uncertainty includes probabilistic uncertainty as a special
case: if, for example, we have no idea whether the bit $b$\ was generated by
process $P$ or process $Q$, then for all we know, $b$ might \textit{also} have
been generated by choosing between $P$ and $Q$ with some arbitrary probabilities.

Under the two rules above, the disjunction (*) can be replaced by $p\in\left[
0.1,0.5\right]  $. \ More generally, it's easy to see that our states will
always be nonempty, convex regions of the probability simplex: that is,
nonempty sets $S$\ of probability distributions that satisfy $\alpha
\mathcal{D}_{1}+\left(  1-\alpha\right)  \mathcal{D}_{2}\in S$\ for all
$\mathcal{D}_{1},\mathcal{D}_{2}\in S$ and all $\alpha\in\left[  0,1\right]
$. \ Such a set $S$ can be used to calculate upper and lower bounds on the
probability $\Pr\left[  E\right]  $\ for any event $E$. \ Furthermore, there's
no redundancy in this description: if $S_{1}\neq S_{2}$, then it's easy to see
that there exists an event $E$\ for which $S_{1}$\ allows a value of
$\Pr\left[  E\right]  $\ not allowed by $S_{2}$\ or vice versa.

One might worry about the \textquotedblleft converse\textquotedblright\ case:
probabilistic uncertainty over different states of Knightian uncertainty.
\ However, I believe this case can be \textquotedblleft expanded
out\textquotedblright\ into Knightian uncertainty about probabilistic
uncertainty, like so:%
\[
\frac{\left(  A\text{ OR }B\right)  +\left(  C\text{ OR }D\right)  }%
{2}=\left(  \frac{A+C}{2}\right)  \text{ OR }\left(  \frac{A+D}{2}\right)
\text{ OR }\left(  \frac{B+C}{2}\right)  \text{ OR }\left(  \frac{B+D}%
{2}\right)  .
\]
By induction, any hierarchy of probabilistic uncertainty about Knightian
uncertainty about probabilistic uncertainty about... etc.\ can likewise be
\textquotedblleft collapsed,\textquotedblright\ by such a procedure, into
simply a convex set of probability distributions.

The quantum case, I think, follows exactly the same lines, except that now,
instead of a convex set of probability distributions, we need to talk about a
convex set of density matrices. \ Formally, an $n$\textit{-dimensional
freestate} is a nonempty set $S$ of $n\times n$\ density matrices such that
$\alpha\rho+\left(  1-\alpha\right)  \sigma\in S$\ for all $\rho,\sigma\in
S$\ and all $\alpha\in\left[  0,1\right]  $. \ Once again, there is no
redundancy involved in specifying our knowledge about a quantum system in this
way. \ The argument is simply the following: for all nonempty convex sets
$S_{1}\neq S_{2}$, there either exists a state $\rho\in S_{1}\setminus S_{2}%
$\ or a state $\rho\in S_{2}\setminus S_{1}$. \ Suppose the former without
loss of generality. \ Then by the convexity of $S_{2}$, it is easy to find a
pure state $\left\vert \psi\right\rangle $\ such that $\left\langle \psi
|\rho|\psi\right\rangle \notin\left\{  \left\langle \psi|\sigma|\psi
\right\rangle :\sigma\in S_{2}\right\}  $.

\end{document}